\documentclass[prd,10pt,tightenlines,floatfix,showpacs,preprintnumbers,nofootinbib,eqsecnum]{revtex4-1}

\usepackage{graphicx}
\usepackage{geometry}
\usepackage{color}
\usepackage{mathtools}
\usepackage{amsmath}
\usepackage{amssymb}
\usepackage{amsfonts}
\usepackage{afterpage}
\usepackage{caption}
\usepackage{tikz}
\usepackage{enumerate}
\usepackage{physics}
\usepackage{bbm}

\usepackage{simplewick}
\usepackage{cancel}
\usepackage{hyperref}

\usetikzlibrary{shapes.misc}

\renewcommand{\sec}[1]{\mbox{Sec.~\ref{sec:#1}}}
\newcommand{\eq}[1]{\mbox{Eq.~\eqref{eq:#1}}}
\newcommand{\fig}[1]{\mbox{Fig.~\ref{fig:#1}}}
\newcommand{\tab}[1]{\mbox{Tab.~\ref{tab:#1}}}

\newcommand{\beq}{\begin{equation}}
\newcommand{\beql}[1]{\begin{equation}\label{eq:#1}}
\newcommand{\eeq}{\end{equation}}

\renewcommand{\L}{Lagrangian}

\renewcommand{\H}{Hamiltonian}
\newcommand{\eom}{equation of motion}
\newcommand{\PHI}{\text{ I}\!\!\!\!\raisebox{.3ex}{$\infty$}}

\newcommand{\fsl}[1]{{\ooalign{\(#1\)\cr\hidewidth\(/\)\hidewidth\cr}}}
\newenvironment{itemise}{\begin{itemize}}{\end{itemize}}
\tikzset{cross/.style={cross out, draw=black, minimum size=2*(#1-\pgflinewidth), inner sep=0pt, outer sep=0pt},
cross/.default={4pt}}
\newcommand{\circled}[1]{\raisebox{.5pt}{\textcircled{\raisebox{-.9pt} {#1}}}}

\begin{document}

\title{\texorpdfstring{
  Thermal Field Theory in real-time formalism:\\
  concepts and applications for particle decays}{
  Thermal Field Theory in real-time formalism: 
  concepts and applications for particle decays
  }
}

\author{Torbj\"orn Lundberg}
\email{torbjorn.lundberg@thep.lu.se}

\author{Roman Pasechnik}
\email{Roman.Pasechnik@thep.lu.se}

\affiliation{
\\
{\sl
Department of Astronomy and Theoretical Physics, Lund
University, SE-223 62 Lund, Sweden
}}

\begin{abstract}
\vspace{0.5cm}
This review represents a detailed and comprehensive discussion of the Thermal Field Theory (TFT) 
concepts and key results in Yukawa-type theories. We start with a general pedagogical introduction 
into the TFT in the imaginary- and real-time formulation. As phenomenologically relevant 
implications, we present a compendium of thermal decay rates for several typical reactions calculated 
within the framework of the real-time formalism and compared to the imaginary-time results 
found in the literature. Processes considered here are those of a neutral \mbox{(pseudo)scalar} 
decaying into two distinct \mbox{(pseudo)scalars} or into a fermion-antifermion pair. 
These processes are extended from earlier works to include chemical potentials and 
distinct species in the final state. In addition, a \mbox{(pseudo)scalar} emission off 
a fermion line is also discussed. These results demonstrate the importance of
thermal effects in particle decay observables relevant in many phenomenological 
applications in systems at high temperatures and densities.
\end{abstract}


\maketitle

%%%%%%%%%%%%%%%%%%%%%%%%%%%%%%%%%%%%%%%%%%%%%
\section{Introduction}
\label{sec:introduction}
\setcounter{equation}{0}
%%%%%%%%%%%%%%%%%%%%%%%%%%%%%%%%%%%%%%%%%%%%%

The standard theoretical treatment of fundamental fields and their interactions primarily uses the language of quantum field theory (QFT), see for example \cite{PeskinSchroeder1995,Weinberg:1995mt,Zee:2003mt}. This treatment has historically been extremely successful in describing the behaviour of all the known particles that constitute the Standard Model of particle physics. However, the theoretical framework of this \emph{zero-temperature theory} does not naturally incorporate effects of the medium relevant at high temperatures and densities \cite{ZinnJustin:2002ru}. 

Currently, there is a limited understanding of particle interactions in hot and dense systems despite an enormous work done by several authors and intended to develop a universal formalism for treating the medium-induced phenomena both in and out of equilibrium. The problem of thermal dynamics has engaged physicists for a long time, and early attempts of analysis were made by, for example, Bloch~\cite{Bloch1932}. A pioneer in the realm of \emph{thermal (quantum) field theory} (TFT) was Matsubara~\cite{Matsubara1955} who developed the imaginary-time, or Matsubara, formalism that describes equilibrated systems. This treatment has great resemblance to zero-temperature QFT in the form of the propagators, the diagrammatic structure of the perturbative expansion, and the self-energies. It differs, however, in the way of how time is treated considering it as a purely imaginary quantity. Shortly after the developments by Matsubara, Kubo~\cite{Kubo1957} and Martin and Schwinger~\cite{MartinSchwinger1959} provided an important relation between propagators, the so-called Kubo-Martin-Schwinger (KMS) condition, which holds for thermal propagators in equilibrium. Further important developments of thermal theories with equilibrated media, such as an explicit consideration of real-time evolution, were made by Keldysh~\cite{Keldysh1965}. As a framework for a first principle analysis, \emph{thermo field dynamics} was largely developed by Matsumoto et al.~\cite{MatsumotoNakanoUmezawaManciniMarinaro1983} while a detailed investigation of QFTs at finite temperatures for real times was made by Niemi and Semenoff~\cite{NiemiSemenoff1984}. Comprehensive outlines of the path-integral treatment of quantum fields in a thermal medium may be found in~\cite{NiemiSemenoff1984, KobesKowalski1986, LandsmanvanWeert1987}, and much of the theoretical concepts presented in this review draws upon that work. Three years before the work of Kobes and Kowalski~\cite{KobesKowalski1986}, Weldon~\cite{Weldon1983} provided self-energy calculations performed using the Matsubara formalism and presented a condensed and clear overview and interpretation of the \emph{thermal decay rates}. Discontinuities of Green's functions, important for the decay rates have also been studied in detail by Kobes and Semenoff
~\cite{KobesSemenoff1985, KobesSemenoff1986}. Further studies of thermal $N$-point function, important for higher order processes have been performed by e.g. Evans~\cite{Evans1992}
  
This review aims to give a detailed and comprehensive overview of basic concepts and implications of finite-temperature effects, and thereby providing an insight on the importance of implementing the TFT framework for a consistent and universal description of high to low temperature and density phenomena. In particular, we present an explicit calculation of typical thermal effects on observable decay rates for several generic processes found in Yukawa-type field theories. In the following \sec{elements}, an overview of the underlying theory is presented starting from the statistical approach introduced in QFT by Wagner~\cite{Wagner1991}. This general approach is valid in as well as out of equilibrium. Then, the equilibrium propagator is presented, followed by an overview of the Matsubara formalism, \sec{ImaginaryTimeFormalism}, and the real-time formulation, \sec{RealTimeFormalism}. Having the real-time propagators and self-energies established, the concept and definition of the thermal decay rate is presented in \sec{ThermalDecayRates}. In \mbox{Secs.~\ref{sec:pStpSpS}-\ref{sec:FtpSF}}, thermal decay rates are presented in the real-time formalism for several processes. Here, we present such examples as a \mbox{(pseudo)scalar} particle decaying into two distinct \mbox{(pseudo)scalars}, \sec{pStpSpS}, a \mbox{(pseudo)scalar} decaying into a fermion-antifermion pair, \mbox{Secs.~\ref{sec:StFaF}-\ref{sec:pStFaF}}, and finally in \sec{FtpSF}, a \mbox{(pseudo)scalar} emission off a fermion line is discussed. In \sec{outlook}, a short summary is given.

%%%%%%%%%%%%%%%%%%%%%%%%%%%%%%%%%%%%%%%%%%%%%%%%%%%
\section{Elements of Thermal Quantum Field Theory}
\label{sec:elements}
\setcounter{equation}{0}
%%%%%%%%%%%%%%%%%%%%%%%%%%%%%%%%%%%%%%%%%%%%%%%%%%%

This section provides the theoretical formulation of TFT with the purpose
of introducing the reader to the relevant methodology that will then
further be used in explicit calculations of thermal decay rates. In order 
to obtain a diagrammatic formulation of TFT, the Green's function techniques 
are deployed in terms of a thermodynamic extension of the Wick's theorem. This
technique is then used for analysis for two-body decay rates in the thermal
medium that are related to the corresponding self-energies.

%%%%%%%%%%%%%%%%%%%%%%%%%%%%%%%%%%%%%%%%%%%%%%%%%%%
\subsection{Statistical treatment of initial states}
\label{sec:StatMech}
%%%%%%%%%%%%%%%%%%%%%%%%%%%%%%%%%%%%%%%%%%%%%%%%%%%

The underlying principle of TFT will be outlined in this subsection. The discussion 
is mainly based on the broad and comprehensive article by Wagner~\cite{Wagner1991} and 
the initial formulas are valid in equilibrium as well as out of equilibrium for arbitrary
initial distributions. This general treatment precedes a discussion of equilibrium
theories, e.g. the Matsubara formalism or the real-time formalism.

%%%%%%%%%%%%%%%%%%%%%%%%%%%%%%%%%%%%%%%%%%%%%%%%%%%
\subsubsection{Operator expectation values}
\label{sec:expectation}
%%%%%%%%%%%%%%%%%%%%%%%%%%%%%%%%%%%%%%%%%%%%%%%%%%%

Experimentally measurable observables in QFT are usually expressed as expectation values 
of certain operators evaluated at any given time. An arbitrary measurement is defined 
by two characteristic properties that can be introduced as follows.

The first property is the preparation of the initial state which fully specifies 
the system at some initial time $t_\text{in}$. It is reasonable to assume that 
no initial state can be determined exactly; rather, one should adopt the view that 
the initial state can be prepared up to a probability distribution $\rho$ over pure 
states so that the initial preparation of any system results in a \emph{mixed quantum 
state}. Mixed states are statistical ensembles of pure states where the ensemble 
specifies some lack of knowledge of the system e.g. due to noise, entanglement with 
larger systems etc. Using the language of statistical quantum mechanics, the distribution
$\rho$ assigns a weight \mbox{$\rho(n) \in [0, 1]$} to the pure states $\ket{n}$ such that 
the weight describes a fraction of the ensemble in each state. The weights are normalised
according to \mbox{$\sum_n \rho(n) = 1$} and, in the Fock space spanned by $\ket{n}$, they
provide a definition of the \emph{density matrix}:
\beq
  \hat\rho(t_\text{in}) = \textstyle\sum_n \ket{n(t_\text{in})} \rho(n) 
  \bra{n(t_\text{in})} \,.
\eeq
This expression describes a classical probability distribution over the pure states. Note
that any state, with pure or mixed initial preparation, may be described by this
formalism\footnote{A pure state $\ket{\Psi}$ has $\hat\rho = \ket{\Psi}\bra{\Psi}$ and 
one may in this case note the idempotency $\hat\rho^2 = \hat\rho$.}. Using statistical
mechanics, the operator expectation value (the observable) is given by
\begin{align}
  \ev*{\hat{\mathcal O}(t_\text{in})} & = \tr\big[ \hat\rho(t_\text{in}) 
  \hat{\mathcal O}(t_\text{in}) \big]
  \nonumber
  \\ & =
  \textstyle\sum_n \mel{n(t_\text{in})}{\rho(n) 
  \hat{\mathcal O}(t_\text{in})}{n(t_\text{in})}
  \label{eq:OperatorExpectationValue}
\end{align}
at the initial time $t_\text{in}$ for an observable $\mathcal O$. Hence, the initial
preparation of a system is equivalent to the determination of
$\hat{\mathcal\rho}(t_\text{in})$.

With the initial state now dealt with, the second property to consider is the time evolution
of the system. This process is specified by the \H\ \mbox{$\hat{\mathcal H}(t)$}, and, 
in the Heisenberg picture, the system evolves according to the Heisenberg \eom:
\beql{HeisenbergEoM}
  i\dv{t}\hat{\mathcal O}_{\mathcal H}(t)
  =
  \comm*{\hat{\mathcal O}_{\mathcal H}(t)}{\hat{\mathcal H}(t)}
  +
  i\pdv{t}\hat{\mathcal O}_{\mathcal H}(t) \,.
\eeq
The states are time-independent in this picture and specified fully at $t_\text{in}$. 
A consequence of the above \eom\ is that the time-dependent expectation value, 
the \emph{one-point function}, is given by
\beq
  \label{eq:ThermalTrace}
  \ev*{\hat{\mathcal O}_{\mathcal H}(t)}
  =
  \textstyle\sum_n \mel{n(t_\text{in})}
  {\rho(n) \hat{\mathcal O}_{\mathcal H}(t)}{n(t_\text{in})} \,.
\eeq
The subscript $\mathcal H$ will be dropped from now on.

%%%%%%%%%%%%%%%%%%%%%%%%%%%%%%%%%%%%%%%%%%%%%%%%%%%
\subsubsection{Expansion of the density operator}
\label{sec:densityOperator}
%%%%%%%%%%%%%%%%%%%%%%%%%%%%%%%%%%%%%%%%%%%%%%%%%%%

The initial density operator for an arbitrary experiment may be a multi-particle operator. 
The most general proof of Wick's theorem was provided by Danielewicz~\cite{Danielewicz1984}. 
This work implemented the necessary condition that the expectation value must be taken over
non-interacting operators with respect to a \mbox{one-particle} density operator in order for 
the theorem to hold exactly. However, Wagner~\cite{Wagner1991} provided an appropriate expansion 
of correlation functions that allows the use of Wick's theorem. This expansion is two-fold and
considers both an \emph{arbitrary} (potentially multi-particle) initial density operator and 
the time evolution operator. Let us discuss such an expansion in more detail.

A generic density operator may always be expressed in an exponential form due to the physical
requirements of hermiticity and positivity. This form is realised by the introduction of 
an operator $\hat{\mathcal B}$ defined at the initial time $t_\text{in}$. This is done 
in complete analogy to the formalism developed by Matsubara~\cite{Matsubara1955} so that
\beql{InitialDensityOperator}
     \hat\rho(t_\text{in})
     =
     \frac{1}{Z}
     \exp\ \!\! [-\lambda\hat{\mathcal B}(t_\text{in}) ]
     \,,
     \qquad
     Z
     =
     \tr\exp\ \!\! [-\lambda\hat{\mathcal B}(t_\text{in})]
     \,,
  \eeq
which is the starting point of a theory that describes thermal systems.
Note that for the grand canonical ensemble \mbox{$\hat{\mathcal B} = \hat{\mathcal H} -
\mu\hat{\mathcal N}$} and \mbox{$\lambda = \beta$}, with $\mu$ being the chemical potential,
$\hat{\mathcal N}$ -- the particle number operator, and \mbox{$\beta = T^{-1}$} -- inverse 
temperature, the formalism of Matsubara is recovered. The exponential form of $\hat\rho$ 
closely resembles the Boltzmann distribution which emerges in any system in thermal equilibrium. 

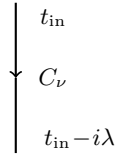
\begin{figure}
     \centering
     \begin{tikzpicture}[baseline=-2.5pt]
         \draw[->,thick] (0,1) -- (0,0);
         \draw[-,thick] (0,0) -- (0,-1);
         \node at (.5,0) {$C_\nu$};
         \node at (.5,.8) {$t_\text{in}$};
         \node at (.85,-.8) {$t_\text{in} \!-\! i\lambda$};
     \end{tikzpicture}
     \caption{ 
         The generalised thermal integration contour of Matsubara.
         For the choice of the thermal density operator in the exponential form, 
         the end-point is \mbox{$t_\text{in} - i\beta$} with $\lambda = \beta$. 
         \label{fig:MatsubaraContour}
     }
\end{figure}

One may further define the \mbox{one-particle} density operator $\hat\rho_0$ as follows
  \beq
  \label{eq:1ParticleDensityOperator}
     \hat\rho_0
     =
     \frac{1}{Z_0} \exp\ \!\! [-\lambda\hat B_0] \,,
     \qquad
     Z_0 = \tr\exp\ \!\! [-\lambda\hat B_0] \,.
  \eeq
Here, the general \mbox{one-particle} operator $\hat B_0$ is extracted from $\hat{\mathcal B}$ 
leaving the residual operator \mbox{$\hat B'$}: \mbox{$\hat B' \equiv \hat{\mathcal B} - \hat B_0$}.
Since $\hat B_0$ is an analogue to the free one-particle \H\ of the zero-temperature theory, 
a \emph{generalised interaction picture} can be defined with respect to the time-independent 
operator in the Schr\"odinger picture using $\hat B_0$ so that
  \beq
     \hat{\mathcal O}_{B_0}(\tau)
     =
     \exp\ \!\! [i\tau\hat B_0 ]
     \hat{\mathcal O}_S(t_0)
     \exp\ \!\! [-i\tau\hat B_0 ]
     \label{eq:Schr-pic}
  \eeq
for some parameter $\tau$. Comparing the translation operators (exponentials) on the right-hand side of \eq{Schr-pic} to that of \eq{1ParticleDensityOperator}, one arrives at a connection between $\tau$ and $\lambda$ shown below. Making use of this generalised interaction picture to express the residual of $\hat{\mathcal B}$ after the extraction of $\hat B_0$, Wagner \cite{Wagner1991} defined an expansion of the multi-particle \mbox{$\hat\rho$} in terms of the \mbox{one-particle} $\hat\rho_0$. From \eq{InitialDensityOperator}, we have
  \beq
      \hat\rho
      =
      \frac{Z_0}{Z} \hat\rho_0 \hat S_{C_\nu}(\tau, 0) \,,
      \qquad
      \tau = -i\lambda \,.
  \eeq
Here, the parameter $\tau$ is purely imaginary for real $\lambda$ and it defines an imaginary contour of evolution (or integration, in what follows) $C_\nu$ in the complex plane, see \fig{MatsubaraContour}. The new operator $\hat S_{C_\nu}$, a generalisation of the time evolution operator of the zero-temperature theory, was introduced above purely by trivial extraction and takes the form
  \beq
     \hat S_{C_\nu}(\tau, 0)
     =
     \exp\ \!\! [i\tau\hat B_0]
     \exp\ \!\! [-i\tau\hat {\mathcal B}] \,.
  \eeq
This operator satisfies
  \beq
     i\frac{\partial}{\partial \tau}
     \hat S_{C_\nu}(\tau, 0)
     =
     {\hat B'}_{B_0}(\tau) \hat S_{C_\nu}(\tau, 0)
  \eeq
and, hence, it can be formally integrated as
  \beq
  \label{eq:MatsubaraEvolutionOperator}
     \hat S_{C_\nu}(\tau, 0)
     =
     T_{C_\nu} \exp[
         -i\!\int_0^\tau \! \dd{\tau'}
             \hat B_{B_0}'(\tau')
     ] \,.
  \eeq
The exponential on the right-hand side has been contour-ordered along the vertical contour segment 
that goes from \mbox{0 to $-i\lambda$} according to
  \beq
     T_{C_\nu}
     \hat{\mathcal O}_1(\tau_1)
     \hat{\mathcal O}_2(\tau_2)
     =
     \begin{cases}
         \hat{\mathcal O}_1(\tau_1)
         \hat{\mathcal O}_2(\tau_2)
         \quad
         \tau_1 \geq \tau_2 \text{ on } C_\nu\,,
         \\
         \eta\hat{\mathcal O}_2(\tau_2)
         \hat{\mathcal O}_1(\tau_1)
         \quad
         \tau_1 < \tau_2 \text{ on } C_\nu\,.
     \end{cases}
  \eeq
The sign expressed by $\eta = +1 \, (-1)$ accounts for the case of commuting (anticommuting) operators. 
As mentioned above, $\hat B'$ has been expressed in the generalised interaction picture with respect 
to $\hat B_0$. Wagner~\cite{Wagner1991} thereby provides an arbitrary expectation value 
at $t_\text{in}$ as follows
  \beql{OneParticleExpectation}
     \ev*{\hat{\mathcal O}(t_\text{in})}
     =
     \frac{Z_0}{Z}\textstyle\sum_{n_0}
         \mel{n_0(t_\text{in})}
             {\rho_0 \hat S_{C_\nu}
             \hat{\mathcal O}(t_\text{in})}
             {n_0(t_\text{in})} \,.
  \eeq
This has the same form as \eq{ThermalTrace} if one regards \mbox{$\big(Z_0/Z\big) \hat S_{C_\nu} \hat{\mathcal O}(t_\text{in})$} as a non-interacting operator (i.e.\ expressed in the generalised interaction picture) to be averaged with respect to $\hat\rho_0$. Notably, $\hat\rho_0$ is a \mbox{\emph{one-particle}} density operator defined at the initial time $t_\text{in}$. As a consequence, Wick's theorem holds for each term in the expansion of $\hat S_{C_\nu}$ following the criteria presented by Danielewicz~\cite{Danielewicz1984}. The emerging contour of integration is the vertical line in \fig{MatsubaraContour}. The formalism outlined above is valid for any system that is defined by a distribution in the exponential (thermal) form \eq{InitialDensityOperator} at some initial point. Quite generally, the formalism has been shown~\cite{Wagner1991} to hold for a wide family of systems since the multi-particle initial distribution may always, at least formally, be expressed in terms of quantities that obey the Wick's theorem.
%%%%%%%%%%%%%%%%%%%%%%%%%%%%%%%%%%%%%%%%%%%%%%%%%%%%%%%%%%%%%%
%%%%%%%%%%%%%%%%%%%%%%%%%%%%%%%%%%%%%%%%%%%%%%%%%%%%%%%%%%%%%%

%%%%%%%%%%%%%%%%%%%%%%%%%%%%%%%%%%%%%%%%%%%%%%%%%%%%%%%%%%
\subsubsection{Real-time evolution operator}
\label{sec:evolutionOperator}
%%%%%%%%%%%%%%%%%%%%%%%%%%%%%%%%%%%%%%%%%%%%%%%%%%%%%%%%%%

The generalised time evolution operator, formally defined by \eq{MatsubaraEvolutionOperator}, transforms 
an operator into its average along the contour of \fig{MatsubaraContour}. In this section, the definition 
of this operator will be extended in order to incorporate evolution along the real-time axis thereby allowing 
for physical time dependence. This development is necessary in order to work out the so-called 
\emph{real-time formalism} in which real-time quantities may be extracted directly.

\begin{figure}
  \centering
  \begin{tikzpicture}[scale=2.5]
    %axis
    \draw[->] (0,-1) -- (0,1);
    \node at (0.2, 1) {$\imaginary\ t$};
    \draw[->] (-2,0) -- (3,0);
    \node at (2.9, -0.1) {$\real\ t$};
    %C1
    \node at (-1, 0.2) {$t_\text{in}$};
    \draw[->,thick] (-1,0.07) -- (1,0.07);
    \node at (0.7, 0.2) {$C_1$};
    \draw[-,thick] (1,0.07) -- (2.5,0.07);
    %arc
    \node at (2.5, 0.2) {$t_\text{fi}$};
    \draw[thick] (2.5,-0.07) arc (-90:90:0.07cm);
    %C2
    \draw[->,thick] (2.5,-0.07) -- (1.1,-0.07);
    \node at (1.3, -0.2) {$C_2$};
    \draw[-,thick] (1.1,-0.07) -- (-0.93,-0.07);
    \draw[thick] (-0.93,-0.07) arc (90:180:0.07cm);
    %C3
    \draw[->,thick] (-1,-0.14) -- (-1,-0.50);
    \node at (-0.6, -0.4) {$C_3 = C_\nu$};
    \draw[-,thick] (-1,-0.5) -- (-1, -0.9);
    \node at (-1.3, -0.9) {$t_\text{in} - i\lambda$};
  \end{tikzpicture}
\caption{A partly real-time contour of Keldysh \mbox{$C = C_1 \cup C_2 \cup C_3$}. The contour is extended 
along the real axis going through the time arguments of any \mbox{$n$-point} correlation function. 
The times $t_\text{in}$ and $t_\text{fi}$ are arbitrary and may be suitably chosen for a given 
system under the condition that all time arguments of the observable are incorporated. Note that 
the apparent offset from the real axis is superficial; $C_1$ and $C_2$ lie exactly on top of each 
other and the radii of the arcs connecting the contour segments are limited to zero.
\label{fig:KeldyshContour}}
\end{figure}
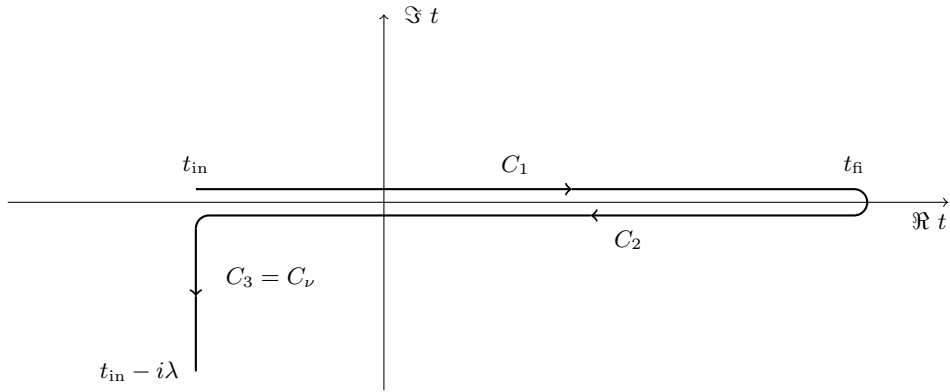

Wagner~\cite{Wagner1991} and Danielewicz~\cite{Danielewicz1984} made an observation that the time evolution 
operator must be expressed as a non-interacting operator in order for the Wick's theorem to hold. However, 
the general time-dependent \H\ contains \mbox{multi-particle} interactions. This issue was resolved 
in the previous section in \eq{OneParticleExpectation} for the vertical contour in \fig{MatsubaraContour} 
since this expression provides an expansion in which the Wick's theorem holds for each term in $\hat S_{C_\nu}$. 
In general, in order to compute observables one is interested in the resulting \emph{\mbox{$n$-point} correlation
functions}. The latter can be generalised in the TFT as follows
  \beql{nPointfunction}
      \ev*{\hat{\mathcal O}(t_1, t_2, \ldots, t_n)}
     =
     \ev*{
         \hat{\mathcal O}_1(t_1)
         \hat{\mathcal O}_2(t_2)
         \cdots
         \hat{\mathcal O}_n(t_n)
     } \,,
  \eeq
and then extended further to directly incorporate the real-time arguments as opposed to the purely imaginary 
temporal dependence along the $C_\nu$ contour discussed in the previous section. Keldysh~\cite{Keldysh1965} 
developed an expansion of the time evolution operator that is valid on the real axis. As shown in 
\fig{KeldyshContour}, a contour segment $C_1$ that goes up to some largest time $t_\text{fi}$ on the real 
axis is added. A second segment $C_2$ goes back to $t_\text{in}$, again along the real axis, before 
the piece $C_3$ goes down to \mbox{$t_\text{in}-i\lambda$.} The segments on top of the real axis 
go through all temporal arguments of the \mbox{$n$-point} function of interest. Note that 
the apparent offset from the real axis in the figure is purely for display purposes: both contour 
pieces $C_1$ and $C_2$ lie exactly on top of the axis.
%%%%%%%%%%%%%%%%%%%%%%%%%%%%%%%%%%%%%%%%%%%%%%%%%%%%%%%%%%%%%%
%%%%%%%%%%%%%%%%%%%%%%%%%%%%%%%%%%%%%%%%%%%%%%%%%%%%%%%%%%%%%%

The contour ordering $T_C$ may be defined along the extended Keldysh contour. In complete analogy 
to the previous section, the notion of $\hat{\mathcal B}$ may be extended by the introduction of
  \beq
     \hat{\mathcal K}(t)
     =
     \begin{cases}
         \hat{\mathcal H}(t)
         \qquad t\in\mathbb{R} \text{ on } C_1 \cup C_2\,,
         \\
         \hat{\mathcal B}(t)
         \qquad t=\tau \text{ on } C_\nu \,.
     \end{cases}
  \eeq
The new operator $\hat{\mathcal K}(t)$ is defined as the real-time \H\ of a given theory for real-time
arguments and the generalised \H\ discussed in the previous section on the vertical contour piece. The
procedure may be followed through introducing the \mbox{one-particle} $\hat K_0$ and residual 
$\hat K'$ operators that are defined over the entire contour, similarly to $\hat{\mathcal K}$ 
(see the analogous definition of $\hat B'$, $\hat B_0$ following \eq{1ParticleDensityOperator}). 

The \mbox{two-point} function may be written as
\beql{General2PointFunction}
  \ev*{\hat{\mathcal O}(t,t')}
  =
  \frac{Z_0}{Z} \, {}_{0} \! \ev*{T_C \hat S_C \hat O_{K_0}(t,t')}_0\,.
\eeq
The subscripts of the brackets here indicate that the trace has to be taken with respect to the 
\mbox{one-particle} density operator $\hat\rho_0$ defined in terms of $\hat K_0$. Besides, $T_C$ orders 
the operators along the entire contour $C$ and the time evolution operator
\beql{generalTimeEvolution}
     \hat S_C
     =
     T_C \exp[-i\!\int_C \dd{\tau} \hat K'_{K_0}(\tau)]
\eeq
is expressed in the generalised interaction picture with respect to $K_0$ analogously to \sec{densityOperator}. 
Hence, a formulation of an arbitrary system, valid also out of equilibrium, exists in terms of the series 
expansion of $\hat S_C$. For such series, the thermodynamic version of Wick's theorem holds \cite{Danielewicz1984}. 
Extracting the physical quantities from such an expansion is unfortunately anything but a straightforward procedure 
since it is clear that the series contains terms other than real-time correlation functions as well. Several authors 
have analysed the full expansion out of equilibrium, see for example~\cite{Keldysh1965, ChouSuHaoYu1985, Wagner1991, GarnyMuller2009, Millington2014}. In particular, Wagner~\cite{Wagner1991} provides a systematic series expansion 
of the contour-ordered two-point correlation function for fermion fields and sets up the framework in which one 
could attempt to find a solution to the generalised Dyson's equation.

From this point on, this work will focus on approaches to thermal equilibrium theories, mainly developing 
their imaginary- and real-time formulations, with the main goal of presenting the decay rates for processes 
involving neutral scalars and Dirac fermions as practically relevant example calculations in the TFT.

%%%%%%%%%%%%%%%%%%%%%%%%%%%%%%%%%%%%%%%%%%%%%%%%%%%%%%%%%%%%%%%%
\subsection{Field operators in thermal theory}
\label{sec:operatorProperties}
%%%%%%%%%%%%%%%%%%%%%%%%%%%%%%%%%%%%%%%%%%%%%%%%%%%%%%%%%%%%%%%

As was elaborated above in \sec{densityOperator}, the statistical approach to TFT produces a complex-time dependence 
through the formal equivalence between time and inverse temperature. The expectation values of operators that are 
needed for the evaluation of the corresponding observables are expressed above as traces over operators with 
a statistical weight implemented by the density matrix. This subsection provides relevant details on properties 
of the field operator and its modes in complete analogy to the zero-temperature theory.

%%%%%%%%%%%%%%%%%%%%%%%%%%%%%%%%%%%%%%%%%%%%%%%%%%%%%%%%%%%%%%%%
\subsubsection{KMS condition and ladder operators}
\label{sec:ladderKMS}
%%%%%%%%%%%%%%%%%%%%%%%%%%%%%%%%%%%%%%%%%%%%%%%%%%%%%%%%%%%%%%%%

 The trace interpretation enforce a periodicity condition on the $n$-point functions of \eq{nPointfunction}. Let us discuss this condition in more detail.
  
Consider, as an example, the simplest class of $n$-point functions -- the statistical average with respect 
to the canonical density matrix, \mbox{$\hat\rho = \exp\,[-\beta\hat H]$}, of $N$ scalar field operators expressed 
in the Heisenberg picture. In the case of a one-particle \H, the average procedure results in
  \beq
    {}_{0} \langle \hat\phi(t_1)\hat\phi(t_2)\cdots\hat\phi(t_N) \rangle_0 =
    \frac{1}{Z}\tr\big[\hat\phi(t_1)\hat\phi(t_2)\cdots\hat\phi(t_N)e^{-\beta\hat H}\big] \,.
  \eeq
  Using the cyclic property of the trace, it is straightforward to show that
  \begin{align}
  \label{eq:KMS}
     {}_{0} \langle
     \hat\phi(t_1)\hat\phi(t_2)\cdots\hat\phi(t_N)
     \rangle_0
     & =
     \frac{1}{Z}\tr\big[
         \hat\phi(t_2)\cdots\hat\phi(t_N)
         e^{-\beta\hat H}\hat\phi(t_1)
     \big] \nonumber
     \\ & =
     \frac{1}{Z}\tr\big[
         \hat\phi(t_2)\cdots\hat\phi(t_N)
         \hat\phi(t_1 + i\beta)e^{-\beta\hat H}
     \big] \nonumber
     \\ & =
     {}_{0} \langle
         \hat\phi(t_2)\cdots\hat\phi(t_N)
         \hat\phi(t_1 + i\beta)
     \rangle_0 \,.
  \end{align}
%  In the second to last equality, an important identification was made. 
It was noted already in 1932 by Bloch~\cite{Bloch1932} that the sandwiched field operator that appears 
in the above calculation can \emph{formally} be represented as a result of temporal evolution due to 
the Heisenberg \eom\ of \eq{HeisenbergEoM},
  \beq
     \hat\phi(t) = e^{it\hat H}\hat\phi(0)e^{-it\hat H} \,.
  \eeq
The condition of \eq{KMS} is the well-known Kubo-Martin-Schwinger (KMS) condition \cite{Kubo1957, MartinSchwinger1959} 
that holds for all thermal $n$-point functions.

Now, any field operator $\hat\phi(x)$ of a free theory, as a function of the space-time variable $x$, 
may be expanded 
in the basis of creation and annihilation (ladder) operators. For a bosonic (fermionic) field, these operators 
(anti)commute. In the case of a bosonic field operator, the commutation relations of the ladder operators 
may be expressed according to
  \beql{commutationRelation}
     \comm{\hat a_k}{\hat a_l^\dagger} = \delta_{kl}
  \eeq
for modes $k$, $l$ and the commutator vanishes for any other combination of operators. 
Introducing the mode functions $f_k(x)$, $f_k^*(x)$, the free field can be expanded as
  \beq
     \hat\phi(x)
     =
     \sum_k\Big[
         f_k(x)\hat a_k + f_k^*(x)\hat a_k^\dagger
     \Big] \,.
  \eeq
The mode functions of the free massive field are the solutions to the Klein-Gordon equation 
$(\partial^2 + m^2)f_k(x) = 0$ which, for spinless bosons in a finite volume $V=L^3$, read
  \beql{ScalarModeFunction}
     f_k(x)
     =
     \frac{1}{\sqrt{2\omega_kV}}
     e^{i\vb{k}\cdot\vb{x}}e^{-i\omega_kt}
  \eeq
with $\omega_k = \sqrt{\vb{k}^2 + m^2}$ denoting the energy of the $k$th mode, and
$\vb{k} = 2\pi\vb{n}/L$ where $\vb{n} = \mqty(n_x & n_y & n_z)^\text{T}$ is 
the unit 3-vector along the particle momentum. As usual, the dynamics of the field 
operator above is governed by the free field \H\
  \beq
     \hat H_0 = \int \! \dd^3 x \, \Big[
         \tfrac{1}{2}\hat\pi^2
         +
         \tfrac{1}{2}\big( \nabla\hat\phi \big)^2
         +
         \mathcal V(\hat\phi) \Big] =
     \tfrac{1}{2}\sum_k \omega_k \big[
         \hat a_k \hat a^\dagger_k
         +
         \hat a^\dagger_k \hat a_k
     \big] \,.
  \eeq
with the conjugate momentum $\pi(x) = \partial_0\hat\phi(x)$ and the free-field potential 
\mbox{$\mathcal{V} = -\tfrac{1}{2}m^2\hat\phi(x)$}.
  
In a theory described by the free-field \H, the thermal average of any combination of ladder 
operators may be evaluated. In analogy to the Heisenberg picture, one may define the shifted 
operator as
  \beq
     \hat a^\dagger_k(i\beta)
     =
     e^{-\beta\hat H_0} \hat a^\dagger_k e^{\beta\hat H_0}
     =
     \hat a^\dagger_k e^{\beta\omega_k} \,,
  \eeq
where the second equality is extracted from the commutator 
\mbox{$\comm{\hat H_0}{\hat a^\dagger_k} = \omega_k\hat a^\dagger_k$.} 
Making use of the KMS condition, one notices that
  \beql{ladderKMS}
     {}_{0} \langle \hat a^\dagger_k \hat a_l \rangle_0
     =
     {}_{0} \langle
         \hat a_l \hat a^\dagger_k(i\beta)
     \rangle_0
     =
     {}_{0} \langle \hat a_l \hat a^\dagger_k \rangle_0 \,
     e^{-\beta\omega_k} \,.
  \eeq
Applying the commutation relation of \eq{commutationRelation} results in
  \beq
     {}_{0} \langle \hat a^\dagger_k \hat a_l \rangle_0
     =
     {}_{0} \langle
         (\hat a^\dagger_k \hat a_l + \delta_{kl})
     \rangle_0 \,
     e^{-\beta\omega_k} \,.
  \eeq
From this, one obtains
  \beql{introducingThermalFunction}
     {}_{0} \langle \hat a^\dagger_k \hat a_l \rangle_0
     =
     \frac{\delta_{kl}}{e^{\beta\omega_k} - 1}
     =
     n(\omega_k) \delta_{kl} \,,
  \eeq
where $n(\omega_k)$ is the \emph{thermal distribution function} of Bose-Einstein statistics.
It is a straightforward exercise to show, in a similar fashion, that
  \begin{align}
     {}_{0} \langle \hat a_k \hat a^\dagger_l \rangle_0
     & =
     \big[ n(\omega_k) + 1 \big]\delta_{kl},
     \\
     {}_{0} \langle \hat a_k \hat a_l \rangle_0
     & =
     0,
     \\
     {}_{0} \langle
         \hat a^\dagger_k \hat a^\dagger_l
     \rangle_0
     & =
     0 \,.
  \end{align}
  
Adding the case of anticommuting ladder operators, the thermal distribution generalises to
\beql{thermaldistribution}
     n(\omega_k) = \frac{1}{e^{\beta \omega_k} - \eta} \,, \qquad \eta = \pm 1\,,
\eeq
for commuting ($\eta = +1$) and anti-communiting ($\eta = -1$) operators.
The latter guarantees that the fields satisfy the periodicity condition imposed 
by the trace interpretation. 

%%%%%%%%%%%%%%%%%%%%%%%%%%%%%%%%%%%%%%%%%%%%%%%%%%%%%%%%%%%%%%%
 \subsubsection{Thermodynamic Wick's theorem}
 \label{sec:Wick}
%%%%%%%%%%%%%%%%%%%%%%%%%%%%%%%%%%%%%%%%%%%%%%%%%%%%%%%%%%%%%%%

The most general proof of the Wick's theorem can be found in Ref.~\cite{Danielewicz1984}. 
Following the basic set up introduced above, one may deploy the same strategy as used for deriving \eq{introducingThermalFunction} and examine the general expression
  \beql{CreationAnnihilationAverage}
     {}_{0} \langle
         \hat a^\dagger_{k_1}
         \cdots
         \hat a^\dagger_{k_i}
         \hat a_{l_j} \cdots \hat a_{l_1}
     \rangle_0 \,.
  \eeq
It is a simple procedure of applying the commutation relation of \eq{commutationRelation} together 
with the KMS condition \eq{ladderKMS} to show that this expression vanishes if \mbox{$i \neq j$.} 
If the number of creation and annihilation operators is equal, this expression allows for a simple 
proof of the thermodynamic version of the Wick's theorem.
  
Indeed, applying the KMS condition once one gets
  \begin{align*}
     {}_{0} \langle
         \hat a_{k_1}^\dagger
         \cdots
         \hat a_{k_i}^\dagger \hat a_{l_j}
         \cdots
         \hat a_{l_1}
     \rangle_0
     & =
     {}_{0} \langle
         \hat a_{k_2}^\dagger
         \cdots
         \hat a_{k_i}^\dagger \hat a_{l_j}
         \cdots
         \hat a_{l_1} \hat a_{k_1}^\dagger(i \beta)
     \rangle_0
     \\ & =
     {}_{0} \langle
         \hat a_{k_2}^\dagger
         \cdots
         \hat a_{k_i}^\dagger \hat a_{l_j}
         \cdots
         \hat a_{l_1} \hat a_{k_1}^\dagger
     \rangle_0
     \, e^{-\beta \omega_{k_1}}.
  \end{align*}
Then, moving the exponential factor to the left-hand side and commuting the creation operator 
$\hat a_{k_1}^\dagger$ through the series of annihilation operators \mbox{$\hat a_{l_j}$}, 
one finds
  \begin{align*}
     &
     {}_{0} \langle
         \hat a_{k_1}^\dagger
         \cdots
         \hat a_{k_i}^\dagger \hat a_{l_j}
         \cdots
         \hat a_{l_1}
     \rangle_0
     \, e^{+\beta \omega_{k_1}}
     =
     {}_{0} \langle
         \hat a_{k_2}^\dagger
         \cdots
         \hat a_{k_i}^\dagger \hat a_{l_j}
         \cdots
         \hat a_{l_1}
         \hat a_{k_1}^\dagger
     \rangle_0
     \\ & \qquad =
     {}_{0} \! \left\langle
         \hat a_{k_2}^\dagger
         \cdots
         \hat a_{k_i}^\dagger \hat a_{l_j}
         \cdots
         \hat a_{l_2}
         \Big(
             \hat a_{k_1}^\dagger \hat a_{l_1}
             +
             \delta_{k_1 l_1}
         \Big)
     \right\rangle_{\!\!0}
     \\ & \qquad =
     {}_{0} \! \left\langle
         \hat a_{k_2}^\dagger
         \cdots
         \hat a_{k_i}^\dagger \hat a_{l_j}
         \cdots
         \hat a_{l_3}
         \Big(
             \hat a_{k_1}^\dagger \hat a_{l_2}
             +
             \delta_{k_1 l_2}
         \Big)
         \hat a_{l_1}
     \right\rangle_{\!\!0}
     +
     \delta_{k_1 l_1}
     {}_{0} \langle
         \hat a_{k_2}^\dagger
         \cdots
         \hat a_{k_i}^\dagger \hat a_{l_j}
         \cdots
         \hat a_{l_2}
     \rangle_0
     \\ & \qquad = \ldots =
     {}_{0} \langle
         \hat a_{k_1}^\dagger
         \cdots
         \hat a_{k_i}^\dagger \hat a_{l_j}
         \cdots
         \hat a_{l_1}
     \rangle_0
     \\ & \qquad \qquad \qquad \qquad +
     \delta_{k_1 l_j}
     {}_{0} \langle
         \hat a_{k_2}^\dagger
         \cdots
         \hat a_{k_i}^\dagger \hat a_{l_j-1}
         \cdots
         \hat a_{l_1}
     \rangle_0
     \\ & \qquad \qquad \qquad \qquad \qquad +
     \ldots
     +
     \\ & \qquad \qquad \qquad \qquad \qquad \qquad +
     \delta_{k_1 l_1}
     {}_{0} \langle
         \hat a_{k_2}^\dagger
         \cdots
         \hat a_{k_i}^\dagger \hat a_{l_j}
         \cdots
         \hat a_{l_2}
     \rangle_0 \,.
  \end{align*}
Here, the first term is the average of interest and it can be moved to the left-hand side 
where it combines to gives the thermal function. Collecting all terms proportional to the \mbox{$\delta$-functions results in}
  \beq
     {}_{0} \langle
         \hat a_{k_1}^\dagger
         \cdots
         \hat a_{k_i}^\dagger \hat a_{l_j}
         \cdots
         \hat a_{l_1}
     \rangle_0
     \Big( e^{+\beta \omega_{k_1}} - 1 \Big)
     =
     \sum_{m=1}^j
         \delta_{k_1 l_m}
         {}_{0} \langle
             \hat a_{k_2}^\dagger
             \cdots
             \hat a_{k_i}^\dagger \hat a_{l_j}
             \cdots
             \cancel{\hat a_{l_m}}
             \cdots
             \hat a_{l_1}
         \rangle_0 \,.
  \eeq
Moving the thermal factor to the right-hand side, one identifies the Bose-Einstein distribution such that \eq{CreationAnnihilationAverage} becomes
  \beq
     {}_{0} \langle
         \hat a_{k_1}^\dagger
         \cdots
         \hat a_{k_i}^\dagger \hat a_{l_j}
         \cdots
         \hat a_{l_1}
     \rangle_0
     =
     \sum_{m=1}^j
         {}_{0} \langle
             \hat a_{k_1}^\dagger \hat a_{l_m}
         \rangle_0
         \,
         {}_{0} \langle
             \hat a_{k_2}^\dagger
             \cdots
             \hat a_{k_i}^\dagger \hat a_{l_j}
             \cdots
             \cancel{\hat a_{l_m}}
             \cdots
             \hat a_{l_1}
         \rangle_0 \,.
  \eeq
The rightmost factor in the sum is expanded in a similar fashion using \mbox{$\Bqty{\hat a_{m_n}}_{n=1}^{j-1} \equiv \Bqty{\hat a_{l_i}}_{i=1}^{j} \Big \backslash \hat a_{l_m}$}. Then,
  \beq
     {}_{0} \langle
         \hat a_{k_2}^\dagger
         \cdots
         \hat a_{k_i}^\dagger \hat a_{l_j}
         \cdots
         \cancel{\hat a_{l_m}}
         \cdots
         \hat a_{l_1}
     \rangle_0
     =
     \sum_{n=1}^{j-1}
         {}_{0} \langle
             \hat a_{k_2}^\dagger \hat a_{m_n}
         \rangle_0
         \,
         {}_{0} \langle
             \hat a_{k_3}^\dagger
             \cdots
             \hat a_{k_i}^\dagger \hat a_{m_{j-1}}
             \cdots
             \cancel{\hat a_{m_n}}
             \cdots
             \hat a_{m_1}
         \rangle_0 \,.
  \eeq
Hence, by induction, the full average
  \mbox{
     $\langle
         \hat a_{k_1}^\dagger \cdots \hat a_{k_i}^\dagger
         \hat a_{l_j} \cdots \hat a_{l_1}
     \rangle_0$
  } 
reduces to
  \begin{align}
     &
     {}_{0} \langle
         \hat a_{k_1}^\dagger
         \cdots
         \hat a_{k_i}^\dagger \hat a_{l_j}
         \cdots
         \hat a_{l_1}
     \rangle_0
     = \nonumber
     \\ & =
     \sum_{m=1}^j
         \sum_{n=1}^{j-1}
             \ldots
                 \sum_{r=1}^{j-(i-1)}
                     {}_{0} \langle
                         \hat a_{k_1}^\dagger \hat a_{l_m}
                     \rangle_0 \,
                     {}_{0} \langle
                         \hat a_{k_2}^\dagger \hat a_{m_n}
                     \rangle_0
                     \cdots
                     {}_{0} \langle
                         \hat a_{k_i}^\dagger \hat a_{q_r}
                     \rangle_0
                     \,
                     {}_{0} \langle
                         \hat a_{h_{j-(i-1)}}
                         \cdots
                         \cancel{\hat a_{q_r}}
                         \cdots
                         \hat a_{h_1}
                     \rangle_0 \,,
  \end{align}
for \mbox{$j > i$}. Note, the rightmost factor in the nested sum must vanish since 
the average of a product of annihilation operators is zero. Similarly for \mbox{$j < i$},
  \begin{align}
     &
     {}_{0} \langle
         \hat a_{k_1}^\dagger
         \cdots
         \hat a_{k_i}^\dagger \hat a_{l_j}
         \cdots
         \hat a_{l_1}
     \rangle_0
     = \nonumber
     \\ & =
     \sum_{m=1}^j
         \sum_{n=1}^{j-1}
             \ldots
             \sum_{q=1}^2
                 {}_{0} \langle
                     \hat a_{k_1}^\dagger \hat a_{l_m}
                 \rangle_0
                 \,
                 {}_{0} \langle
                     \hat a_{k_2}^\dagger \hat a_{m_n}
                 \rangle_0
                 \cdots
                 {}_{0} \langle
                     \hat a_{k_{j-1}}^\dagger \hat a_{p_q}
                 \rangle_0
                 \,
                 {}_{0} \langle
                     \hat a_{k_j}^\dagger \hat a_{q_r}
                 \rangle_0
                 \,
                 {}_{0} \langle
                     \hat a_{k_{j +1}}^\dagger
                     \cdots
                     \hat a_{h_i}^\dagger
                 \rangle_0, \nonumber
     \\ & =
     0 \,.
  \end{align}
Hence, it can be concluded that
  \beq
     {}_{0} \langle
         \hat a_{k_1}^\dagger
         \cdots
         \hat a_{k_s}^\dagger \hat a_{l_s}
         \cdots
         \hat a_{l_1}
     \rangle_0
     =
     \sum_{m=1}^s
         \sum_{n=1}^{s-1}
             \ldots
             \sum_{q=1}^2
                 {}_{0} \langle
                     \hat a_{k_1}^\dagger \hat a_{l_m}
                 \rangle_0
                 \,
                 {}_{0} \langle
                     \hat a_{k_2}^\dagger \hat a_{m_n}
                 \rangle_0
                 \cdots
                 {}_{0} \langle
                     \hat a_{k_{s-1}}^\dagger \hat a_{p_q}
                 \rangle_0
                 \,
                 {}_{0} \langle
                     \hat a_{k_s}^\dagger \hat a_{q_1}
                 \rangle_0 \,.
  \eeq
So, any average over an equal number of creation and annihilation operators may be expanded 
as a sum over the product of every possible combination of two-operator averages. All other 
combinations of creation/annihilation operators have a vanishing thermal average. This is 
the essence of the thermodynamical Wick's theorem.

%%%%%%%%%%%%%%%%%%%%%%%%%%%%%%%%%%%%%%%%%%%%%%%%%%%%%%%%%%%%%%%%
\subsubsection{Thermo field dynamics}
\label{sec:thermoFieldDynamics}
%%%%%%%%%%%%%%%%%%%%%%%%%%%%%%%%%%%%%%%%%%%%%%%%%%%%%%%%%%%%%%%%

In Secs.~\ref{sec:ladderKMS}-\ref{sec:Wick}, an operator formalism for treating 
dynamics of the fundamental fields has been introduced. As seen already in \sec{evolutionOperator}, the physical 
operators of TFT can be defined on the real-time contour shown in \fig{KeldyshContour}. The theory gets simplified 
significantly if one considers a thermal system in equilibrium. The study of thermal operators acting on the states 
of the equilibrated Hilbert space is referred to as \emph{thermo fields dynamics}. This approach was investigated 
first by Umezawa et al (see e.g.~Refs.~\cite{UmezawaMatsumotoTachiki1982, MatsumotoNakanoUmezawaManciniMarinaro1983, MatsumotoNakanoUmezawa1984}) 
and the idea results in a doubling of the Hilbert space. Thermo field dynamics has been further developed by 
Fujimoto et al~\cite{FujimotoMorikawaSasaki1986} and incorporates, for example, a thorough structure of 
the thermal vacuum as well as provides a fundamental understanding of gauge theories in TFT~\cite{Das1997}.

The emergent property of the doubling of degrees of freedom is present also in the path-integral formulation of the 
real-time formulation. This makes the evaluation of dynamical thermal quantities, e.g. the full thermal propagator, 
cumbersome in comparison to the case of the zero-temperature theory. However, both thermo field dynamics and its 
path-integral version provide the means for evaluating the observables without considering the imaginary times, 
a property which aids the physical interpretation. We refer the interested reader to several books written on 
this topic, see for example~\cite{UmezawaMatsumotoTachiki1982, Das1997}.

%%%%%%%%%%%%%%%%%%%%%%%%%%%%%%%%%%%%%%%%%%%%%%%%%%%%%%%%%%%%%%%%
\subsection{Thermal two-point functions and propagators}
\label{sec:Thermaltwo-pointCorrAndProps}
%%%%%%%%%%%%%%%%%%%%%%%%%%%%%%%%%%%%%%%%%%%%%%%%%%%%%%%%%%%%%%%%

Knowing the properties of the thermal trace from \sec{operatorProperties}, the first non-trivial 
correlation, the two-point function, may be discussed in some detail. Its explicit form will be 
obtained and a comparison to the corresponding quantity from the zero-temperature theory is 
made straightforward. Besides, the thermal propagator of the Klein-Gordon theory is described, 
to be further generalised to charged anti-commuting multi-component fields in later sections.

%%%%%%%%%%%%%%%%%%%%%%%%%%%%%%%%%%%%%%%%%%%%%%%%%%%%%%%%%%%%%%%%
\subsubsection{Thermal two-point correlation functions}
\label{sec:Thermaltwo-pointCorr}
%%%%%%%%%%%%%%%%%%%%%%%%%%%%%%%%%%%%%%%%%%%%%%%%%%%%%%%%%%%%%%%%

Owing to the discussion in previous sections, the tools have been outlined in order to calculate the two-point 
correlation function of a free neutral scalar field which will be used as an example for thermal calculations 
later on. First, let us define the two-point correlation functions ordered in the temporal arguments according to
\begin{align}
     C^> (x, x')
     =
     {}_{0} \langle \hat \phi(x) \hat \phi(x') \rangle_0 \,,
\qquad     
     C^< (x, x')
     =
     {}_{0} \langle \hat \phi(x') \hat \phi(x) \rangle_0 \,.
\end{align}
Now, the first one can be computed by expansion of the field in terms 
of the ladder operators as
\beq
     C^> (x, x')
     =
     \raisebox{-13pt}{\scriptsize{$0$}} \!\!\!
     \left \langle
     \left(
         \sum_k
             \left[
                 f_k(x) \hat a_k
                 +
                 f_k^*(x) \hat a_k^\dagger
             \right]
     \right)
     \left(
         \sum_ l
             \left[
                 f_l(x') \hat a_l
                 +
                 f_l^*(x') \hat a_l^\dagger
             \right]
     \right)
     \right \rangle_{\!\!\! 0} \,.
\eeq
Only averages over an equal number of creation and annihilation operator will contribute 
to this expression. Hence,
  \begin{align}
    C^> (x, x')
    & =
    \sum_{k,l}
	  \left[
		f_k(x) f_l^*(x')
		{}_{0} \langle
			\hat a_k \hat a_l^\dagger
		\rangle_0
		+
		f_k^*(x) f_l(x')
		{}_{0} \langle
			\hat a_k^\dagger \hat a_l
		\rangle_0
	  \right] \nonumber
    \\ & =
    \sum_{k}
	  \left[
		- f_k(x) f_k^*(x') n(-\omega_k)
		+
		f_k^*(x) f_k(x') n(\omega_k)
	  \right] \,.
  \end{align}
Substituting the expression for the mode functions $f_k$ from \eq{ScalarModeFunction}, we obtain
  \beq
     C^>(x, x')
     =
     \frac{1}{V}
     \sum_k
         \frac{1}{2 \omega_k}
         \bigg[
             - e^{-i \big(
                 \omega_k(t-t')
                 -
                 \vb k \cdot (\vb x - \vb {x'})
             \big)}
             n(-\omega_k)
             +
             e^{+i \big(
                 \omega_k(t-t')
                 -
                 \vb k \cdot (\vb x - \vb {x'})
             \big)}
             n(\omega_k)
         \bigg] \,.
  \eeq
In the \emph{thermodynamic limit}\footnote{Keeping the density of states constant, the replacement
     $\frac{1}{V}\sum_k
     \to
     \int \! \frac{\dd^3 k}{(2\pi)^3}$
  is valid.}, an integral replaces the sum~\cite{LandsmanvanWeert1987} so that
  \beq
     C^>(x, x')
     =
     \int \! \frac{\dd[3] k}{(2 \pi)^3}
         \frac{1}{2 \pi} \frac{\pi}{\omega_k}
         \bigg[
             - n(-\omega_k)
             e^{-i \omega_k(t-t')}
             e^{+ i \vb k \cdot (\vb x - \vb {x'})}
             +
             n(\omega_k)
             e^{+i \omega_k(t-t')}
             e^{-i \vb k \cdot (\vb x - \vb {x'})}
         \bigg] \,.
  \eeq
In order to produce an integral in \mbox{four-space} over $k$, the following identity 
may be employed
  \beq
     n(\mp\omega_k) e^{\mp i \omega_k(t-t')}
     =
     \int \! \dd k_0 \,
         n(-k_0) \, e^{-i k_0 (t-t')}
         \delta(k_0 \mp \omega_k) \,,
  \eeq
so that the correlation function becomes
  \begin{align}
     C^> (x, x')
     & =
     \int \! \frac{\dd[4] k}{(2 \pi)^4}
         \frac{\pi}{\omega_k}
         \bigg[
             -n(-k_0)
             e^{-i k_0 (t-t')}
             e^{+ i \vb k \cdot (\vb x - \vb {x'})}
             \delta(k_0 - \omega_k)
             \nonumber
             \\ & \hspace{2.3cm} +
             n(-k_0)
             e^{-i k_0 (t-t')}
             e^{-i \vb k \cdot (\vb x - \vb {x'})}
             \delta(k_0 + \omega_k)
         \bigg] \nonumber
     \\ & =
     \int \! \frac{\dd[4] k}{(2 \pi)^4}
         \frac{\pi}{\omega_k}
         \bigg[
             -\delta(k_0 - \omega_k)
             +
             \delta(k_0 + \omega_k)
         \bigg]
         n(-k_0) e^{-i k \cdot (x-x')} \,.
  \end{align}
Since \mbox{$\omega_k = \omega_{-k}$} by definition of $\omega_k$, the second integral could be 
recast by a change of variables according to \mbox{$\vb k \rightarrow - \vb k$}. Hence, 
the exponential could be taken out for \mbox{$k = (k_0, \vb k)$}. Now, by identifying
  \beq
     n(-k_0)
     =
     -\left[ 1 + n(k_0) \right]
  \eeq
and by defining the \emph{free spectral density}
  \beql{spectralDensity}
     \rho_0(k)
     =
     \frac{\pi}{\omega_k}
     \bigg[
         \delta(k_0 - \omega_k) - \delta(k_0 + \omega_k)
     \bigg] \,,
  \eeq
the correlation function finally transforms to
  \beql{TwoPointScalarCorrelation-1}
     C^> (x, x')
     =
     \int \! \frac{\dd[4] k}{(2 \pi)^4}
         e^{-i k \cdot (x - x')}\tilde C^> (k) \,, \qquad
         \tilde C^> (k) = \rho_0(k) \left[ 1 + n(k_0) \right] \,.
  \eeq
Note that the free spectral density $\rho_0(k)$ here should not be confused with the initial 
density matrix.
  
Similarly, the Fourier transform of \mbox{$C^< (x, x')$} may be found as
  \beql{TwoPointScalarCorrelation-2}
     C^< (x, x')
     =
     \int \! \frac{\dd[4] k}{(2 \pi)^4}
         e^{-i k \cdot (x - x')} \tilde C^< (k) \,,
  \eeq
for which for following relation holds
  \beq
  \label{eq:CorrelationScalarRelation}
     \tilde C^> (k) = e^{k_0 \beta} \tilde C^< (k) \,.
  \eeq
The latter can be straightforwardly proven by utilising the KMS condition.

Let us now consider the average of the commutator
\begin{align}
     {}_{0} \langle
         [\hat \phi(x) , \hat \phi(x')]
     \rangle_0
     & =
     \int \! \frac{\dd[4] k}{(2 \pi)^4} \,
         e^{-i k \cdot (x-x')}
         \left[ \tilde C^> (k) - \tilde C^< (k) \right] \,.
\end{align}
Making use of the relation between the two Fourier transforms in \eq{CorrelationScalarRelation}, 
one finds that
\beq
\label{eq:ScalarCommutator}
     {}_{0} \langle
         [\hat \phi(x) , \hat \phi(x')]
     \rangle_0
     =
     \int \! \frac{\dd[4] k}{(2 \pi)^4}
         e^{-i k \cdot (x-x')} \rho_0(k) \,,
\eeq
utilising the fact that $\left[ 1 + n(k_0) \right]\left[ 1 - e^{- k_0 \beta} \right] \equiv 1$.
Such an average be used in analogy to the zero-temperature theory e.g. in order to define the thermal 
retarded and advanced propagators as follows
\begin{align}
     iD^R(x-x')
     & =
     \Theta\big(x^0 - {x'}^0\big) \,
     {}_{0} \langle
         [\hat \phi(x) , \hat \phi(x')]
     \rangle_0 \,,
     \\
     iD^A(x-x')
     & =
     -\Theta\big({x'}^0 - x^0\big) \,
     {}_{0} \langle
         [\hat \phi(x) , \hat \phi(x')]
     \rangle_0 \,,
\end{align}
where $\Theta(x)$ is the standard Heaviside step function.

%%%%%%%%%%%%%%%%%%%%%%%%%%%%%%%%%%%%%%%%%%%%%%%%%%%%%%%%%%%%%%%%
\subsubsection{Thermal propagators}
\label{sec:Thermal-props}
%%%%%%%%%%%%%%%%%%%%%%%%%%%%%%%%%%%%%%%%%%%%%%%%%%%%%%%%%%%%%%%%

In analogy to the correlation function \mbox{$\langle\hat{\mathcal O}(t_1, t_2,\ldots, t_n)\rangle$} 
of \eq{nPointfunction}, one may define the \emph{thermal Green's functions} as
\beql{thermalScalarGreen}
     G(x_1, x_2, \ldots, x_n)
     =
     \langle T \{
         \hat \phi(x_1) \hat \phi(x_2)
         \cdots
         \hat \phi(x_n)
     \} \rangle
\eeq
that are the time-ordered thermal averages of products of the field operators $\big\{ \hat \phi(x_i) \big\}$ 
as prescribed by \eq{General2PointFunction}.

As an important non-trivial example, the two-point thermal Green's function for the free theory reads
\beq
     G_0 (x, x')
     \overset{\text{free}} \equiv
     {}_{0} \langle T \{
         \hat \phi(x) \hat \phi(x')
     \} \rangle_0 \,,
\eeq
and in order for \mbox{$G_0 (x, x')$} to be a Green's function of the theory, it must obey 
the Klein-Gordon \eom
\beq
\label{eq:KGThermalGreensFunction}
     (\partial^2 + m^2) G_0(x,x') = -i \delta^{(4)}(x-x') \,.
\eeq
Using the definition of \mbox{$G_0 (x, x')$}, one identifies the thermal 
propagator of the free field
\begin{align}
     iD(x-x')
     & =
     \Theta(t - t')
     {}_{0} \langle \hat \phi(x) \hat \phi(x') \rangle_0
     +
     \Theta(t' - t)
     {}_{0} \langle \hat \phi(x') \hat \phi(x) \rangle_0
     \\ & \equiv
     \Theta(t - t')
     iD^>(x-x')
     +
     \Theta(t' - t)
     iD^<(x-x') \,,
\end{align}
which should itself be a Green's function of the Klein-Gordon equation, i.e.
\beq
     (\partial^2 + m^2) D(x-x') = \delta^{(4)}(x-x') \,.
\eeq

In order to demonstrate this explicitly, let the derivative act on the variable $x$ as will be 
denoted by \mbox{$\partial_{(x)}^2$} while treat $x'$ as a constant shift. Then,
\begin{align}
     (\partial_{(x)}^2 + m^2) D(x-x')
     & =
     (\partial_{(x)}^2 + m^2)
     \Big \{
         \Theta(t-t') D^>(x,x') + \Theta(t' - t) D^<(x,x')
     \Big \}
     \nonumber
     \\ & =
     \big[ \partial_{(x)}^2 \Theta(t-t') \big] D^>(x,x')
     +
     \big[ \partial_{(x)}^2 \Theta(t'-t) \big] D^<(x,x')
     \nonumber
     \\ & \quad +
     2 \Big \{
         \big[ \partial_{(x) \mu} \Theta(t-t') \big]
         \big[ \partial_{(x)}^\mu D^>(x,x') \big]
         \nonumber
         +
         \big[ \partial_{(x) \mu} \Theta(t'-t) \big]
         \big[ \partial_{(x)}^\mu D^<(x,x') \big]
     \Big \}
     \nonumber
     \\ & \quad +
     \Theta(t\!-\!t')
     \big[ (\partial_{(x)}^2 + m^2) D^>(x,x') \big]
     +
     \Theta(t'\!-\!t)
     \big[ (\partial_{(x)}^2 + m^2) D^<(x,x') \big] \,.
\end{align}
Using the identity \mbox{$\dd \Theta(t)/\dd t = \delta(t)$}
and rewriting the terms with second-order derivatives as
\begin{align}
     \big[ \partial_{(x)}^2 \Theta (t-t') \big] D^>(x,x')
     & =
     \partial_{(x)}^\mu
     \Big \{
         \big[ \partial_{(x) \mu} \Theta(t-t') \big]
         D^>(x,x')
     \Big \}
     -
     \big[ \partial_{(x) \mu} \Theta(t-t') \big]
     \big[ \partial_{(x)}^\mu D^>(x,x') \big]
     \nonumber
     \\ & =
     \partial_{(x)}^0
     \Big \{ \delta(t-t') D^>(x,x') \Big \}
     -
     \delta(t-t') \big[ \partial_{(x)}^0 D^>(x,x') \big]
\end{align}
  and
\beq
     \big[ \partial_{(x)}^2 \Theta(t'-t) \big] D^<(x,x')
     =
     \partial_{(x)}^0
     \Big \{ - \delta(t-t')  D^<(x,x') \Big \}
     +
     \delta(t-t') \big[ \partial_{(x)}^0 D^<(x,x') \big].
\eeq
we notice that the terms proportional to $\delta(t-t')$ disappear. 
Besides, one observes that
\begin{align}
     (\partial_{(x)}^2 + m^2) D^>(x,x')
     & =
     (\partial_{(x)}^2 + m^2)
     \int \! \frac{\dd^{4} k}{(2 \pi)^4} \,
         e^{-i k \cdot (x-x')} \tilde D^>(k)
     \nonumber
     \\ & =
     \int \! \frac{\dd^{4} k}{(2 \pi)^4} \,
         e^{-i k \cdot (x-x')} [-k^2 + m^2] \tilde D^>(k) = 0 \,,
\end{align}
due to \mbox{$\tilde D^>(k) \propto \delta(k^2 - m^2)$}, and the same holds true for the term containing 
\mbox{$(\partial_{(x)}^2 + m^2) D^<(x,x')$}. Hence, the full expression is straightforwardly evaluated to
\begin{align}
     (\partial_{(x)}^2 + m^2) D (x-x')
     & =
     \delta(t-t') \partial_{(x)}^0
     \Big \{ D^>(x,x') - D^<(x,x') \Big \}
     \nonumber
     \\ & \hspace{-0.5cm}
     \overset{\text{\eq{ScalarCommutator}}}{=}
     \delta(t-t') \partial_{(x)}^0
     \int \! \frac{\dd^{4} k}{(2 \pi)^4} \,
         e^{-i k \cdot (x-x')} \rho_0(k)
     \nonumber
     \\ & = - i \delta^{(4)}(x-x') \,,
\end{align}
thus, completing the proof.

%%%%%%%%%%%%%%%%%%%%%%%%%%%%%%%%%%%%%%%%%%%%%%%%%%%%%%%%%%%%%%%%%%%%
\subsubsection{Spectral representation}
\label{sec:Spectralrep}
%%%%%%%%%%%%%%%%%%%%%%%%%%%%%%%%%%%%%%%%%%%%%%%%%%%%%%%%%%%%%%%%%%%%

A special form of the thermal propagator \mbox{$D(x-x')$} is known as its \emph{spectral representation} 
in terms of the spectral density defined above in \eq{spectralDensity}. The propagator introduced in the previous 
section may be re-written as an integral in momentum space as
\begin{align}
     D(x-x')
     & =
     \Theta(t-t') D^>(x,x') + \Theta(t'-t) D^<(x,x')
     \nonumber
     \\ & \overset{\text{KMS}}{=}
     \Theta(t-t')
     \int \! \frac{\dd^{4} k}{(2 \pi)^4} \,
         e^{-i k \cdot (x-x')} \rho_0(k)[1 + n(k_0)]
     \nonumber
     \\ & \qquad +
     \Theta(t'-t)
     \int \! \frac{\dd^{4} k}{(2 \pi)^4} \,
         e^{-i k \cdot (x-x')} e^{-k_0 \beta}
         \rho_0(k)[1 + n(k_0)] \,.
\end{align}
  Note that \mbox{$e^{-k_0 \beta}[1 + n(k_0)] = n(k_0)$,} recovering the spectral representation 
  for a bosonic scalar field (see e.g.~Ref.~\cite{DolanJackiw1974})
\beq
\label{eq:ScalarPropagator}
     D(x-x')
     =
     \int \! \frac{\dd^{4} k}{(2 \pi)^4} \,
         e^{-i k \cdot (x-x')} \rho_0(k)
         \Big \{ \Theta(t-t') + n(k_0) \Big \} \,, \qquad \eta=+1 \,.
\eeq

For practical applications in Feynman diagrams computation, it is instructive to turn into 
momentum space and to compare the Fourier-transform of the thermal propagator to its zero-temperature
counterpart. To recover an explicit form of the Fourier transform of \mbox{$D(x-x')$}, one starts 
with \eq{ScalarPropagator} and writes
\beq
\label{eq:tildeDk}
     \tilde D(k)
     =
     \int \! \frac{\dd^4 q}{(2 \pi)^4} \,
         \rho_0(q)
         \int \, \dd^4 x \,
         \Theta(t-t') e^{-i (q-k) \cdot (x-x')}
     +
     \rho_0(k) n(k_0) \,.
\eeq
Using the following representation of the Heaviside function,
\beq
     \Theta(t-t')
     =
     -
     \frac{1}{2 \pi i}
     \lim_{\epsilon \to 0^+}
     \int_\mathbbm{R} \! \dd \tau \,
         \frac{1}{\tau + i \epsilon} e^{- i \tau (t-t')},
\eeq
the integral in \eq{tildeDk} may be re-written as
\begin{align}
     \tilde D(k)
     & =
     i \lim_{\epsilon \to 0^+}
     \int \! \frac{\dd s_0'}{2 \pi} \,
         \frac{1}{s_0' + i \epsilon}
         \int \! \frac{\dd^4 q}{(2 \pi)^4} \,
             \rho_0(q)
             \int \! \dd^4 x \,
                 e^{-i s_0' (t-t')}
                 e^{-i (q-k) \cdot (x-x')}
     +
     \rho_0(k) n(k_0)
     \nonumber
     \\ & =
     i \lim_{\epsilon \to 0^+}
     \int \! \frac{\dd k_0'}{2 \pi} \,
         \frac{\rho_0(k_0', \vb k)}
             {k_0 - k_0' + i \epsilon}
     +
     \rho_0(k) n(k_0) \,, \label{eq:FTScalarPropagator}
\end{align}
where \mbox{$k_0' = k_0 - s_0'$}. This rather compact expression for the thermal propagator has a straightforward 
interpretation. The first term contains no reference to the thermal 
(medium) parameter $\beta$ and resembles the propagator of the zero-temperature theory. The last term contains 
the thermal distribution function $n(k_0)$ that characterises the medium in equilibrium through the temperature parameter 
$T\equiv \beta^{-1}$ and for the Bose-Einstein (Fermi-Dirac) distribution governing the statistics of bosons (fermions). 
Note, this function vanishes when \mbox{$T\to0$.} The thermal propagator has therefore naturally been split into 
thermal and non-thermal terms.

%%%%%%%%%%%%%%%%%%%%%%%%%%%%%%%%%%%%%%%%%%%%%%%%%%%%%%%%%%%%%%%
\subsection{Contour path-integral formulation}
\label{sec:ContourPathIntegralFormulation}
%%%%%%%%%%%%%%%%%%%%%%%%%%%%%%%%%%%%%%%%%%%%%%%%%%%%%%%%%%%%%%%

From the outline of the TFT given above, together with some illuminating examples, it has become clear that in order to perform 
explicit calculations, one must specify the initial thermal density. Focusing on the case of density operators that describe 
the thermal systems in equilibrium, this subsection overviews the path-integral formulation of such systems starting 
from the density distribution of the grand canonical ensemble. Then, the \emph{generating functional} is discussed before 
the perturbative expansion of the thermal path-integral formalism is considered. The general propagator of TFT 
is discussed and the KMS condition is then restated in the context of path integral formulation.

\mbox{One-particle} quantities will be assumed initially since it has been shown above that the thermal functions of multi-particle systems may be recast into a series expansion over one-particle quantities and much of the theory from this point on is initially 
introduced for free particles. Interactions are thereafter introduced perturbatively. The material presented here is primarily 
based on an extensive discussion by Landsman and van Weert in Ref.~\cite{LandsmanvanWeert1987}.

%%%%%%%%%%%%%%%%%%%%%%%%%%%%%%%%%%%%%%%%%%%%%%%%%%%%
\subsubsection{Grand canonical ensemble}
\label{sec:GrandCanonicalEnsemble}
%%%%%%%%%%%%%%%%%%%%%%%%%%%%%%%%%%%%%%%%%%%%%%%%%%%%

In thermal equilibrium, the \H\ $\hat{\mathcal H}$ is time-independent, and a system of charged particles 
is characterised by the density operator of the grand canonical ensemble
\beq
     \hat\rho
     =
     \exp\ \!\![
         -\Phi
         -\textstyle \sum_a \alpha_a \hat Q_a
         -\beta \hat{\mathcal H}
     ] \,,
     \qquad
     \Phi
     =
     \log\tr[
         - \textstyle \sum_a \alpha_a \hat Q_a
         - \beta \hat {\mathcal H}
     ] \,.
\eeq
Here, $\hat Q_a$ are the conserved charges with $a\in\{1, A\}$, and $\Phi$ is the thermodynamical potential 
related to the partition function through \mbox{$\Phi = -\log Z$}. The charge operators $\hat Q_a$ and 
the Lagrange multipliers $\alpha_a$ may be related to the number density operator and the chemical 
potential, respectively, as
\beq
     \hat N_a = \frac{1}{V}\ev*{\hat Q_a} \,,
     \qquad
     \alpha_a = -\beta\mu_a\,,
\eeq
where $V$ is the volume of the system and $\mu_a$ corresponds to the chemical potential related 
to each type of charge $a$. As seen in the previous section, the form of $\hat\rho$ determines 
the expectation value of any \mbox{$n$-point} function\footnotemark.

\footnotetext{This work presents the formalisms and results in the rest frame of the medium, for simplicity. 
A Lorentz-covariant formulation of the grand canonical ensemble has been introduced 
by Niemi and Semenoff in Ref.~\cite{NiemiSemenoff1984}.}

%%%%%%%%%%%%%%%%%%%%%%%%%%%%%%%%%%%%%%%%%%%%%%%%%%%%%%%%
\subsubsection{Generating functional}
\label{sec:GeneratingFunctional}
%%%%%%%%%%%%%%%%%%%%%%%%%%%%%%%%%%%%%%%%%%%%%%%%%%%%%%%%

Analogous to zero-temperature QFT and to the scalar-field theory example in \eq{thermalScalarGreen}, 
the \mbox{$n$-point} correlation functions (or thermal Green's functions) of arbitrary (commuting or anticommuting) 
field operators $\hat\PHI$ can be defined as
\beq
     G_C(x_1, x_2, \ldots, x_n)
     =
     \ev*{
         T_C \hat\PHI(x_1)\hat\PHI(x_2)\cdots\hat\PHI(x_n)
     } \,.
\eeq
An important detail brushed over in the example calculations above is the fact that the thermal Green's functions 
depend explicitly on the choice of integration contour $C$. This is evident mainly in \eq{generalTimeEvolution}. 
The contour of integration may be chosen quite arbitrarily but it becomes restricted for diagrammatic formalisms 
that require the thermodynamic version of Wick's theorem to hold.

The thermal Green's functions can be generated by the functional
\beql{gen-funl-operator}
     Z[j]
     =
     Z[0]\ev{
         T_C\exp[i\!\int_C \! \dd^4 x \, j(x) \hat\PHI(x)]
     }
\eeq
expressed in terms of the \mbox{c-number} sources \mbox{$j(x)$} through functional differentiation as
\beql{ThermalGreensFunctions}
     G_C(x_1, x_2, \ldots, x_n)
     =
     \frac{1}{Z[0]}
     \frac{\delta^n Z[j]}{
         i\delta j(x_1)\cdots i\delta j(x_n)
     }
     \Bigg\vert_{j=0} \,.
\eeq
The normalisation here is \mbox{$Z[0] = \exp[\Phi(\beta, V)]$.}

The generating functional of \eq{gen-funl-operator} contains all \mbox{$n$-point} functions \mbox{$G_C(x_1, \ldots, x_n)$}. 
In analogy to the zero-temperature theory, all \emph{disconnected diagrams} can be factorised out by considering only 
the \emph{connected Green's functions} \mbox{$G_C^{(n)}(x_1, \ldots, x_n)$} defined by
\beql{ConnectedGreensFunctions}
     G_C^{(n)}(x_1, \ldots, x_n)
     =
     \frac{\delta^n W[j]}{
         i\delta j(x_1) \cdots i\delta j(x_n)
     }
     \Bigg\vert_{j=0},
     \qquad
     W[j] = \log Z[j] \,.
\eeq
The connected Green's functions satisfy the following \emph{the cluster property:}
\beq
     \lim_{\lvert x_i - x_j \rvert \to \infty}
     G_C^{(n)}(x_1, \ldots, x_n)
     =
     0 \,,
\eeq
which may be verified by direct evaluation. That is, the connected pieces fall off to zero when the distance 
between the field points increases. The expansion of $W[j]$
\beql{Cumulant}
     W[j]
     =
     \log Z[0]
     +
     \sum_{n=1}^\infty
         \frac{i^n}{n!}
         \int\!\!\cdots\!\!\int \dd^4 x_1 \cdots \dd^4 x_n
         \,\,
         G_C^{(n)}(x_1, \ldots, x_n)
         \cdot
         j(x_1) \cdot \ldots \cdot j(x_n)
\eeq
is often referred to as the \emph{cumulative (or cumulant) expansion} in statistical mechanics. The resulting series trivially 
generates the connected Green's functions, i.e. the sum of all diagrams that are fully connected. As a consequence 
of the logarithmic relation between $Z[j]$ and $W[j]$, the contribution from \emph{vacuum bubble diagrams}, $Z[0]$, 
factorises and may be absorbed by the normalisation of the Green's functions. An illustration of the order-by-order 
cancellation of \emph{disconnected} bubble diagrams can be found in~\ref{sec:bubble-example}.

Time-ordering procedure on the contour, $T_C$, may be explicitly expressed by parametrising $C$ according to 
\mbox{$t = z(\tau)$}, with $\tau$ being real and monotonically increasing along the contour. Through the definition 
of the \emph{contour step function}
\beql{ContourStepFunction}
     \Theta_C(t - t') = \Theta(\tau - \tau') \,,
\eeq
the \mbox{two-point} function, taken as an example, reads
\beql{GeneralTwoPointFunction}
     G_C(x, x')
     =
     \Theta_C(t-t')\ev*{\hat\PHI(x) \hat\PHI(x')}
     +
     \eta\Theta_C(t'-t)\ev*{\hat\PHI(x') \hat\PHI(x)} \,,
\eeq
where the sign factor $\eta = \pm1$ for commuting/anticommuting (bosonic/fermionic) field operators. Weldon~\cite{Weldon1983} 
showed that this expression is well behaved only if the imaginary component of the contour $C$ does not increase with $\tau$. 
Feynman, Matthews and Salam~\cite{FeynmanHibbs1965, MatthewsSalam1955} provided the first proof of this statement in the context 
of the path-integral formulation.

It should be mentioned at this point that the terms \emph{scalar}, \emph{boson} and \emph{fermion} will be used extensively throughout 
the following theoretical outline below, as well as \emph{commuting} and \emph{anticommuting} fields and operators. Some of those labels 
are used in a somewhat different context by various authors and hereby we would like to clarify on how such terms are used within this work. 

The term `scalar' intends to reference a scalar quantity both in Lorentz and in any internal space. Hence, a `scalar field' associates 
a single value to each point in space-time but \emph{it may be either a commuting or an anticommuting field (the latter being a Grassmann
variable)} in this review. A `boson' refers to an operator that obeys canonical \emph{commutation} relations. Hence, a `scalar boson' 
is the specification of a scalar (i.e. not a vector) field that obeys Bose-Einstein statistics. The use of the term `fermionic scalar' 
in this review might induce some confusion for the reader; the label refers to a scalar field that anticommutes (which is not a vector 
in a given space). This distinction of scalars is introduced simply because it is an easier exercise and a common approach in the literature 
to present the initial theory in terms of either commuting or anticommuting scalar (neutral one-component) fields. The formalism is then 
later extended to include fields that carry Lorentz or spin components. Hence, proper care must be taken when constructing the \L\ so as 
to only include field types with the proper structure in Lorentz and spin space respecting the relevant symmetries of the theory. 
The term `fermion' is applied along its most common usage to Dirac spinors that carry components in spin space.

%%%%%%%%%%%%%%%%%%%%%%%%%%%%%%%%%%%%%%%%%%%%%%%%%%%%%%%%%%%%%%%%
\subsubsection{Path-integral formulation}
\label{sec:pathIntegralFormulation}
%%%%%%%%%%%%%%%%%%%%%%%%%%%%%%%%%%%%%%%%%%%%%%%%%%%%%%%%%%%%%%%%

The functional that generates the \mbox{$n$-point} functions was determined in the previous section as a thermal expectation value 
over operators (\eq{gen-funl-operator}). In the path-integral formulation, the state vectors that span the Fock space introduced 
in \sec{StatMech} can be used as a basis of \emph{coherent states} that are eigenstates of the annihilation component of 
the field operator. The discrete modes are then exchanged for the continuous \mbox{three-space} variable\footnote{For more details, 
we refer the interested reader to any suitable textbook on QFT, e.g. Ref.~\cite{PeskinSchroeder1995}}, $\vb{x}$. This coherent state 
basis may be chosen to express the field states $\ket{\phi(\vb x); 0}$ at time $t=0$ that are eigenstates of the field operator
\mbox{$\hat\phi(x)$}:
\beql{EigenstateEquation}
     \hat\phi(x)\ket{\phi(\vb x);0}
     =
     \phi(x)\ket{\phi(\vb x);0} \,,
     \qquad
     \hat\phi(x) = e^{i\hat Ht}\hat\phi(0,\vb x)e^{-i\hat Ht} \,.
\eeq
The theory of \emph{neutral bosonic} field considered in this subsection provides a simple example with vanishing 
chemical potential. The coherent states generalised to the continuum picture can be used to express 
the \emph{functional integration measure} and further to express the thermal trace in the generating 
functional as
\beql{ExplicitTraceFunctional}
     Z[j]
     =
     \int \! [\mathcal D \varphi]
     \ev{e^{-\beta\hat H} T_C\exp[i\!\int_C \! \dd^4 x \,
         j(x)\hat\phi(x)]}{\varphi;t_\text{in}} \,.
\eeq
Note that the bra and ket states here must be taken at equal times to uphold the trace interpretation.

Given the time evolution of the field operator in \eq{EigenstateEquation}, the states absorb 
the exponential through
\beq
     \ket{\phi(\vb x);t} = e^{i\hat H t}\ket{\phi(\vb x);0} \,.
\eeq
The action of the canonical density operator inside the trace of \eq{ExplicitTraceFunctional} can then analogously 
be interpreted as a complex shift in time, i.e.
\beq
     \bra{\varphi;t_\text{in}}e^{-\beta\hat H}
     =
     \bra{\varphi;t_\text{in}\!-\!i\beta} \,.
\eeq
This, so far, purely formal equivalence of the thermal density operator to the time-evolution operator was first noted by
Bloch~\cite{Bloch1932} already in 1932 and from this observation follows the introduction of the complex temporal contour 
discussed in \sec{StatMech}. Note specifically that the eigenvalue of the field operator, $\varphi$, remains in order for 
the trace interpretation to hold. The functional measure \mbox{$[\mathcal D\varphi]$} is such as it may only pick out 
the fields with this periodicity condition over \mbox{$i\beta$} so that \mbox{$\varphi(t_\text{in},\vb x) = 
\varphi(t_\text{in} - i\beta,\vb x)$}. With the formal observation above, the shifted trace can be restated in terms 
of some arbitrary initial and final times $t_\text{i}$, $t_\text{f}$ with the aid of the Feynman-Salam-Matthews 
(FSM) formula~\cite{FeynmanHibbs1965, MatthewsSalam1955},
\begin{align}
\label{eq:FSM}
     \mel*{
         \varphi_\text{f}(\vb x);t_\text{f}
     }{T_CF[\hat\phi]}{
         \varphi_\text{i}(\vb x);t_\text{i}
     }
     & =
     \mathcal N' \iint \! [\mathcal D \phi] [\mathcal D \pi] \,
         F[\phi]
         \exp[
             i\int_{t_\text{i}}^{t_\text{f}} \dd t \int_V \dd^3 x \, 
                 \big\{
                     \pi(x) \dot{\phi}(x) - \mathcal H(x)
                 \big\} 
         ]
     \nonumber
     \\ & =
     \mathcal N \int \! [\mathcal D \phi] \,
         F[\phi]e^{iS[\phi]} \,,
\end{align}
if the path-integral on the right-hand side is taken over all \mbox{c-number} fields that satisfy the boundary conditions
\beq
  \phi(t_\text{i}, \vb x) = \varphi_\text{i}(\vb x) \quad \text{and} \quad 
  \phi(t_\text{f}, \vb x) = \varphi_\text{f}(\vb x) \,.
\eeq

  In fact, \eq{FSM} remains valid for arbitrary initial and final times under the single restriction that the imaginary component of the integration contour $C$ connecting $t_\text{i}$ and $t_\text{f}$ may not increase along the direction of the contour~\cite{LandsmanvanWeert1987}. Hence, $C$ must go downwards in the complex plane or extend in a parallel direction to the real axis. When specifying the theory to one with no derivative couplings, the normalisation $\mathcal N'$ in \eq{FSM} may absorb the Gaussian integration over the conjugate momentum so that \mbox{$\mathcal N' \to \mathcal N$}. The action is then
\beq
     S[\phi]
     =
     \int_{t_\text{i}}^{t_\text{f}} {\dd t} \int \! \dd^3 x \,
         \mathcal L(x)
         \underset{\text{by }C.}{
             \underset{\text{connected}}{
                 \underset{
                     \text{For } t_\text{i}, t_\text{f}
             }{\to}}
         }
     \int_C \! \dd^4 x \, \mathcal L(x) \,,
\eeq
and hence the generating functional becomes
\begin{align}
\label{eq:FullGeneratingFunctionalScalar}
     Z[j]
     & =
     \mathcal N
     \int \! [\mathcal D \phi] \,
         \exp[
             i\!\int_C \! \dd^4 x \,
                 \Big( \mathcal L(x) + j(x)\phi(x) \Big)
         ]
     \nonumber
     \\ & =
%     Z[0]\expval{\exp[i\!\int_C \! \dd^4 x \, j(x)\phi(x)]} \,, \quad
      Z[0]\expval{\exp\big[ i \, j\!\cdot\!\phi \big]} \,, \quad
        j\!\cdot\!\phi \equiv i\int_C \! \dd^4 x \, j(x) \phi(x) \,,
\end{align}
in terms of the field $\phi(x)$ and its source $j(x)$. Here, $t_\text{i} = t_\text{in}$ 
and $t_\text{f} = t_\text{in} - i\beta$ are connected via the contour $C$. 

Expressed above as a path-integral in terms of fields rather than operators, the generating 
functional accounts for the time ordering on the contour. Further, the averaging procedure 
has been reinterpreted in the path-integral formulation to be taken with respect to the action 
as a statistical weight. The ill-defined\footnotemark\ normalisation \mbox{$Z[0]$} 
cancels out in the thermal Green's functions of \eq{ThermalGreensFunctions}. This quantity 
contains all vacuum bubbles and is briefly discussed in terms of diagrams in \sec{bubble-example}.

  \footnotetext{The functional integration $\int [\mathcal D \phi]$ represents the summation over all 
  sufficiently differentiable and integrable functions that must be taken into account when considering 
  all paths. However, the total space of all functions is vastly larger and there exists no rigorous 
  method of picking out only the physically relevant functions in the path-integral. The main strategies 
  include pulling back to a discrete formulation, as is done in the field of lattice theories, or 
  to normalise the theory to a reference problem that is analytically solvable.}

In order to apply the tools of perturbation theory, one first splits the \L\ into its free and interaction 
parts as \mbox{$\mathcal L = \mathcal L_0 + \mathcal L_\text{I}$,} such that the prefactor $Z[0]$ 
in \eq{FullGeneratingFunctionalScalar} factorises into free, $Z_0[0]$, and interaction, $Z_\text{I}[0]$ parts. 
In order to obtain the Feynman rules and a diagrammatic series, the $Z_\text{I}[0]$ then undergoes 
a perturbative expansion into a series over the interaction term i.e.
\beq
     \exp[i\int_C \dd^4x \, \mathcal L_\text{I}(x)]
     \equiv
     \exp\big[ iS_\text{I}[\phi] \big]
     =
     \sum_{n=0}^\infty
         \frac{1}{n!}\big( iS_\text{I}[\phi] \big)^n \,.
\eeq
By observing that the expansion series can be obtained by functional differentiation \mbox{w.r.t.} 
the source \mbox{$j(x)$}, the argument of $\mathcal S_\text{I}$ may be replaced by the functional
differential, and the exponentiated interaction term, no longer a functional of $\phi$, can be 
taken out of the functional integral. The full \mbox{$Z[j]$} can thus be rewritten in analogy 
to that in the zero-temperature theory as \beql{GeneratingFunctionalSeries}
     Z[j]
     =
     \exp[
         i\!\int_C \! \dd^4 x \,
             \mathcal L_\text{I}
             \Big[ \frac{\delta}{i\delta j(x)} \Big]
     ]
     Z_0[j] \,.
\eeq
Note, however, that the series expansion is formally presented in full generality w.r.t. 
a given contour $C$ and the individual terms in this expansion may only have a clear interpretation once 
a particular choice for the contour $C$ is specified.

%%%%%%%%%%%%%%%%%%%%%%%%%%%%%%%%%%%%%%%%%
\subsection{Thermal propagators}
\label{sec:ThermalPropagators}
%%%%%%%%%%%%%%%%%%%%%%%%%%%%%%%%%%%%%%%%%

In this subsection, the propagator of a free neutral boson will be connected to the \mbox{two-point} correlation function. 
The latter is of great importance for practical computations and it is found by the evaluation of expectation values in the form
\beq
  \ev*{\hat{\mathcal O}(t, t')} = \ev*{\hat{\mathcal O}_1(t) \hat{\mathcal O}_2(t')} \,,
\eeq
while making use of the time-ordering procedure on the contour, see \sec{GeneratingFunctional}. This section will 
further generalise the discussion of thermal propagators to include anticommuting charged multi-component fields 
and we follow closely the notation and procedure provided by Landsman and van Weert~\cite{LandsmanvanWeert1987}.

%%%%%%%%%%%%%%%%%%%%%%%%%%%%%%%%%%%%%%%%%%%%%%%%%%%%%%%%%%%%%%%%
\subsubsection{Propagator of the free neutral scalar boson}
\label{sec:ScalarPropagator}
%%%%%%%%%%%%%%%%%%%%%%%%%%%%%%%%%%%%%%%%%%%%%%%%%%%%%%%%%%%%%%%%

In the special case of a free scalar particle with no derivative couplings\footnote{No such couplings are assumed to exist 
in the theory in order for the integration over the canonical momentum variable in the generating functional 
to be analytically performed. As an immediate consequence, the free scalar-field action reduces to
     \mbox{$
     S
     =
     \int \! \dd^4 x \, \frac{1}{2}\Big(
         \partial_\mu\phi\partial^\mu\phi - m^2\phi^2
     \Big)
     \to
     \int \! \dd^4 x \,
         \frac{1}{2}
         \phi\Big( \! -\partial^2 - m^2 \Big)\phi $} \,, up to an inessential total derivative term.
  }, here a neutral \mbox{($\mu = 0$)} boson with the free Klein-Gordon \L\
\beq
     \mathcal L_0(x)
     =
     \tfrac{1}{2}
     \phi(x)\big( \!-\!\partial^2 - m^2 \big)\phi(x) \,,
\eeq
the \emph{free} generating functional in \eq{FullGeneratingFunctionalScalar} may be 
rewritten by means of a change of variables
\begin{align}
\label{eq:Changeofvariables}
     \phi(x)
     \to
     \phi(x) - \int _C \! \dd^4 x'  \, D_C(x-x')j(x')
\end{align}
as follows
\begin{align}
\label{eq:FreeGeneratingFunctionalScalar}
     & Z_0[j]
     =
     Z_0[0] \exp[
         -\frac{i}{2}\int_C \! \dd^4 x \int_C \! \dd^4 x' \,
             j(x)D_C(x-x')j(x')
     ] \,.
\end{align}
The shift in \mbox{$\phi(x)$} is chosen carefully in order to complete the square of the free \L. The emerging 
\emph{thermal bosonic propagator} \mbox{$D_C(x-x')$} is defined in relation to the differential operator 
of the \eom\, and it satisfies 
\beql{PropagatorBoundaryCondition}
  K\!(i\partial) D_C(x-x') = \delta_C(x-x') \qquad \text{for} \qquad K\!(i\partial)\phi(x) = 0 \,.
\eeq
For a neutral scalar boson, \mbox{$K\!(i\partial) \equiv -(\partial^2 + m^2)$} is the usual Klein-Gordon operator. 
The contour \mbox{$\delta$-function} may be defined as \mbox{$\delta_C(t - t') = 
\left( \pdv{z}{\tau} \right)^{\!\!-1}\delta(\tau - \tau')$} with $\tau$ parametrising 
the contour $C$ (recall the definition \mbox{$t \equiv z(\tau)$}). 

Therefore, the propagator is related to the free two-point Green's function as follows
\beq
     G_{0C}(x, x') \equiv iD_C(x-x') \,.
\eeq
Generalising the discussion above in \sec{Spectralrep}, it can also be rewritten 
in the \emph{spectral representation},
\beql{NeutralScalarPropagator}
     iD_C(x-x')
     =
     \int \! \frac{\dd^4 k}{(2\pi)^4} \,
         \rho_0(k) e^{-ik\cdot(x-x')}
         \big[ \Theta_C(t-t') + n(k_0) \big] \,, \qquad \eta = +1 \,,
\eeq
where $n(k_0)$ is the thermal distribution function defined in \eq{thermaldistribution}.
This result can be found by a generalisation of the expression first derived by Mills in Ref.~\cite{Mills1969} 
for the non-relativistic case and then by Dolan and Jackiw in Ref.~\cite{DolanJackiw1974} for ordinary time ordering. 
Here, the \emph{spectral density} for the free neutral bosonic field reads
\beql{FreeSpectralDensity}
     \rho_0(k)
     =
     2\pi \text{ sign}(k_0)\delta\big( k^2 \!- m^2 \big) \,.
\eeq
Note that this quantity is not related to the distribution found in \sec{StatMech} and 
will not appear from this point on other than in its explicit exponential form.

%%%%%%%%%%%%%%%%%%%%%%%%%%%%%%%%%%%%%%%%%%%%%%%%%%%%%%%%%%%%%%
\subsubsection{Generating functional for an arbitrary free field}
\label{sec:GenGeneratingFunc}
%%%%%%%%%%%%%%%%%%%%%%%%%%%%%%%%%%%%%%%%%%%%%%%%%%%%%%%%%%%%%%

Consider a general multi-component covariant complex field (bosonic or fermionic) described by the operator
\mbox{$\hat\PHI_\alpha^i(x)$} \cite{LandsmanvanWeert1987}. Here, $\alpha$ is the index of the Lorentz representation 
of the field, e.g. a spinor or vector index and $i$ is an index of a representation of an internal symmetry or a field
generation index, i.e. not related to space-time transformations. Let the operator transform under some representation 
of the Lorentz group \mbox{$\mathcal J_{\alpha\beta}[\Lambda]$}. The field is considered to be charged under a given 
set of internal symmetries and the conserved charges are denoted as $q_a^{ij}$ so that
\beq
     \comm{\hat Q_a}{\hat\PHI_\alpha^i(x)}
     =
     -q_a^{ij}\hat\PHI_\alpha^j(x) \,.
\eeq
The chemical potential for the fields charged under multiple symmetries is generally given by the summed quantity
\mbox{$\mu = \sum_a \mu_a q_a$}; see the grand canonical ensemble in \sec{GrandCanonicalEnsemble} 
applied to the eigenstates of $\hat Q_a$, such that
\beq
     \sum_a\mu_a\hat Q_a \vert \PHI \rangle
     =
     \sum_a\mu_a q_a \vert \PHI \rangle \,.
\eeq
Generalising the plain Klein-Gordon case discussed above, it is reasonable to start with 
a free \L\ in the following quadratic form
\beq
     \mathcal L_0(x)
     =
     \bar\PHI_\alpha^i(x)
     K_{\alpha\beta}^{ij}(i\partial, \mu)
     \PHI_\beta^j(x) \,.
\eeq
Here, the \emph{conjugated or adjoint field} is defined as\footnote{The matrix $\mathcal A_{\alpha\beta}$ ensures 
invariance of the \L\ under Lorentz transformations of $\PHI$. In the case when $\alpha$ is an index of the Dirac 
spinor representation, \mbox{$\mathcal A \equiv \gamma^0$}.}: \mbox{$\bar\PHI_\alpha^i \equiv (\PHI^\dagger \! 
\mathcal A)_\alpha^i$}. Besides, \mbox{$K_{\alpha\beta}^{ij}(i\partial, \mu)$} is the differential operator 
of the free field in the medium which generally depends on the chemical potential. The free thermal Green's 
functions that arise from this \L\ can be obtained from the generating functional. In the path-integral 
formulation, this functional may be written in terms of two source fields
\beql{GeneralGeneratingFunctional}
     Z_0[\bar\jmath, j]
     =
     \mathcal N
     \int \! [\mathcal D\bar\PHI] \int \! [\mathcal D\PHI] \,
         \exp[i\int_C \! \dd^4 x \, \Big(
             \bar\PHI K \PHI + e^{i\mu t} \bar\jmath\PHI
             +
             e^{-i\mu t} \bar\PHI j
         \Big)] \,,
\eeq
where $j(x)$ is the source of $\PHI$ while $\bar\jmath(x)$ is the source of $\bar\PHI$, and the fields are fully 
contracted over all discrete indices, e.g. \mbox{$\bar\jmath\PHI = \bar\jmath_\alpha^{\hspace{2pt}i}(x)\PHI_\alpha^i(x)$}.
The $\mu$-dependence has been shifted away from $K$ to the source terms by means of a change of variables introduced 
by Weldon~\cite{Weldon2007}, under which the fields and the differential operator are transformed as
\beq
  \PHI \to e^{i\mu t}\PHI, \qquad \bar\PHI \to e^{-i\mu t}\bar\PHI \,, \qquad 
  K(i\partial, \mu) \to K(i\partial) \,,
\eeq
thus treating the field and its adjoint counterpart as independent variables.

%%%%%%%%%%%%%%%%%%%%%%%%%%%%%%%%%%%%%%%%%%%%%%%%%%%%%%%%%%%%%%%%%%%%%%%%%
\subsubsection{Thermal propagator and KMS condition: a generic treatment}
\label{sec:GenthermpropandKMS}
%%%%%%%%%%%%%%%%%%%%%%%%%%%%%%%%%%%%%%%%%%%%%%%%%%%%%%%%%%%%%%%%%%%%%%%%%

The functional integral of \eq{GeneralGeneratingFunctional} is defined in order to generate the thermal Green's 
functions (\mbox{$n$-point} functions) under functional differentiation and therefore it must represent 
the thermal trace of \eq{ThermalTrace}:
\beq
  \tr[e^{-\beta(\hat H - \mu\hat Q)} T_C\exp[i\!\int_C \! \dd^4 x \, \big( \bar\jmath\PHI + \bar\PHI j \big)]] \,.
\eeq
In order to recover the trace interpretation of the generating functional, the functional measure should be taken 
only over \mbox{c-number} fields $\PHI$, $\bar\PHI$ such that the following periodicity condition is satisfied 
over a period \mbox{$i\beta$} such that\footnote{
     This periodicity condition ensures preservation of the trace interpretation and without it the functional 
     integral would no longer generate the trace:
     \mbox{$
         \bra{\PHI; t_\text{in} \!-\! i\beta}
         \hat{\mathcal O}
         \ket{\PHI;t_\text{in}}
         \equiv
         \bra{\PHI;t_\text{in}}
         \hat{\mathcal O}
         \ket{\PHI; t_\text{in}}
    $}. 
  }
\beql{FieldBoundaryCondition}
     \PHI_\alpha^i(t_\text{in} \!-\! i\beta, \vb x)
     =
     \eta e^{\beta\mu} \PHI_\alpha^i(t_\text{in}, \vb x) \,.
\eeq
The periodicity leads to the \emph{Kubo-Martin-Schwinger condition} (KMS) on the propagator which 
was introduced by Kubo~\cite{Kubo1957}, implemented by Martin and Schwinger~\cite{MartinSchwinger1959} 
and is presented in what follows.

The free generating functional of \eq{GeneralGeneratingFunctional} can be rewritten in terms of the 
\emph{general free thermal propagator} by a shift of the field and its adjoint analogously to the case 
of the scalar field in \sec{ScalarPropagator}. Contracted over all indices, the result is
\beq
     Z_0[\bar\jmath, j]
     =
     Z_0[0,0] \exp[
         -i\!\int_C \! \dd^4 x \! \int_C \! \dd^4 x' \, 
             \bar\jmath_\alpha^{\,i}(x)
             D_{\!C\,\alpha\beta}^{\phantom{C}ij}(x-x')
             j_\beta^j(x')
     ]
\eeq
with
\beq
     Z_0[0,0]
     =
     \mathcal N 
     \int \! [\mathcal D\bar\PHI] \int \! [\mathcal D\PHI] \,
         \exp[i\!\int_C \! \dd^4 x \, \bar\PHI K \PHI] \,.
\eeq
In order to complete the square and extract the propagator \mbox{$D_{\!C\,\alpha\beta}^{\phantom{C}ij} 
\equiv \big(K^{-1}\big)_{\alpha\beta}^{ij}$,} the thermal propagator must satisfy the differential equation
\beql{PropagatorEquation}
     K_{\alpha\beta}^{ij}(i\partial)
     D_{\!C\,\beta\gamma}^{\phantom{C}jk}(x-x')
     =
     \delta_{\alpha\gamma} \delta^{ik} \delta_C(x-x') \,.
\eeq
The generating functional must be invariant under the introduced shifts of the fields. This condition fixes 
the solution of the differential equation so that the propagator can be related to the \mbox{two-point function} as
\begin{align}
     iD_{C\,\alpha\beta}^{\phantom{C\,}ij}(x-x')
     & \equiv
     G_{0C\,\alpha\beta}^{\phantom{0C\,}ij}(x, x')
     =
     {}_{0} \!
     \ev*{T_C\hat\PHI_\alpha^i(x)\hat{\bar\PHI}_\beta^j(x')}_0
     =
     \ev*{\PHI_\alpha^i(x){\bar\PHI}_\beta^j(x')} \,.
\end{align}
This is a special case of the general relation \eq{General2PointFunction} with the final expectation value 
expressed in the path-integral formalism over the fields rather than the operators, analogously to
\eq{FullGeneratingFunctionalScalar}. Guided by the definition of the time ordering along the contour $T_C$, 
the free propagator may be split using the contour step-function of \eq{ContourStepFunction} so that
\beq
     D_{C\,\alpha\beta}^{\phantom{C\,}ij}(x-x')
     =
     \Theta_C(t-t')D_{\phantom{>\,}\alpha\beta}^{>\,ij}(t-t')
     +
     \Theta_C(t'-t)D_{\phantom{<\,}\alpha\beta}^{<\,ij}(t-t') \,.
\eeq
Let us compare this expression to \eq{GeneralTwoPointFunction} for the general \mbox{$n$-point} function. 
The imposed boundary condition on the field of \eq{FieldBoundaryCondition} enforces a boundary condition 
on the propagator that relates its advanced and retarded components. As a consequence, invariance of 
the path-integral over the time shift by $i\beta$ holds. This way, we arrive at the KMS condition 
in the following generic form,
\beql{KMSCondition}
     D_{\phantom{>\,}\alpha\beta}^{>\,ij}(t_\text{in} - i\beta)
     =
     \eta e^{-\beta\mu}
     D_{\phantom{<\,}\alpha\beta}^{<\,ij}(t_\text{in}) \,,
\eeq
which is a proper generalisation of the condition \eq{KMS} previously introduced 
in the case of a neutral scalar field. This condition is important for thermal theories as relates several 
thermal quantities in the \emph{real-time formalism} (see \sec{RealTimeFormalism}).

Note that any relativistic, multi-component field \mbox{$\PHI_\alpha^i(x)$} should satisfy 
the Klein-Gordon equation. Therefore, assuming a multi-mass \emph{Klein-Gordon divisor} \mbox{$d_{\alpha\beta}^{ij}(i\partial)$} to exist according to Ref.~\cite{Landsman1986} 
and further references therein, we have
\beq
     d_{\alpha\beta}^{ij}(i\partial)
     K_{\beta\gamma}^{jk}(i\partial)
     =
     K_{\alpha\beta}^{ij}(i\partial)
     d_{\beta\gamma}^{jk}(i\partial)
     =
     \delta_{\alpha\gamma}\delta^{ik}
     \prod_l \big( \!-\! \partial^2 - m_l^2 \big) \,.
\eeq
The $m_l$'s represent the mass spectrum and can be inferred from the structure 
of $K$~\cite{Wightman1978, Landsman1986}. For an extended and comprehensive 
discussion regarding the mass spectrum, see the lecture notes by Wightman~\cite{Wightman1978}. 

The solution to \eq{PropagatorEquation} simply becomes
\beql{ArbitraryPropagator}
     D_{C\,\alpha\beta}^{\phantom{C\,}ij}(x-x')
     =
     d_{\alpha\beta}^{ij}(i\partial_x) D_C(x-x')
\eeq
in terms of the scalar propagator encountered earlier. A slight reservation should be made 
regarding the form of \mbox{$D_C(x-x')$}. The solution presented in \eq{NeutralScalarPropagator} 
of \sec{ScalarPropagator} is that of a \emph{neutral} scalar boson. When incorporating anticommuting 
as well as charged fields \mbox{($\mu \neq 0$)} this propagator is modified slightly to become
\beql{GeneralScalarPropagator}
     iD_C(x-x')
     =
     \int \! \frac{\dd^4 k}{(2\pi)^4} \,
         \rho_0(k) e^{-ik\cdot(x-x')} \big[ \Theta_C(t-t')
     +
     \eta n(\omega_+) \big]
\eeq
with the distribution function
\beq
     n(\omega_\pm) = \frac{1}{e^{\omega_\pm} - \eta} \,.
\eeq
The $\pm$ corresponds to charged particles/antiparticles through the definition 
\mbox{$\omega_\pm = \beta(k_0 \mp \mu$).} 
Generally, the spectral density of a free field can be found as
\beq
     \rho_0(k) = i\,\text{Disc} \prod_l \frac{1}{k^2 - m_l^2}.
\eeq
Here, the discontinuity introduced by Ref.~\cite{Landsman1986} and further discussed in 
Ref.~\cite{LandsmanvanWeert1987} is defined over the real axis as
\mbox{$\text{Disc}\,f(k_0) = f(k_0 + i\epsilon) - f(k_0 - i\epsilon)$} in the limit 
\mbox{$\epsilon \to 0^+$}. 

Note, the results of this section rely on the fact that the contour step-function commutes with 
the Klein-Gordon divisor. Further, the charge operators are assumed to commute with $\mathcal L_0$ 
so that the matrices $q_a$ commute with both \mbox{$K(i\partial)$} and \mbox{$d(i\partial)$} 
\cite{LandsmanvanWeert1987}. 

Let us now summarise the most important points from the above discussion as follows:
\begin{itemise}
	\item Time integration is to be taken along a contour $C$ that begins at an arbitrary initial 
	time $t_\text{in}$ and goes down to the final time \mbox{$t_\text{in} \!-\! i \beta$} with the restriction 
	that the imaginary component of the contour can not increase\footnote{Recall that the contour is a consequence 
	of the trace interpretation of the path-integral formulation with one of the states shifted by the thermal 
	density operator. The imposed restriction on the imaginary part of the contour is required for the application 
	of the FSM-formula of \eq{FSM}.}.
	\item The propagator \mbox{$D_C(x - x')$} depends explicitly on the choice of contour.
\end{itemise}
These points are valid for all thermal propagators as a consequence of the very general expression 
of \eq{ArbitraryPropagator}. Since the quantities used in practical calculations depend explicitly 
on the choice of the contour $C$, it is imperative to choose a formalism where the physical quantities 
defined at real times can straightforwardly be extracted and in such a way that the final result 
does not depend on this choice of $C$. Two commonly used formalisms that provide reliable methods 
to extract such quantities will be presented in the following.

%%%%%%%%%%%%%%%%%%%%%%%%%%%%%%%%%%%%%%%%%%%%%%%%%%%%%%%%%%%%%%%%
\section{Imaginary-time formalism}
\label{sec:ImaginaryTimeFormalism}
\setcounter{equation}{0}
%%%%%%%%%%%%%%%%%%%%%%%%%%%%%%%%%%%%%%%%%%%%%%%%%%%%%%%%%%%%%%%%

The thermal formulation of QFT presented in the sections above formally resembles to a high degree the well-known 
QFT at zero temperature. The main distinguishing property is that the explicit form of the thermal \mbox{$n$-point}
function (Green's function) depends on the choice of integration contour $C$ in the complex temporal plane. 
This contour connects some arbitrary initial time $t_\text{in}$ with the final time \mbox{$t_\text{in} \!-\! i\beta$}. 
The latter time point emerges as a consequence of the formal interpretation of the thermal equilibrium distribution
operator as an evolution operator. In the equilibrium case, the contour that connects the initial and final time 
points can be arbitrarily chosen with the single restriction that the imaginary component of the contour must 
not increase when integrating along the contour. In this section, a specific choice of contour will be considered 
in order to make explicit calculations possible. This discussion will be followed by a diagrammatic expansion 
of the generating functional in order to derive the Feynman rules on this contour.

%%%%%%%%%%%%%%%%%%%%%%%%%%%%%%%%%%%%%%%%
\subsection{Matsubara contour}
\label{sec:Matsubaracont}
%%%%%%%%%%%%%%%%%%%%%%%%%%%%%%%%%%%%%%%%

The simplest possible choice of the integration contour has been proposed by Matsubara in Ref.~\cite{Matsubara1955}. 
It constitutes a straight vertical line that connects $t_\text{in}$ and \mbox{$t_\text{in} - i\beta$}, 
cf.~\fig{MatsubaraContour} with \mbox{$\lambda = \beta$}. The time variable may then be parametrised on a simple and 
purely imaginary form as \mbox{$t = i\tau$} with \mbox{$\tau \in [0, \beta]$} being a real variable. This particular 
choice has the advantage that it generates a perturbative expansion that contains the diagrams recognised 
from the corresponding zero-temperature theory (vacuum bubbles, connected diagrams, 1PI, etc.)~\cite{Das1997}. 
This section will mainly be concerned with a single scalar field with \mbox{$\mu = 0$} as an illuminating example, 
but the results may be generalised to a theory with charged fields in a non-trivial representation of 
both Lorentz and internal symmetries, see Ref.~\cite{LandsmanvanWeert1987, Das1997}. 

Through a suitable parametrisation of time, the action may be written as an integral 
in \emph{Euclidean} space:
\beq
     -\!S_\text{E} = i\int_C \! \dd^4 x \, \mathcal L(x)
     =
     -\int_0^\beta \! \dd \tau \int \! \dd^3 x \,
         \mathcal L_\text{E}(x) \,,
\eeq
where the Euclidean \L\ reads
\begin{align}
     \mathcal L_\text{E}(\tau, \vb x)
     & =
     \tfrac{1}{2}
     \phi(\tau,\vb x)
     \big( \!\!-\!\partial_\tau^2 - \nabla_{\vb x}^2 + m^2 \big)
     \phi(\tau, \vb x)
     +
     \mathcal V(x)
     \nonumber
     \\ & =
     \tfrac{1}{2}
     \phi(\tau,\vb x) K_\text{E} \, \phi(\tau, \vb x)
     +
     \mathcal V(x) \,.
\end{align}
The \mbox{$n$-point} 
Green's functions may be obtained by functional differentiation w.r.t. a source \mbox{$j(\tau, \vb x)$} 
of the generating functional
\beq
     Z[j]
     =
     \int \! [\mathcal D\phi] \, e^{-S_\text{E} + j\cdot\phi},
     \qquad
     j\!\cdot\!\phi
     =
     \int_0^\beta \! \dd \tau \int \! \dd^3 x \,
         j(\tau, \vb x)\phi(\tau, \vb x) \,.
\eeq
Since the complex shift of the temporal variable is always present in theories 
with an equilibrated medium, TFT is in a sense intrinsically Euclidean. Let us discuss
the basic ingredients of such a Euclidean formulation of Matsubara.

%%%%%%%%%%%%%%%%%%%%%%%%%%%%%%%%%%%%%%%%%%%%%%%
\subsection{Imaginary-time propagator}
\label{sec:Imagtimeprop}
%%%%%%%%%%%%%%%%%%%%%%%%%%%%%%%%%%%%%%%%%%%%%%%

As stated in \eq{FreeGeneratingFunctionalScalar}, the free generating functional \mbox{$Z_0[j]$} 
(with \mbox{$\mathcal V(x) = 0$}) may be written in terms of the free propagator by completion 
of the square. In the Matsubara formalism, the result takes the following form
\beq
     Z_0[j]
     =
     Z_0[0]
     \exp[\tfrac{1}{2}j\!\cdot\!K_\text{E}^{-1}\!\cdot\!j] \,.
\eeq
The inverse of the differential operator must satisfy the boundary condition of \eq{PropagatorBoundaryCondition} 
in order to fix a proper solution of the differential equation. The solution is the \emph{thermal propagator of Matsubara}
\mbox{$K_\text{E}^{-1} = \Delta(\tau-\tau', \vb x - \vb x')$} which is given by \eq{GeneralScalarPropagator} over 
the chosen contour\footnote{On the Matsubara contour, the propagator may be expressed in terms of the real variable 
$\tau$ so that \mbox{$t-t' \to \tau - \tau'$} and \mbox{$\Theta_C(t-t') \to \Theta(\tau-\tau')$.}}.
The propagator is periodic in $\tau$ over a period $\beta$ as imposed by the KMS condition (\eq{KMSCondition}), 
which can be straightforwardly verified. 

Completion of the square of the free \L\ renders the generating functional in the form
\beq
     Z_0[j]
     =
     Z_0[0] \exp[
         \frac{1}{2}
         \iint_0^\beta \!\! \dd \tau \, \dd \tau' \!
         \iint \! \dd^3 x \, \dd^3 x' \,
             j(\tau, \vb x)
             \Delta(\tau-\tau', \vb x - \vb x')
             j(\tau', \vb x')
     ] \,.
\eeq
Comparing with \eq{FreeGeneratingFunctionalScalar}, the \emph{Euclidean propagator} of a neutral scalar 
may be found as
\beq
     iD_\text{E}(-i\tau, \vb x) \equiv \Delta(\tau, \vb x) \,.
\eeq
Note the purely imaginary temporal coordinate of the Euclidean propagator here. Thus, \mbox{$Z_0[j]$} shown 
above generates all the Green's functions with imaginary time arguments.

The above discussion is analogous to the zero-temperature theory, with the following two differences:
\begin{itemise}
    \item the temporal integration is reduced to the periodic interval \mbox{$[0, \beta]$}: \\
\mbox{$i\int {\dd t} \to \int_0^\beta {\dd\tau}$};
    \item the appropriate propagator of a free scalar field is no longer the Feynman propagator 
but the thermal propagator of Matsubara: \\
\mbox{$D_\text{F}(x-x') \to \Delta(\tau-\tau', \vb x-\vb x')$.}
\end{itemise}

Due to the periodicity condition in $\tau$ applied to the bosonic fields (antiperiodicity for fermionic fields), 
and therefore on the thermal propagator, over $\beta$, the Matsubara propagator may be written in Fourier space 
as a sum over discrete frequencies $\omega_n$. Transformation of \eq{NeutralScalarPropagator} using the Matsubara 
contour leads to
\beql{MatsubaraDeltaPropagator}
     \Delta(\tau-\tau', \vb x - \vb x')
     =
     \frac{1}{\beta}\sum_n \int \! \frac{\dd^3 k}{(2\pi)^3} \,
         e^{
             -i\omega_n(\tau-\tau')
             +
             i\vb k \cdot (\vb x - \vb x')
         }
         \frac{1}{\omega_n^2 + \omega_{\vb k}^2}
\eeq
with \mbox{$\omega_n = 2n\pi/\beta$} (bosonic), \mbox{$\omega_n = (2n+1)\pi/\beta$} (fermionic), 
and \mbox{$\omega_{\vb k}^2 = \abs{\vb k}^2 + m^2$}. Note that the mode energy here reads \mbox{$\omega_{\vb k} \equiv 
\omega(\abs{\vb k})$}, meaning that $\vb k$ denotes an argument of the function rather than an index to be summed over. 
The $\omega_n$'s are the \emph{Matsubara frequencies}~\cite{Matsubara1955, LandsmanvanWeert1987, Das1997, HoScherrer2015}
and they parametrise an infinite number of poles of the momentum-space propagator located on the imaginary axis provided that the Fourier transform is
\beql{ImaginaryTimePropagator}
     \tilde\Delta(i\omega_n, \vb k)
     =
     \frac{1}{\omega_n^2 + \omega_{\vb k}^2}
     \equiv
     \int\!
     \frac{\dd k_0}{2\pi}\frac{\rho_0(k)}{k_0 - i\omega_n} \,.
\eeq
A more general multi-mass version of Matsubara propagator including charged fields 
was presented by Ref.~\cite{LandsmanvanWeert1987}. In order to obtain an expression 
analogous to the zero-temperature theory, define \mbox{$k_0 = i\omega_n$} 
in \eq{MatsubaraDeltaPropagator}. Then
\beql{ExplicitEuclideanPropagator}
  iD_E(x-x')
  =
  -\frac{1}{\beta}\sum_n \! \int \! \frac{\dd^3 k}{(2\pi)^3} \,
    e^{-k\cdot(x-x')}\frac{1}{k^2 - m^2} \,,
\eeq
where the energy $k_0$ is purely imaginary.

%%%%%%%%%%%%%%%%%%%%%%%%%%%%%%%%%%%%%%%%%%%%%%%%%%%%
\subsection{Imaginary-time Feynman rules}
\label{sec:ImagFeynmanRules}
%%%%%%%%%%%%%%%%%%%%%%%%%%%%%%%%%%%%%%%%%%%%%%%%%%%%

In the path-integral formulation, the generating functional may be expanded to yield the Feynman 
rules of the imaginary-time formalism through a perturbative series. Following the reasoning of
\sec{ScalarPropagator}, the Euclidean action may be split into a free part that is quadratic 
in the field and an interaction part so that \mbox{$S_\text{E} = S_0 + S_\text{I}$}. 
The generating functional of \eq{FullGeneratingFunctionalScalar} may then be expanded as
\begin{align}
     Z[j]
     & =
     \mathcal N \int\![\mathcal D\phi] \,
         e^{-S_0} e^{-S_\text{I} + j\cdot\phi}
     \nonumber
     \\ & =
     \mathcal N \int\![\mathcal D\phi] \,
         e^{-S_0} \bigg\{
             1
             +
             \big( \!-\!S_\text{I} + j\!\cdot\!\phi \big)
             +
             \tfrac{1}{2}
             \big( \!-\!S_\text{I} + j\!\cdot\!\phi \big)^2
             +
             \ldots
     \bigg\} \,.
\end{align}
Each term in this series may be reinterpreted as a statistical average with respect 
to the weight $\exp[-S_0]$ so that
\beql{MatsubaraPertubationSeries}
     Z[j]
     =
     Z_0[0]\bigg\{
         1
         +
         \expval*{(-S_\text{I} + j\!\cdot\!\phi)}_0
         +
         \tfrac{1}{2}
         \expval*{(-S_\text{I} + j\!\cdot\!\phi)^2}_0
         +
         \ldots
     \bigg\} \,.
\eeq
The Feynman rules are directly extracted from this series. For the specific case 
of $\phi^l$ ($l=3,4,\dots$) interacting theory with a coupling constant $\kappa$, the rules are 
given as follows~\cite{LandsmanvanWeert1987, Kapusta1989}:
\begin{itemise}
     \item Draw diagrams using Wick contractions and determine symmetry factors.
     \item Define \mbox{$k_0 = i\omega_n$} and assign a propagator \mbox{$\tilde\Delta(k)$} 
     to each line and a factor of $-\kappa$ to each vertex $i$.
     \item Impose energy-momentum conservation through
     \mbox{$\beta(2\pi)^3 \, \delta_{n0} \delta^{(3)}\big( \textstyle\sum_i \vb k_i \big)$}
     with \mbox{$n = \sum_i n_i$}. A global conservation factor
     \mbox{$\beta(2\pi)^3\delta_{00}\delta^{(3)}(0)$}
     may be separated out corresponding to the exclusion of one vertex.
     \item Sum and integrate over all internal energies and momenta $k_i$:
     \mbox{$\tfrac{1}{\beta}\sum_n\int \! \frac{\dd^3 \vb k_i}{(2\pi)^3}$.}
\end{itemise}
Again, one should note that the diagrammatic structure is identical to that 
of the zero-temperature theory. Further, the definition of the cumulant in \eq{Cumulant}
allows for the extraction of all connected diagrams with at least one external leg so that
\beql{MatsubaraCumulant}
     \log Z[j]
     =
     \log Z[0]
     +
     \expval*{e^{-S_\text{I}}(e^{j\cdot\phi} - 1)}_\text{con} \,.
\eeq

%%%%%%%%%%%%%%%%%%%%%%%%%%%%%%%%%%%%%%%%%%%%%%%%%%%%%%%%%%%%%%%%
\subsection{Matsubara frequency summation}
\label{sec:Matsubarafrequency}
%%%%%%%%%%%%%%%%%%%%%%%%%%%%%%%%%%%%%%%%%%%%%%%%%%%%%%%%%%%%%%%%

  This subsection contains a short discussion on practical calculations in Matsubara theory. Every term in the perturbation theory series of \mbox{$\log Z[j]$} may be expanded as a product of Matsubara propagators due to the thermodynamic Wick's theorem. Such propagator products must generally be summed up over Matsubara frequencies. An example of Matusbara summation from $\phi^3$-theory is the propagator product of the scalar loop diagram with one propagator for each internal line in the loop,
\begin{align}
     \frac{1}{\beta}\sum_n
         \tilde \Delta(i\omega_n, \vb{q})
         \tilde \Delta(i\omega_{m-n}, \vb{k} - \vb{q})
     & =
     \frac{1}{\beta}\sum_n
         \frac{1}{\omega_n^2 + \omega_{\vb q}^2}
         \cdot
         \frac{1}{\omega_{m-n}^2 + \omega_{\vb k - \vb q}^2}
     \nonumber
     \\ & =
     \frac{1}{\beta}\sum_n
         \frac{1}{4 \omega_{\vb q} \omega_{\vb k - \vb q}}
         \sum_{s, s'}
             \frac{s}{i\omega_n + s\omega_{\vb q}}
             \cdot 
             \frac{s'}{
                 i\omega_m - i\omega_n
                 +
                 s'\omega_{\vb k - \vb q}
             } \,,
\end{align}
with $\omega_{m-n} \equiv \omega_{m} - \omega_n$. The summation corresponds to the integration over time 
in the zero-temperature theory and may be arbitrarily complicated. A general scheme should therefore 
be provided in order to perform the summation over any number of propagators in the product. 
  
Generically, summation over Matsubara frequencies $\omega_n$ comes in the following form
\beql{Matsum}
     S_\eta
     =
     \frac{1}{\beta} \sum_n g(i\omega_n)
\eeq
with the function \mbox{$g(i\omega_n)$} being a product of arbitrarily many Matsubara 
propagators, and $\eta$ indicates bosonic or fermionic frequencies i.e.
\begin{align}
     \eta = +1:
     & \qquad
     \omega_n = \frac{2n\pi}{\beta} \,,
     \nonumber \\
     \eta = -1:
     & \qquad 
     \omega_n = \frac{(2n + 1) \pi}{\beta} \,.
\end{align}
In the case of a single propagator
  \mbox{$\tilde \Delta(i\omega, \vb k)$}, the function $g$ reads
\beq
     g(i\omega_n) = \frac{1}{\omega_n^2 + \omega_{\vb k}^2} \,.
\eeq
Note, $S_\eta$ converges if \mbox{$g(z=i\omega_n)$} approaches 0 
when \mbox{$\abs{z} \to \infty$} faster that $z^{-1}$.

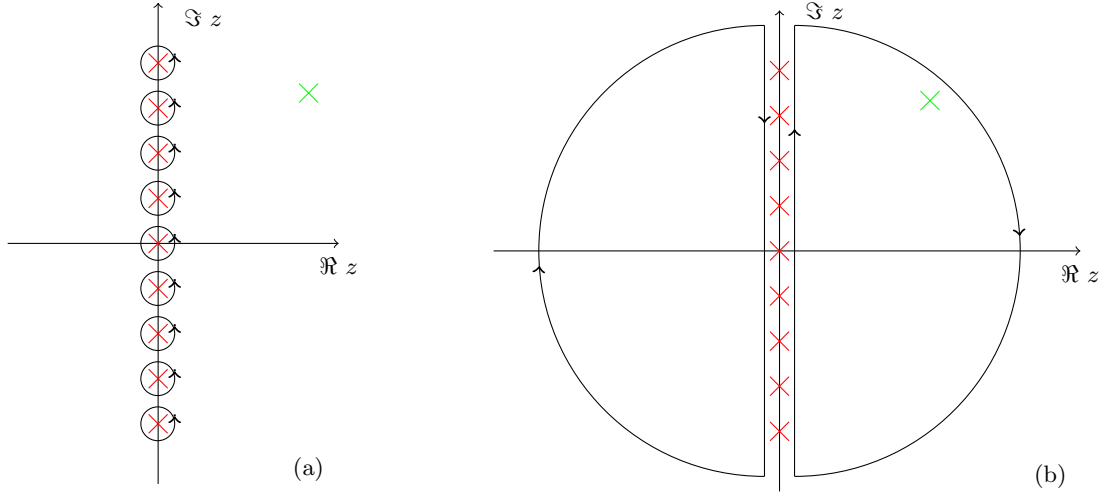
\begin{figure}
    \begin{minipage}[c]{0.35\textwidth}
    \begin{tikzpicture}[scale=2.0]
		\draw [->] (-1,0) -- (1.2,0);
		\draw [->] (0,-1.6) -- (0,1.6);
		\node at (0.3, 1.5) {$\imaginary\ z$};
		\node at (1.2, -0.15) {$\real\ z$};
		\foreach \y in {-4,...,4}
			\draw (0, 0.3*\y) node[cross, red] {};
		\foreach \y in {-4,...,4}
			\draw (0.11, 0.3*\y) arc (0:360:3.2pt);
		\foreach \y in {-4,...,4}
			\draw [thick, ->] (0.11, 0.3*\y+0.05) -- (0.11, 0.3*\y+0.05);
		\draw (1, 1) node[cross, green] {};
		\draw (1, -1.5) node {(a)};
	\end{tikzpicture}
    \end{minipage}
    \hfill
    \begin{minipage}[c]{0.57\textwidth}
    \begin{tikzpicture}[scale=2.0]
		\draw [->] (-1.9,0) -- (2.0,0);
		\draw [->] (0,-1.6) -- (0,1.6);
		\draw (0.1, -1.5) arc (-90:90:1.5cm);
		\draw (0.1, -1.5) -- (0.1, 1.5);
		\draw (-0.1, 1.5) arc (-270:-90:1.5cm);
		\draw (-0.1, -1.5) -- (-0.1, 1.5);
		\node at (0.3, 1.6) {$\imaginary\ z$};
		\node at (2.0, -0.15) {$\real\ z$};
		\foreach \y in {-4,...,4}
			\draw (0, 0.3*\y) node[cross, red] {};
		\draw (1, 1) node[cross, green] {};
		\draw [thick, ->] (0.1, 0.8) -- (0.1, 0.81);
		\draw [thick, ->] (1.596, 0.11) -- (1.597, 0.1);
		\draw [thick, ->] (-0.1, 0.86) -- (-0.1, 0.85);
		\draw [thick, ->] (-1.597, -0.11) -- (-1.596, -0.1);
		\draw (1.8, -1.5) node {(b)};
	\end{tikzpicture}
    \end{minipage}
	\caption{Poles of the product $g(z)h_\eta(z)$. The crosses on the imaginary axis (red) are the infinite 
	number of poles of the weighting function $h_\eta(z)$. The cross at complex $z$ (green) symbolises 
	the poles of $g(z)$. The contour in (a) may be deformed as shown in (b) \cite{Kapusta1989, Nieto1995}. 
	The poles of the weight function indicated in red may be used to evaluate the sum of the residues 
	indicated in green. 
	\label{fig:Poles}}
\end{figure}

In order to evaluate $S_\eta$, one may deploy calculus of residues in a procedure described by, for instance, 
Refs.~\cite{Kapusta1989, Nieto1995}. The Matsubara sum \eq{Matsum} may be treated as the sum of residues 
of simple poles at $i\omega_n$ and, thus, may be turned into a closed integral. A suitable weighting function
\mbox{$h_\eta(z)$}, different for the bosonic and fermionic cases, that has simple poles at every instance 
of $i\omega_n$ should be introduced to accomplish this. Then,
\beq
     S_\eta
     =
     \frac{1}{\beta} \sum_n g(i\omega_n)
     =
     \frac{1}{2 \pi i \cdot \beta} \oint \! \dd z \,
         g(z) \cdot h_\eta(z) \,.
\eeq
The poles of \mbox{$h_\eta(z)$} are illustrated in Fig.~\ref{fig:Poles}. A number of crosses on the imaginary 
axis in the figure indicate the poles of the appropriately constructed \mbox{$h_\eta(z)$} 
at $i\omega_n$. The contour picks up the residues of all these poles giving rise to $S_\eta$.
%%%%%%%%%%%%%%%%%%%%%%%%%%%%%%%%%%%%%%%%%%%%%%%%%%%%%%%%
%%%%%%%%%%%%%%%%%%%%%%%%%%%%%%%%%%%%%%%%%%%%%%%%%%%%%%%%

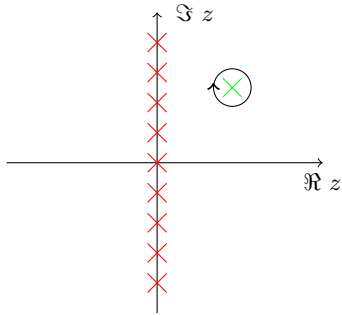
\begin{figure}
	\centering
	\begin{tikzpicture}[scale=1.0]
		\draw [->] (-2,0) -- (2.2,0);
		\draw [->] (0,-2) -- (0,2);
		\draw (1.25, 1) arc (0:360:0.25cm);
		\node at (0.5, 2) {$\imaginary\ z$};
		\node at (2.2, -0.25) {$\real\ z$};
		\foreach \y in {-4,...,4}
			\draw (0, 0.4*\y) node[cross, red] {};
		\draw (1, 1) node[cross, green] {};
		\draw [thick, ->] (0.75, 1.05) -- (0.75, 1.06);
	\end{tikzpicture}
	\caption{Deformation of the contour. This procedure may be carried through 
	given that $g(z)$ has no poles on the imaginary axis \cite{Kapusta1989}. 
	\label{fig:gPoles}}
\end{figure}

To perform the Matsubara summation, the contour lines are deformed to enclose the poles of \mbox{$g(z)$}, 
see Fig.~\ref{fig:gPoles}. The sum $S_\eta$ is then the sum of the residues of \mbox{$g(z)h_\eta(z)$} 
over all poles of \mbox{$g(z)$}:
\beq
     S_\eta
     =
     -\frac{1}{\beta}
     \sum_{z_0 \in \{ g(z) \text{ poles} \}}
         \Res [g(z_0) h_\eta(z_0)] \,.
\eeq
The leading negative sign is due to the clockwise orientation of the contour.
%%%%%%%%%%%%%%%%%%%%%%%%%%%%%%%%%%%%%%%%%%%%%%%%%%%%%%%
%%%%%%%%%%%%%%%%%%%%%%%%%%%%%%%%%%%%%%%%%%%%%%%%%%%%%%%%

%%%%%%%%%%%%%%%%%%%%%%%%%%%%%%%%%%%%%%%%%%%%%%%%%%%%%%%%%%%%%%%%
\subsubsection{Choice of \texorpdfstring{$h_\eta(z)$}{TEXT}}
\label{sec:heta}
%%%%%%%%%%%%%%%%%%%%%%%%%%%%%%%%%%%%%%%%%%%%%%%%%%%%%%%%%%%%%%%%

The choice of a suitable Matsubara weighting function \mbox{$h_\eta(z)$} is a somewhat 
delicate matter. In order 
to guarantee simple poles at \mbox{$z = i \omega_n$}, for example, for bosonic frequencies 
\mbox{$\omega_n = 2 \pi n/\beta$}, any of the following weighting functions may be utilised:
\begin{align}
     h_+^{(1)}(z)
     & =
     \beta[1 + n_\text{B}(z)] \,,
     \nonumber \\
     h_+^{(2)}(z)
     & =
     \beta n_\text{B}(z) \,.
\end{align}
Note, $h_+^{(1)}(z)$ may be used to control convergence in the half-plane with 
\mbox{$\real\ z < 0$}, while \mbox{$h_+^{(2)}(z)$} controls convergence 
in the right half-plane where \mbox{$\real\ z > 0$}.

Adding the two functions to provide convergence in the entire complex plane yields
\beq
     S_\eta
     =
     \frac{1}{\beta} \sum_n g(i\omega_n)
     =
     \frac{1}{2 \pi i} \oint \! \dd z \,
         g(z) \cdot \frac{1}{2} \coth \frac{\beta z}{2} \,,
\eeq
and this relation will (by means of the contour deformation) further reduce down to
\beq
     S_\eta
     =
     -\frac{1}{2} \sum_{z_0 \in \{ g(z) \text{ poles} \}}
         \Res [ g(z_0) \cdot \coth \left( \frac{\beta z_0}{2} \right)] \,.
\eeq

%%%%%%%%%%%%%%%%%%%%%%%%%%%%%%%%%%%%%%%%%%%%%%%%%%%%%%%%%%%%%%%%
\subsubsection{Example: tadpole diagram}
\label{sec:tadexample}
%%%%%%%%%%%%%%%%%%%%%%%%%%%%%%%%%%%%%%%%%%%%%%%%%%%%%%%%%%%%%%%%

In the case of the tadpole diagram in $\phi^4$-theory
\beq
     g(z)
     =
     \sum_{s=\pm 1} 
         \frac{1}{2 \omega_{\vb k}}
         \frac{s}{z + s \omega_{\vb k}}\,.
\eeq
This function clearly has simple poles at \mbox{$z = -s \omega_{\vb k}$} and the two residues 
(one for each term in the sum) are \mbox{$\Res g(z_0) = s/2 \omega_{\vb k}$} due to the structure 
of the Laurent series of \mbox{$g(z)$}. Hence,
\begin{align}
     S_\eta
     & =
     -\frac{1}{2} \sum_{z_0 \in \{ g(z) \text{ poles} \}}
         \Res [ g(z_0) \cdot \coth \left( \frac{\beta z_0}{2} \right)]
     \nonumber \\ & =
     -\frac{1}{2} \sum_s
     \frac{s}{2 \omega_{\vb k}}
     \cdot
     \coth \left( \frac{-s \beta \omega_{\vb k}}{2} \right) \,.
\end{align}
Employing the following identities,
\begin{align}
     \coth(sz)
     & =
     s \coth (z)
     \\
     \coth\left( \frac{\beta z}{2} \right)
     & =
     [n(z) - n(-z)] \,, \label{eq:coshidentities}
\end{align}
the above equation simplifies to
\beq
     S_\eta
     =
     \frac{1}{2 \omega_{\vb \kappa}}
     [1 + 2n(\omega_{\vb \kappa})] \,.
\eeq

%%%%%%%%%%%%%%%%%%%%%%%%%%%%%%%%%%%%%%%%%%%%%%%%%%%%%%%%%%%%%%%%
\subsubsection{Example: eye diagram}
\label{sec:eyeexample}
%%%%%%%%%%%%%%%%%%%%%%%%%%%%%%%%%%%%%%%%%%%%%%%%%%%%%%%%%%%%%%%%

In the case of the eye-diagram of $\phi^3$-theory discussed further down in the context 
of thermal decay rates, the sum $S_\eta$ appears to be somewhat more complicated. In this case,
\beq
     g(z)
     =
     \frac{1}{4 \omega_{\vb q} \omega_{\vb k - \vb q}}
     \sum_{s, s'}
         \frac{s}{z + s\omega_{\vb q}}
         \cdot 
         \frac{s'}{i\omega_n - z + s'\omega_{\vb k - \vb q}} \,.
\eeq
  The poles of \mbox{$g(z)$} are simply the four poles \mbox{$z_0 = -s\omega_{\vb q}$} 
  and \mbox{$z_0 = i\omega_n + s'\omega_{\vb k - \vb q}$}. Hence, accounting for the residues over 
  these poles, the Matsubara frequency sum takes the form
\begin{align}
     S_\eta
     & =
     -\frac{1}{2}
     \sum_{z_0 \in \{ g(z) \text{ poles} \}}
     \Res [
         g(z_0) \cdot \coth \left( \frac{\beta z_0}{2} \right)
     ]
     \nonumber \\ & =
     -\frac{1}{2}
     \frac{1}{4 \omega_{\vb q} \omega_{\vb k - \vb q}}
     \sum_{s, s'} \Bigg \{
         \frac{s \cdot s'}{
             i\omega_n
             -
             (-s\omega_{\vb q})
             +
             s'\omega_{\vb k - \vb q}
         }
         \coth \left( \frac{-s\beta \omega_{\vb q}}{2} \right)
         -
         \nonumber
         \\ & \hspace{3.2cm} -
         \frac{s \cdot s'}{
             (i\omega_n + s'\omega_{\vb k - \vb q})
             +
             s\omega_{\vb q}
         }
         \coth \left(
             \beta
             \frac{i\omega_n + s'\omega_{\vb k - \vb q}}{2}
         \right)
     \Bigg \} \,.
\end{align}
Noticing that
\beq
     \coth \left( \beta \frac{z + i\omega_n}{2} \right)
     =
     \coth \left( \beta \frac{z}{2} \right)
\eeq 
  and taking out signs of the $\coth$-functions in order to cancel the sign 
  in front of the sum, we arrive at
\beq
     S_\eta
     =
     \frac{1}{2}
     \frac{1}{4 \omega_{\vb q} \omega_{\vb k - \vb q}}
     \sum_{s, s'}
     \Bigg \{
		s' \coth \left(
			\frac{\beta \omega_{\vb q}}{2}
		\right)
		+
		s \coth \left(
			\frac{\beta \omega_{\vb k - \vb q}}{2}
		\right)
	\Bigg \}
	\frac{1}{
		i\omega_n
		+ 
		s\omega_{\vb q}
		+
		s'\omega_{\vb k - \vb q})
	} \,.
\eeq
Then, expanding the sum over $s$, $s'$ and using \eq{coshidentities}, 
one gets the final result for the Matsubara summation:
\begin{align}
     S_\eta
     & =
     \frac{1}{4 \omega_{\vb q} \omega_{\vb k - \vb q}}
     \Bigg \{
         [1 + n(\omega_{\vb q}) + n(\omega_{\vb - \vb q})]
         \bigg [
         \frac{1}{i\omega_n  + \omega_{\vb q} + \omega_{\vb k - \vb q}}
         -
         \frac{1}{i\omega_n - \omega_{\vb q} - \omega_{\vb k - \vb q}}
         \bigg ]
         +
         \nonumber
         \\ & \hspace{2.5cm} +
         [n(\omega_{\vb q}) - n(\omega_{\vb - \vb q})]
         \bigg [
         \frac{1}{i\omega_n - \omega_{\vb q} + \omega_{\vb k - \vb q}}
         -
         \frac{1}{i\omega_n + \omega_{\vb q} - \omega_{\vb k - \vb q}}
         \bigg ]
     \Bigg \} \,,
\end{align}
that can be exploited in practical calculations.

%%%%%%%%%%%%%%%%%%%%%%%%%%%%%%%%%%%%%%%%%%%%%%%%%%%%%%%%%%%%%%%%
\section{The real-time formalism}
\label{sec:RealTimeFormalism}
\setcounter{equation}{0}
%%%%%%%%%%%%%%%%%%%%%%%%%%%%%%%%%%%%%%%%%%%%%%%%%%%%%%%%%%%%%%%%

Matsubara found that, by the procedure of analytic continuation of the Euclidean Green's functions, the retarded correlation 
functions with real time-arguments may be obtained in the imaginary-time formalism presented above. However, the evaluation 
of three-momentum integrals over Matsubara propagators summed over the frequencies is by no means straightforward 
for any but the simplest cases. By evaluating the Green's functions with respect to a contour that contains the real axis, in the so-called \emph{real-time formalism}, an analytic continuation is avoided completely. Such a real-time procedure 
is originally due to Schwinger~\cite{Schwinger1961} and Keldysh~\cite{Keldysh1965}. The formalism makes use of a special contour of temporal integration, the \emph{real-time contour} presented in \fig{KeldyshContour}, which has the effect 
of doubling the degrees of freedom. This property, in particular, results in a \mbox{$2\!\cross\!2$-matrix} 
structure of the propagator for a given field to be discussed below.

Other contours have been investigated as well. Particularly, one could mention the suggestion by Evans~\cite{Evans1993} to use an asymmetric contour.

%%%%%%%%%%%%%%%%%%%%%%%%%%%%%%%%%%%%%%%%%%%%%%%%%%%%%%%%%%%%%%%%
\subsection{Keldysh contour}
\label{sec:realtimeContour}
%%%%%%%%%%%%%%%%%%%%%%%%%%%%%%%%%%%%%%%%%%%%%%%%%%%%%%%%%%%%%%%%

\begin{figure}
  \centering
  \begin{tikzpicture}[scale=2.5]
    %axis
    \draw[->] (0,-1) -- (0,1);
    \node at (0.2, 1) {$\imaginary\ t$};
    \draw[->] (-2,0) -- (3,0);
    \node at (2.9, -0.1) {$\real\ t$};
    %C1
    \node at (-1, 0.2) {$t_\text{in}$};
    \draw[->,thick] (-1,0.07) -- (1,0.07);
    \node at (0.7, 0.2) {$C_1 = C_+$};
    \draw[-,thick] (1,0.07) -- (2.5,0.07);
    %C4
    \node at (2.5, 0.2) {$t_\text{fi}$};
    \draw[->,thick] (2.5,0.07) -- (2.5,-0.25);
    \node at (2.3, -0.20) {$C_4$};
    \draw[-,thick] (2.5,-0.20) -- (2.5, -0.40);
    \node at (2.5, -0.55) {$t_\text{fi} - i\sigma$};
    %C2
    \draw[->,thick] (2.5,-0.40) -- (1.1,-0.40);
    \node at (1.3, -0.55) {$C_2 = C_-$};
    \draw[-,thick] (1.1,-0.40) -- (-1,-0.40);
    %C3
    \draw[->,thick] (-1,-0.40) -- (-1,-0.70);
    \node at (-0.8, -0.65) {$C_3$};
    \draw[-,thick] (-1,-0.70) -- (-1, -0.9);
    \node at (-1.3, -0.9) {$t_\text{in} - i\beta$};
  \end{tikzpicture}
  \caption{A generalised Keldysh contour \mbox{$C = C_1 \cup C_2 \cup C_3 \cup C_4$}. The contour is extended 
  along the real axis in order to pick out all the real time-arguments of the \mbox{$n$-point} function 
  of interest. It then goes down by $i\sigma$ before going back in parallel to the real axis. Finally, 
  the contour goes down to the final time \mbox{$t_\text{in} \!-\! i\beta$.} The points $t_\text{in}$ 
  and $t_\text{fi}$ are arbitrary and may be suitably chosen for any given thermal system. In the equilibrium 
  limit, these end-points are taken to $\pm\infty$. 
  \label{fig:GeneralisedKeldyshContour}}
\end{figure}
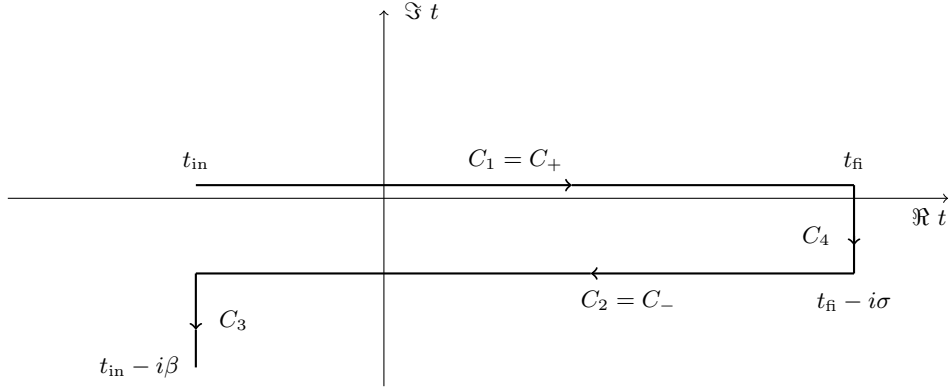

The real-time contour in \fig{KeldyshContour} is a special case of a family of real-time Keldysh-esque
contours~\cite{MatsumotoNakanoUmezawaManciniMarinaro1983, MatsumotoNakanoUmezawa1984} parametrised 
by \mbox{$\sigma \in [0,\beta]$} and presented in \fig{GeneralisedKeldyshContour}. The Keldysh contour 
(or sometimes Kadanoff-Baym-Keldysh contour, see e.g. Ref.~\cite{KemperMoritzFreericksDevereaux2013}) 
is recovered for \mbox{$\sigma = 0$.} An even more general family of contours was presented in 
Ref.~\cite{MatsumotoNakanoUmezawa1984} with pieces going back and forth $N$ times between 
\mbox{$\real\ t_\text{in}$} and \mbox{$\real\ t_\text{fi}$} in parallel to the real axis. However, 
it can be shown that such additional pieces cancel out for \mbox{$N\geq2$} and the fundamental reasons 
for discarding this extended class of contours are laid out in Ref.~\cite{MatsumotoNakanoUmezawa1984}. 
Hence, the contour presented in \fig{GeneralisedKeldyshContour} is the most general real-time contour 
needed to include all \mbox{$n$-point} functions with real arguments, with $\sigma$ being a freedom 
of choice. Commonly, this parameter is set to 0, $\beta$ or $\beta/2$ 
\cite{LandsmanvanWeert1987, KobesKowalski1986, NishikawaMorimatsuHidaka2003, GarnyMuller2009}.
%%%%%%%%%%%%%%%%%%%%%%%%%%%%%%%%%%%%%%%%%%%%%%%%%%%%%%%%%%%%%%%%
%%%%%%%%%%%%%%%%%%%%%%%%%%%%%%%%%%%%%%%%%%%%%%%%%%%%%%%%%%%%

It was argued previously, using Eqs.~(\ref{eq:ScalarPropagator}) and (\ref{eq:ArbitraryPropagator}), that the contour 
propagator depends explicitly on the choice of contour $C$. It is clear from these equations that the contour dependence 
of the propagator enters only through the contour step-function \mbox{$\Theta_C(t-t')$}. In the case of the real-time 
contour shown in \fig{GeneralisedKeldyshContour}, the time coordinates of $x$, $x'$ may be distributed arbitrarily 
over $C$ and several unique two-point functions (propagators) arise. Let us label them as $t_r$, $t_s'$ so that 
\mbox{$t_r \in C_r$}, \mbox{$t_s' \in C_s$} for \mbox{$r,s = \{ 1, 2, 3, 4 \}$}. Then, the different propagators 
may be expressed as
\beq
     D^{rs}(t - t') = D_C(t_r - t_s') \,.
\eeq
Note e.g. that $t_2, t_3$ always are later times than $t_1, t_4$. 
  
The appearance of several distinct propagators is a very general consequence of the statistical formulation 
of the \mbox{two-point} function in \eq{General2PointFunction}. The real-time contour of \fig{KeldyshContour} 
is treated in great detail by Wagner~\cite{Wagner1991}. In an approach valid both in as well as out of equilibrium, 
Wagner organised the \mbox{two-point} functions (Green's functions) in a matrix. In the special case 
of \mbox{$\sigma = 0$}, this matrix takes the following form
\beql{FullMatrixPropagatorOnKeldyshContour}
     \vb D(t, t')
     =
     \mqty(
         D^{11}(t, t')	& D^{12}(t, t')	& D^{13}(t, t') \\
         D^{21}(t, t')	& D^{22}(t, t')	& D^{23}(t, t') \\
         D^{31}(t, t')	& D^{32}(t, t')	& D^{33}(t, t')
     ) \,.
\eeq
Adding the piece $C_4$ \mbox{($\sigma \neq 0$)} is a trivial extension. For any system in equilibrium, 
Landsman and van Weert~\cite{LandsmanvanWeert1987}, Wagner~\cite{Wagner1991} and Das~\cite{Das1997} 
argued that the contour pieces $C_3$, $C_4$ decouple entirely from the parallel pieces due to
\beq
     D^{rs}(t - t') = 0
     \qquad
     \text{if } \big( r=1,2 \, \wedge \, s = 3,4 \big)
     \,\vee\,
     \big( r=3,4 \, \wedge \, s = 1,2 \big)
\eeq
in the limit \mbox{$t_\text{in} \to -\infty$} and \mbox{$t_\text{fi} \to +\infty$}. This may be proven 
by the application of the Riemann-Lebesgue lemma~\cite{KadisonRingrose1983} if the source field is constrained to
\beq
     \lim_{t\to\pm\infty} j(x) = 0 \,.
\eeq
This condition is compatible with the KMS condition~\cite{LandsmanvanWeert1987}. Due to the decoupling 
of the vertical contour segments, the generating functional factorises as
\beql{FactorisedGeneratingFunctional}
     Z[j] = Z[0]Z_{34}[j]Z_{12}[j]
\eeq
  with
\beql{ExplicitFactorisedGeneratingFunctional}
     Z_{rs}[j]
     =
     \exp[i\int_{C_{rs}} \! \dd^4 x \,
         \mathcal L_\text{I}
         \bigg[ \frac{\delta}{i\delta j(x)} \bigg]]
     \exp[-\frac{i}{2} \int_{C_{rs}} \! \dd^4 x \int_{C_{rs}} \! \dd^4 x' \,
         j(x)D_C(x-x')j(x')] \,.
\eeq
Here, \mbox{$C_{rs} = C_r \cup C_s$}. Therefore, the main focus, when considering the real-time Green's functions, 
is the upper-left \mbox{$2\!\cross\!2$}-corner of the matrix in \eq{FullMatrixPropagatorOnKeldyshContour} while the contribution 
from the vertical pieces is treated as a multiplicative factor. This factor only enters the connected Green's functions 
as an additive term in the cumulant expansion of \eq{ConnectedGreensFunctions}.

%%%%%%%%%%%%%%%%%%%%%%%%%%%%%%%%%%%%%%
\subsection{Real-time propagator}
\label{sec:RealTimePropagator}
%%%%%%%%%%%%%%%%%%%%%%%%%%%%%%%%%%%%%%

An explicit form of the real-time thermal propagator, in the case of a neutral scalar particle, can be found by examining
\eq{ExplicitFactorisedGeneratingFunctional} in the light of \eq{FactorisedGeneratingFunctional}. Having factorised out 
the vacuum bubbles \mbox{$Z[0]$} and the vertical contour pieces \mbox{$Z_{34}[j]$}, only the real-time part \mbox{$Z_{12}[j]$} 
will be considered from here on. The two exponentials in $Z_{12}[j]$ can be further factorised with respect to \mbox{$C_1\cup C_2$} 
by conveniently organising the propagator components in a matrix. From now on, the spatial variables will be suppressed. 
At the same time, a new notation will be introduced. The contour pieces $C_1$, $C_2$ in \fig{GeneralisedKeldyshContour} 
are relabelled as $C_+$, $C_-$, respectively, to indicate chronological and antichronological time evolution 
on the different parts of the contour. 
  
The factorisation of the first exponential becomes
\beql{ExponentialFactorisationOne}
     \exp[i\!\int_{C_{+-}} \!\!\! \dd t \,
         \mathcal L_\text{I}\bigg[ \frac{\delta}{i\delta j(t)} \bigg]
     ]
     =
     \exp[i\!\int_{\mathbb R} \! \dd t
     \Bigg(
         \mathcal L_\text{I}
         \bigg[ \frac{\delta}{i\delta j_+(t)} \bigg]
         -
         \mathcal L_\text{I}
         \bigg[ \frac{\delta}{i\delta j_-(t)} \bigg]
     \Bigg)] \,.
\eeq
The relative minus sign is a result of the antichronological evolution of the fields on $C_-$. The two functions here $j_+(t)$, $j_-(t)$ 
are introduced as sources of two separate field components $\phi_+$, $\phi_-$ that exist only on $C_+$, $C_-$, respectively. 
Together, the components form the full real-time field \mbox{$\phi = \mqty(\phi_+ & \phi_-)^\text{T}$}. The physical field 
component $\phi_+$ evolves chronologically and its integration limits are recognised from the zero-temperature field theory. 
The component $\phi_-$ is an unphysical degree of freedom since it evolves antichronologically. This field component can not 
appear on external lines of diagrams generated by the above exponential. Nevertheless, its presence is an unavoidable consequence 
of the Keldysh contour, and contributions from virtual $\phi_-$-fields in loops must be taken into account. Such contributions 
are introduced by the second term \mbox{$\mathcal L_\text{I}[\delta/i\delta j_-]$}. Summing over \mbox{$r,s = \pm$}, the second 
exponential factor can be rewritten as
\beq
     \exp[-\frac{i}{2}\!\iint_{C_{rs}} \!\!\!\! \dd t \, \dd t' \,
         j(t)D_C(t-t')j(t')]
     =
     \exp[-\frac{i}{2}\!\iint_{\mathbb R^2} \!\! \dd t \, \dd t' \,
         j_r(t)D^{rs}(t-t')j_s(t')] \,.
\eeq
The matrix components of the propagator must fulfil
\beq
  \begin{cases}
    D^{(++)}(t-t') \equiv D_C(t-t') \,, \\
    D^{(--)}(t-t') \equiv D_C\big( (t-i\sigma) - (t'-i\sigma) \big) \equiv -\big[ D^{(++)}(t-t') \big]^* \,, \\
    D^{(+-)}(t-t') \equiv D_C\big( t - (t'-i\sigma) \big) \,, \\
    D^{(-+)}(t-t') \equiv D_C\big( (t-i\sigma) - t' \big) \,.
  \end{cases}
\eeq
The right-most side of the second line reflects the negative orientation of $C_2$ and can be shown explicitly from, 
for example, \eq{RealTimeExplicitScalarPropagator}. These components may be used to express the general propagator 
of \eq{ArbitraryPropagator} for any bosonic or fermionic field of arbitrary spin and any charge. The explicit 
solution of \eq{ArbitraryPropagator} for each of the above components gives, after a Fourier transform 
of \eq{ScalarPropagator},
\beql{GeneralFieldPropagatorComponents}
  \begin{dcases}
    i\tilde D_{\alpha\beta}^{(++)}(k)
    =
    \Big[
      \Theta(k_0) i\tilde \Delta_F(k) + \Theta(-k_0)i\tilde \Delta_F^*(k) + \eta\rho_0(k)n(\omega_+)
    \Big] d_{\alpha\beta}(k) \,,
    \\
    i\tilde D_{\alpha\beta}^{(--)}(k) = \big[ i\tilde D^{(++)}(k) \big]^* d_{\alpha\beta}(k) \,,
    \\
    i\tilde D_{\alpha\beta}^{(+-)}(k) = \eta\rho_0(k) e^{\sigma k_0} n(\omega_+) d_{\alpha\beta}(k) \,,
    \\
    i\tilde D_{\alpha\beta}^{(-+)}(k) = \eta e^{\mp\beta\mu}e^{(\beta - 2\sigma)k_0} 
    \cdot i\tilde D_{\alpha\beta}^{(+-)}(k) \,.
  \end{dcases}
\eeq
Here, the internal indices $i$, $j$ have been suppressed, and 
\beq
     i\tilde \Delta_F(k) = \frac{i}{k^2 - m^2 + i\epsilon}
\eeq
is the Feynman propagator of the zero-temperature theory. Recall that the spectral density $\rho_0$ 
is the discontinuity over the real axis of this propagator (\eq{FreeSpectralDensity}) and may therefore 
also be written in terms of $i\tilde \Delta_F$, $i\tilde \Delta_F^*$ as
\beq
     \rho_0(k)
     =
     \text{sign}(k_0)
     \Big[ i\tilde \Delta_F(k) - i\tilde \Delta_F^*(k) \Big] \,.
\eeq
Hence, the propagator matrix may be diagonalised in a basis of zero-temperature functions by means of 
a Bogolyubov transformation. The result is
\beql{DiagonalRealTimePropagator}
  i\vb{\tilde D}_{\alpha\beta}(k)
  =
  \vb{M}_\eta(k)
  \mqty(
    d_{\alpha\beta}(k)i\tilde \Delta_F(k)	& 0 \\
    0	& -d_{\alpha\beta}(k)i\tilde \Delta_F^*(k)
  )
  \vb{M}_\eta(k) \,,
\eeq
with the thermal (Bogolyubov) matrix defined as follows
\beql{ThermalMatrixOfDiagonalisation}
  \vb M_\eta(k)
  =
  \mqty(
    \text{cos\hspace{-1.0pt}[\hspace{-.5pt}h\hspace{-.7pt}]} \hspace{.5pt} (\theta_k)
    &
    \eta e^{\beta\mu/2}e^{-(\beta - 2\sigma)k_0/2}
    \text{sin\hspace{-1.0pt}[\hspace{-.5pt}h\hspace{-.7pt}]} \hspace{.5pt} (\theta_k)
    \\
    e^{-\beta\mu/2}e^{(\beta - 2\sigma)k_0/2}
    \text{sin\hspace{-1.0pt}[\hspace{-.5pt}h\hspace{-.7pt}]} \hspace{.5pt} (\theta_k)
    &
    \text{cos\hspace{-1.0pt}[\hspace{-.5pt}h\hspace{-.7pt}]} \hspace{.5pt} (\theta_k)
  ) \,.
\eeq
Here the thermal angle $\theta_k$ was introduced\footnote{This diagonalisation procedure was first 
formulated by Umezawa, Matsumoto and Tachiki \cite{UmezawaMatsumotoTachiki1982} in the framework 
of thermo field dynamics.} through the functions
\beq
  \begin{dcases}
    \text{sin\hspace{-1.0pt}[\hspace{-.5pt}h\hspace{-.7pt}]} \hspace{.5pt} \theta_k = \sqrt{N(\omega)} \,, \\
    \text{cos\hspace{-1.0pt}[\hspace{-.5pt}h\hspace{-.7pt}]} \hspace{.5pt} \theta_k = 
    \Big[ \Theta(k_0) + \eta\Theta(-k_0) \Big]\sqrt{1 + \eta N(\omega)} \,,
  \end{dcases}
  \quad
  \text{for fermionic[bosonic] fields} \,.
\eeq
Here, the thermal function reads \mbox{$N(\omega) = \Theta(k_0)n(\omega_+) + \Theta(-k_0)n(-\omega_+)$}. 
The square-parenthesis notation conveys the point that for fermions the goniometric functions are considered 
while the hyperbolic functions apply to the scalar case. Note that the propagator is symmetric under simultaneous 
transposition and reversal of momentum, i.e.\ \mbox{$i\tilde D^{rs}(k) = i\tilde D^{sr}(-k)$}, for neutral 
bosons\footnote{Neutral bosons have \mbox{$\eta = +1$} and \mbox{$\mu = 0$}. Note, for the scalar field 
(commuting or anticommuting) the Klein-Gordon divisor \mbox{$d_{\alpha\beta}(k) = 1$}.}.

In the limit of \mbox{$T\to0$}, the Bogolyubov matrix becomes the unit matrix. In such a case, the field components 
completely decouple such that \mbox{$Z_{+-} \to Z_+Z_-$} and the contribution from $\phi_-$ may be fully absorbed 
into the normalisation constant. The diagonalisation of the propagator is particularly useful since any thermal 
dependence may be absorbed into vertices giving rise to a basis where the propagator itself is defined in terms 
of zero-temperature functions. Hence, the real-time formalism and the propagator are intrinsically connected 
to the imaginary-time formalism since they meet when the formalism of Matsubara is analytically continued 
to real time-arguments. This statement is formulated as follows
\beq
  i\tilde D^{(++)}(k)
  =
  i\tilde D(k)
  \equiv
  n(\omega_+)i\Delta(k_0-i\epsilon, \vb k) - [1 + n(\omega_+)]i\Delta(k_0+i\epsilon, \vb k) \,,
\eeq
where $\Delta$ is the analytic continuation of the Matsubara propagator of \eq{MatsubaraDeltaPropagator} down to the real axis.

%%%%%%%%%%%%%%%%%%%%%%%%%%%%%%%%%%%%%%%%%%%%%%%%%%%%
\subsubsection{Real-time scalar propagator(s)}
\label{sec:realtimescalarprop}
%%%%%%%%%%%%%%%%%%%%%%%%%%%%%%%%%%%%%%%%%%%%%%%%%%%%

The thermal decays of interest here involve interactions between fermions and scalar bosons. In this subsection 
and the following one, the real-time free propagators for such fields will be explicitly presented for further references.
  
For the choice of \mbox{$\sigma = 0$,} the components of \eq{GeneralFieldPropagatorComponents} may be written as
\beql{RealTimeExplicitScalarPropagator}
  \begin{dcases}
    i\tilde D^{(++)}(k) = \frac{i}{k^2 \!-\! m^2 \!+\! ik_0\epsilon} + \eta\rho_0(k)n(\omega_+) \,, \\
    i\tilde D^{(--)}(k) = \frac{-i}{k^2 \!-\! m^2 \!-\! ik_0\epsilon} + \eta\rho_0(k)n(\omega_+) \,, \\
    i\tilde D^{(+-)}(k) = \eta\rho_0(k)n(\omega_+) \,, \\
    i\tilde D^{(-+)}(k) = \rho_0(k) e^{\omega_+} n(\omega_+) \,.
  \end{dcases}
\eeq
Here, one recognises the retarded and advanced propagators
\beql{RetardedAdvancedScalarPropagators}
  i\tilde D^R(k) = \frac{i}{k^2 \!-\! m^2 \!+\! ik_0\epsilon} \,, \qquad
  i\tilde D^A(k) = \frac{-i}{k^2 \!-\! m^2 \!-\! ik_0\epsilon} \,,
\eeq
known from the zero-temperature theory. For practical calculations, it is common to rewrite the four components 
in terms of the Feynman propagator (and its conjugate), consequently rewriting also the thermal terms. 
Through the relation
\beql{RetardedPropagatorRelation}
  i\tilde D^R(k) = \Theta(k_0) i\tilde \Delta_F(k) - \Theta(-k_0) i\tilde \Delta_F^*(k)  \,,
\eeq
and by insertion of the explicit spectral density \eq{FreeSpectralDensity}, the components can be found as
\beql{ExplicitRealTimeScalarPropagatorComponents}
  \begin{dcases}
    i\tilde D^{(++)}(k) = \frac{i}{k^2 \!-\! m^2 \!+\! i\epsilon} + \eta 2\pi\delta\big( k^2 \!\!-\! m^2 \big)
      \Big[ n\big( \abs{\omega}_+ \big) + n\big( \abs{\omega}_- \big) \Big] \,,
    \\
    i\tilde D^{(--)}(k) = \frac{-i}{k^2 \!-\! m^2 \!-\! i\epsilon} + \eta 2\pi\delta\big( k^2 \!\!-\! m^2 \big)
      \Big[ n\big( \abs{\omega}_+ \big) + n\big( \abs{\omega}_- \big) \Big] \,,
    \\
    i\tilde D^{(+-)}(k) = \eta 2\pi\delta\big( k^2 \!\!-\! m^2 \big)
      \Big[ n\big( \abs{\omega}_+ \big) - n\big( \!\!-\!\abs{\omega}_- \big) \Big] \,,
    \\
    i\tilde D^{(-+)}(k) = -\eta 2\pi\delta\big( k^2 \!\!-\! m^2 \big)
      \Big[ n\big( \!\!-\!\abs{\omega}_+ \big) - n\big(\abs{\omega}_- \big) \Big] \,.
  \end{dcases}
\eeq
Here, the new thermal distributions introduced for compact notation read
\beq
  n\big( \abs{\omega}_\pm \big)
  =
  \frac{\Theta(\pm k_0)}{e^{\abs{\omega}_\pm} - \eta}, \qquad \abs{\omega}_\pm = \beta(\abs{k_0} \mp \mu) \,.
\eeq
Note especially that, in the case of a neutral boson, one might define
\beq
  n_\text{B}\big( \abs{k_0} \big)
  =
  \Big[ n\big( \abs{\omega}_+ \big) + n\big( \abs{\omega}_- \big) \Big]_{\mu = 0}
  \equiv
  \frac{1}{e^{\beta\abs{k_0}} - 1}
\eeq
and, for further convenience (see \mbox{Secs. \ref{sec:pStpSpS}-\ref{sec:FtpSF}}), 
in the case of charged fermions we have
\beq
  n_{\text{F}/\bar{\text{F}}}(k_0)
  =
  n\big( \abs{\omega}_\pm \big) + n\big( \abs{\omega}_\mp \big)
  \equiv
  \frac{\Theta(k_0)}{e^{\beta(\abs{k_0} \mp \mu)} + 1}
  +
  \frac{\Theta(-k_0)}{e^{\beta(\abs{k_0} \pm \mu)} + 1} \,.
\eeq
Here, the lower sign reflects the opposite charge of antiparticles. 

%%%%%%%%%%%%%%%%%%%%%%%%%%%%%%%%%%%%%%%%%%%%%%%%%%%%%%%%%%%%%%%%%%%%%%%%%%%%%%%%%%%%%
\subsubsection{Real-time spin-\texorpdfstring{$\tfrac{1}{2}$}{TEXT} propagator(s)}
\label{sec:realtimefermionprops}
%%%%%%%%%%%%%%%%%%%%%%%%%%%%%%%%%%%%%%%%%%%%%%%%%%%%%%%%%%%%%%%%%%%%%%%%%%%%%%%%%%%%%

Considering spin-$\tfrac{1}{2}$ fields, the Klein-Gordon divisor carry spin structure, 
and for fermions (upper sign)/antifermions (lower sign) it reads
\beq
     d_{\alpha\beta}(k) \to d_{\text{F}/\bar{\text{F}}}(k)
     =
     \big( \fsl{k} \pm m \big) \,.
\eeq
Recall that \mbox{$\eta = -1$} for Fermi-Dirac statistics. The propagator components may 
then be found from \eq{GeneralFieldPropagatorComponents} as
\beql{ExplicitRealTimeFermionPropagatorComponents}
  \begin{dcases}
    i\tilde S_{\text{F}/\bar{\text{F}}}^{(++)}(k)
    =
    \big( \fsl{k} \pm m \big) \! \bigg\{
      \frac{i}{k^2 \!-\! m^2 \!+\! i\epsilon} - 2\pi\delta\big( k^2 \!\!-\! m^2 \big) \, n_{\text{F}/\bar{\text{F}}}(k_0)
    \bigg\} \,,
    \\
    i\tilde S_{\text{F}/\bar{\text{F}}}^{(--)}(k)
    =
    \big( \fsl{k} \pm m \big) \! \bigg\{
      \frac{-i}{k^2 \!-\! m^2 \!-\! i\epsilon} - 2\pi\delta\big( k^2 \!\!-\! m^2 \big) \, n_{\text{F}/\bar{\text{F}}}(k_0)
    \bigg\} \,,
    \\
    i\tilde S_{\text{F}/\bar{\text{F}}}^{(+-)}(k)
    =
    -2\pi \big( \fsl{k} \pm m \big) \delta\big( k^2 \!\!-\! m^2 \big)
    \Big[ n\big( \abs{\omega}_\pm \big) - n\big( \!\!-\!\abs{\omega}_\mp \big) \Big] \,,
    \\
    i\tilde S_{\text{F}/\bar{\text{F}}}^{(-+)}(k)
    =
    2\pi \big( \fsl{k} \pm m \big) \delta\big( k^2 \!\!-\! m^2 \big)
    \Big[ n\big( \!\!-\!\abs{\omega}_\mp \big) - n\big(\abs{\omega}_\pm \big) \Big] \,.
  \end{dcases}
\eeq
In the simple case of neutral particles, with \mbox{$\mu = 0$}, these components 
reduce to those of Das \cite{Das1997}.

%%%%%%%%%%%%%%%%%%%%%%%%%%%%%%%%%%%%%%%%%%
\subsection{Real-time Feynman rules}
\label{sec:realtimeFeynman}
%%%%%%%%%%%%%%%%%%%%%%%%%%%%%%%%%%%%%%%%%%

It was seen in the previous section that the Keldysh contour in \fig{KeldyshContour} necessarily gives rise 
to a doubling of the degrees of freedom in the real-time thermal theory. This is achieved by separating the field 
into two components, each appearing on the respective segment of the real-time contour. One component is physical 
while the other should be regarded as a \emph{thermal ghost} arising from the choice of the specific contour. 
For notational convenience, a neutral scalar field $\phi$ will be considered in the derivation of the Feynman 
rules below. For illustration, the field is assumed to have an interaction term proportional to \mbox{$\kappa \phi^l$}. 
The generality of the method, however, allows for the treatment of the multicomponent vector field $\PHI_{\alpha}^{i}$ 
introduced above. The derivation of the Feynman rules for the case of $\PHI_{\alpha}^{i}$ is analogous to the case 
of a neutral scalar theory, but with the presence of contraction of Lorentz and internal indices in addition 
to the summation over the thermal propagator components.

\begin{figure}[htbp]
\centering
\begin{minipage}{0.45\textwidth}
     \centering
     \raisebox{-0.5\height}{%
         \includegraphics[scale=0.35]{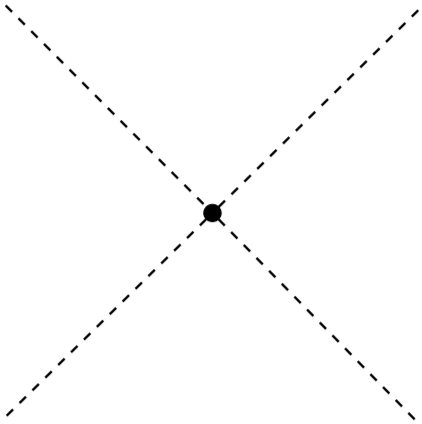}
     }
     $= -i\kappa$
     \\
     \hspace{-1.1cm} (a)
\end{minipage}
\hfill
\begin{minipage}{0.45\textwidth}
     \centering
     \raisebox{-0.5\height}{%
         \includegraphics[scale=0.35]{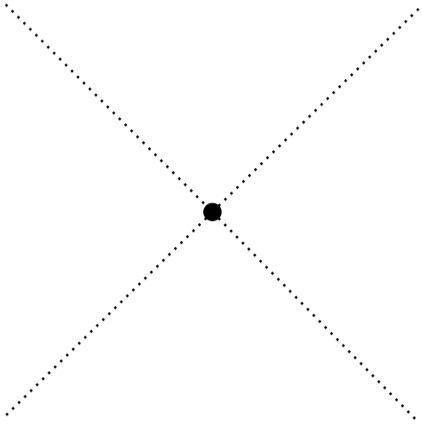}
     }
     $= +i\kappa$
     \\
     \hspace{-1.1cm} (b)
\end{minipage}
\caption{The two real-time vertices of $\phi^4$-theory, appearing on each horizontal 
piece of the real-time contour, respectively. (a) Vertex corresponding to the quartic 
interaction of the physical field $\phi_+$. (b) Vertex corresponding to the quartic 
interaction of the unphysical field $\phi_-$. 
\label{fig:PhiFourRealTimeVertices}}
\end{figure}

It was seen in \sec{RealTimePropagator} that, for a field with no derivative couplings, the relevant factor 
in the generating functional, $Z_{12}[j]$, can be written as
\begin{align}
  & Z_{+-}[j] \to Z[j_+, j_-] \nonumber
  \\ & =
  \exp[i\!\int_{\mathbb R} \! \dd t
    \Bigg(
      \mathcal L_\text{I}\bigg[ \frac{\delta}{i\delta j_+} \bigg]
      -
      \mathcal L_\text{I}\bigg[ \frac{\delta}{i\delta j_-} \bigg]
    \Bigg)
  ]
  \exp[ -\frac{i}{2} \! \iint_{\mathbb R^2} \! \dd t \, \dd t' \,
    j_r(t) D^{rs}(t-t') j_s(t')
  ] \,.
\end{align}
The series expansion of this function generates all the diagrams of the theory in analogy to the zero-temperature case. 
Two distinct vertices appear due to the two interaction terms in the leftmost exponential: one interaction vertex involves 
only the physical field component and one vertex involves only the ghost field. See \fig{PhiFourRealTimeVertices} for 
the vertices of the $\phi^4$-theory as an instructive example. Hence, the field components do not mix in the vertices. 
However, due to non-zero off-diagonal components of the propagator on the Keldysh contour, the fields may propagate 
into each other.

The connected Green's functions of \eq{ConnectedGreensFunctions} are generated by functional differentiation according to
\beq
  G^{(n,m)}(t_1, \ldots, t_n, t_{n+1}, \ldots, t_{n+m})
  =
  \frac{\delta^{n+m} \log Z[j_+, j_-]}{i\delta j_+(t_1)\cdots i\delta j_+(t_n) i\delta j_-(t_{n+1})\cdots i\delta j_-(t_{n+m})}\Bigg\vert_{\substack{j_+=0 \\ j_-=0}}\hspace{-.7cm}  \,.
\eeq
It is clear that, for any value of \mbox{$\sigma \neq 0$}, the \emph{real-time} Green's functions may only be recovered by functional differentiation with respect to the source $j_+$ since the second contour segment includes no real times. It may be proven \cite{LandsmanvanWeert1987} that the real-time Green's functions are independent of the parameter $\sigma$ and, hence, 
they are given by
\beq
  G^{(n,0)}(t_1,\ldots, t_n)
  =
  \frac{\delta^n \log Z[j_+, j_-]}{i\delta j_+(t_1)\cdots i\delta j_+(t_n)}\Bigg\vert_{\substack{j_+=0 \\ j_-=0}} \,, 
  \qquad \forall \sigma \,.
\eeq
The thermal ghost field $\phi_-$ may only propagate in internal lines that connect the vertices. These lines are generated 
by functional differentiation with respect to $j_-$ inside $Z[j_+, j_-]$, see \eq{ExponentialFactorisationOne}.

The full set of Feynman rules can easily be extracted through the generating functional in its path-integral form 
if the \L\ is split into its free and interaction parts. Then, in terms of the action, the generating functional may be written as
\begin{align}
  Z[j_+, j_-] & =
  \int \! [\mathcal D\phi_+] \!\! \int \! [\mathcal D\phi_-] \,
    \exp[ i\int_{\mathbb R} {\dd t}
      \Big\{
        \phi_r K^{rs} \phi_s + \mathcal L_\text{I}(\phi_+) - \mathcal L_\text{I}(\phi_-) + j_r\phi_r
      \Big\}
    ] \nonumber
    \\ & =
    \int \! [\mathcal D\phi_+] \!\! \int \! [\mathcal D\phi_-] \,
      e^{iS_0} \, e^{i(S_\text{I} + j\cdot\phi)} \,,
      \qquad\qquad \text{with } j\!\cdot\!\phi = \int_{\mathbb R} \! \dd t \, j_r\phi_r \,.
\end{align}
Summation over \mbox{$r,s = \pm$} is assumed. Expanding the second exponential, each term may be interpreted 
as an expectation value in the path-integral formalism with respect to a weight $e^{iS_0}$ so that
\begin{align}
  Z[j_+, j_-] & =
  \int \! [\mathcal D\phi_+] \!\! \int \! [\mathcal D\phi_-] \,
    e^{iS_0} \bigg\{
      1 + i\big( S_\text{I} + j\!\cdot\!\phi \big) + \tfrac{i^2}{2}\big( S_\text{I} + j\!\cdot\!\phi \big)^2 + \ldots
    \bigg\} \nonumber
  \\ &=
  Z_0[0,0]\bigg\{
    1 + i\expval*{(S_\text{I} + j\!\cdot\!\phi)}_0 + \tfrac{i^2}{2}\expval*{(S_\text{I} + j\!\cdot\!\phi)^2}_0 + \ldots
  \bigg\}\,.
\end{align}
By means of the cumulant expansion, it is possible to factor out not only the free \mbox{$Z_0[0,0]$} but 
\emph{all} vacuum bubbles \mbox{$Z[0,0]$} so that the cumulant becomes
\beq
  W[j_+,j_-] \equiv \log Z[j_+, j_-]
  =
  \log Z[0,0] + \expval*{e^{iS_\text{I}} (e^{i\hspace{0.9pt} j\cdot\phi} - 1)}_\text{con} \,.
\eeq

The expansion of the final term gives rise to all connected graphs with at least one external leg 
of the physical field $\phi_+$. The real-time Feynman rules are extracted from this series as follows:
\begin{itemise}
  \item Draw diagrams using $\phi_+$-, $\phi_-$-vertices: $n$ external $\phi_+$-lines and arbitrarily many internal $\phi_+$- and $\phi_-$-lines. Determine symmetry factors.
  \item Connect vertices by assigning a propagator: \\
    \mbox{
      $i\tilde D^{rs}(k) = \expval{\phi_r\phi_s}_0 = r$
      \raisebox{0.05\height}{%
         \includegraphics[scale=0.13]{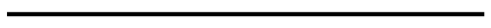}
         }
      $s$
    }.
  \item Assign a factor of $-i\kappa$ to each $\phi_+$-vertex and $i\kappa$ to each $\phi_-$-vertex.
  \item Impose energy-momentum conservation at each vertex through 
  \mbox{$(2\pi)^4 \, \delta^{(4)}\Big( \textstyle\sum_i k_i \Big)$}. 
  A global conservation factor may be separated out as \mbox{$(2\pi)^4 \,\delta^{(4)}(0)$}.
  \item Integrate over each internal momentum $k_i$: \mbox{$\int \! \frac{\dd^4 k}{(2\pi)^4}$} 
  and sum over all internal distributions of $r$, $s$.
\end{itemise}
Comparing the real-time formalism to the Matsubara theory, the propagator has gained a matrix 
structure and, further, two separate types of vertices have appeared.

%%%%%%%%%%%%%%%%%%%%%%%%%%%%%%%%%%%%%%%%%%%%%%%%%%%%%%%%%%%%%%%%
\section{Thermal self-energies}
\label{sec:ThermalSelfEnergies}
\setcounter{equation}{0}
%%%%%%%%%%%%%%%%%%%%%%%%%%%%%%%%%%%%%%%%%%%%%%%%%%%%%%%%%%%%%%%%

This section discusses the expression of the thermal decay rate in terms of the self-energy. The topic 
has been treated by e.g.~Weldon\cite{Weldon1983}, Ho and Scherrer~\cite{HoScherrer2015} and discussions 
on the thermal self-energy has also been advanced by \cite{KobesSemenoff1985,LevinsonBoal1985, KobesSemenoff1986, FujimotoMorikawaSasaki1986, LandsmanvanWeert1987, Das1997, NishikawaMorimatsuHidaka2003, CzarneckiKamionkowskiLeeMelnikov2012}.
Weldon~\cite{Weldon1983} related the self-energy of the Matsubara formalism to a quantity interpreted 
as the thermal decay rate. In the sections below, we will examine the self-energy of Matsubara, 
its analytic continuation and finally the connection of real-time quantities to it. This connection 
has been discussed by several authors, see for example Refs.~\cite{FujimotoMorikawaSasaki1986, LandsmanvanWeert1987, NishikawaMorimatsuHidaka2003}.

%%%%%%%%%%%%%%%%%%%%%%%%%%%%%%%%%%%%%%%%%%%%%%%%%%%%%%%%%%%%%%%%
\subsection{Matsubara self-energies}
\label{sec:Matsubaraselfenergies}
%%%%%%%%%%%%%%%%%%%%%%%%%%%%%%%%%%%%%%%%%%%%%%%%%%%%%%%%%%%%%%%%

The connected two-point function of the Matsubara formalism is obtained by functional differentiation 
of the cumulant in \eq{MatsubaraCumulant}. In order to obtain a perturbative series, and hence the Feynman rules, 
in momentum space, the full generating functional of \eq{MatsubaraPertubationSeries} may be Fourier transformed. 
It may be further manipulated in order to recover a form similar to that of \eq{GeneratingFunctionalSeries}. 
The full propagator appearing in Feynman diagrams is introduced as
\beq
     \tilde G_{(T=0)}(k_1, \ldots, k_n)
     \to
     (-i)^n \tilde G_\text{E}(k_1, \ldots, k_n) 
\eeq
staring from its zero-temperature counterpart. Since global momentum conservation is imposed for all diagrams, 
an overall factor of \mbox{$\beta(2\pi)^3\delta_{n0}\delta\big( \! \sum \vb k_i \big)$} may be extracted, 
leaving the interacting many-body propagator labelled \mbox{$\tilde{\mathcal G}(k)$} by Landsman 
and \mbox{van Weert}~\cite{LandsmanvanWeert1987}. Note here that the propagator carries the discrete 
energies in terms of the Matsubara frequencies. Real-time quantities must be obtained through analytic 
continuation. Analogous to \eq{ImaginaryTimePropagator}, this propagator may be defined in terms of 
the full Matsubara propagator that has been extended away from the discrete frequencies by means 
of analytic continuation: \mbox{$\tilde\Delta(i\omega_n, \vb k) \to \tilde\Delta'(z, \vb k)$}. Hence,
\beq
     \tilde{\mathcal G}(i\omega_n, \vb k)
     =
     \tilde\Delta'(z, \vb k)\big\vert_{z=i\omega_n} \,,
     \qquad
     \tilde\Delta'(z, \vb k)
     =
     \int\!\frac{\dd k_0}{2\pi}\frac{\rho(k)}{k_0 - z} \,.
\eeq

For the full propagator, the free spectral density $\rho_0$ has been replaced by the corresponding full 
quantity defined as \mbox{$i\rho(k) = \text{Disc}\,\tilde\Delta'(k)$.} The procedure of analytic continuation 
allows the propagator to be defined for real energies. This continuation is not unique and that is usually 
resolved by letting \mbox{$\lim_{\abs{z}\to\infty} \tilde\Delta'(z, \vb k) = 0$} and by taking 
\mbox{$\tilde\Delta'(z, \vb k)$} to be analytic away from the real axis~\cite{LandsmanvanWeert1987, DolanJackiw1974} 
which provides the above expression. Thus, we have uniquely
\beq
     \tilde\Delta'(z, \vb k)
     =
     \int_0^{\infty} \! \frac{\dd (k_0')^{\!2}}{2\pi} \,
         \frac{\rho(k_0', \vb k)}{z^2 - {k_0'}^2}
\eeq
provided the property \mbox{$k_0\rho(k) \geq 0$}. This function can be shown to have neither zeroes 
nor poles off the real axis provided that this inequality property holds true~\cite{LandsmanvanWeert1987}. 
Given that the extended propagator is analytic off the real axis, its analytic inverse exists and 
fulfils a Dyson-like equation
\beq
     {\tilde\Delta'(z, \vb k)}^{-1}
     =
     {\tilde\Delta(z, \vb k)}^{-1} + \Pi(z, \vb k) \,,
\eeq
which defines the analytic self-energy. Note the \emph{free} analytic propagator on the right-hand side.

Guided by the above consideration, it is useful to define the thermal Feynman propagator as
\beq
  \tilde\Delta_\text{F}'(k) = -\tilde\Delta'(k_0 + ik_0\epsilon, \vb k) \,.
\eeq
This propagator may be inverted by writing \mbox{$k_0^2 - \abs{\vb k}^2 = m^2$} so that the thermal 
self-energy for the imaginary-time formalism can be extracted:
\beql{AnalyticContinuationFunction}
     {\tilde\Delta_\text{F}}^{'-1}
     =
     k^2 - m^2 - \bar\Pi_\text{F}(k) \,.
\eeq
As in the zero-temperature theory, one may observe that the real part of the self-energy may be absorbed 
as a correction to the mass while its imaginary component will be interpreted as a decay rate~\cite{Weldon1983}.

%%%%%%%%%%%%%%%%%%%%%%%%%%%%%%%%%%%%%%%%%%%%
\subsection{Real-time self-energies}
\label{sec:realtimeselfenergies}
%%%%%%%%%%%%%%%%%%%%%%%%%%%%%%%%%%%%%%%%%%%%

\begin{figure}[htbp]
  \centering
  \begin{equation*}
     \raisebox{-0.45\height}{%
         \includegraphics[scale=0.3]{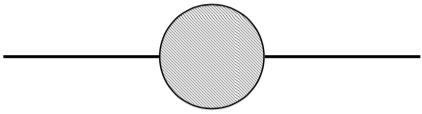}
     }
    \,=\,
    \raisebox{-0.05\height}{%
         \includegraphics[scale=0.15]{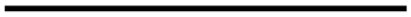}
     }
    \,+\,
    \raisebox{-0.45\height}{%
         \includegraphics[scale=0.3]{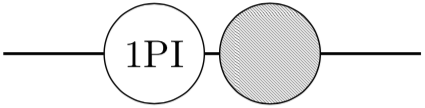}
     }
    \,+\,
    \ldots
  \end{equation*}
  \caption{The graphical interpretation of the Schwinger-Dyson equation. All 1PI insertions are parametrised by the self-energy. By recursive insertion of the full propagator, the 1PI series expansion may be obtained. \label{fig:SchwingerDysonEquation}}
\end{figure}

The full real-time propagator may be assumed to satisfy the Schwinger-Dyson equation 
(see Ref.~\cite{LandsmanvanWeert1987} and references therein) similar to the zero-temperature case:
\beql{RealTimeSchwingerDysonEquation}
     \vb {\tilde {\mathcal D}}_{\alpha\beta}^{rs}(k)
     =
     \vb {\tilde D}_{\alpha\beta}^{rs}(k)
     +
     \big(
         \vb{\tilde D} \vb{\tilde \Pi} \vb{\tilde {\mathcal D}}
     \big)_{\alpha\beta}^{rs}(k) \,.
\eeq
Like in the zero-temperature theory, the real-time self-energy is most often\footnotemark\ the sum of 
all one-particle irreducible (1PI) insertions on the propagator line as visualised in \fig{SchwingerDysonEquation}. 
This self-energy coincides with the self-energy of the many-body propagator provided by 
Ref.~\cite{LandsmanvanWeert1987} in connection to their discussion on the Matsubara theory.
The full real-time components are not independent as a consequence of the KMS condition and due to translational 
invariance \cite{LandsmanvanWeert1987} and they satisfy similar relations as the components of
\eq{GeneralFieldPropagatorComponents}.
  
\footnotetext{A 1PI diagram is defined to be any diagram that cannot be split into two parts by removal of 
a single propagator line. The 1PI blob shown in \fig{SchwingerDysonEquation} is defined as the sum of all 
such diagrams with the eye-diagram (the single loop) being the lowest-order contribution. 
See Ref.~\cite{PeskinSchroeder1995} for a discussion of the 1PI insertion procedure. One must note, 
however, that the properly defined self-energy includes one-particle reducible diagrams as well, e.g. 
tadpole diagrams that can be regulated separately.}
%%%%%%%%%%%%%%%%%%%%%%%%%%%%%%%%%%%%%%%%%%%%%%%%%%%%%%%%%%%%%%%%
%%%%%%%%%%%%%%%%%%%%%%%%%%%%%%%%%%%%%%%%%%%%%%%%%%%%%%%%%%%%%%%%

%%%%%%%%%%%%%%%%%%%%%%%%%%%%%%%%%%%%%%%%%%%%%%%%%%%%%%%%%%%%%%%%
\subsection{Self-energy relations}
\label{sec:SERelations}
%%%%%%%%%%%%%%%%%%%%%%%%%%%%%%%%%%%%%%%%%%%%%%%%%%%%%%%%%%%%%%%%

Viewing the free-field propagator of \eq{GeneralFieldPropagatorComponents} as a special case of the full propagator 
of \eq{RealTimeSchwingerDysonEquation}, the matrix of \eq{ThermalMatrixOfDiagonalisation} also diagonalises 
the full propagator for some function \mbox{$\tilde {\mathcal D}_{\text{F}\alpha\beta}(k)$} so that
\beq
     \tilde{\mathcal D}_{\alpha\beta}(k)
     =
     \vb M_\eta
     \mqty(
         \tilde{\mathcal D}_{\text{F}\alpha\beta}	& 0
         \\
         0	& \tilde{\mathcal D}_{\text{F}\alpha\beta}^*
     )
     \vb M_\eta \,.
\eeq
The new function is assumed to satisfy the Schwinger-Dyson equation that defines 
the self-energy $\bar\Pi_\text{F}$:
  \mbox{$
     \tilde {\mathcal D}_{\text{F}\alpha\beta}(k)
     =
     \tilde D_{\text{F}\alpha\beta}(k)
     +
     \big( \tilde D_\text{F}
     \bar\Pi_\text{F}
     \tilde{\mathcal D}_\text{F} \big)_{\alpha\beta}
  $}.
As an example, the relation between \mbox{$\tilde {\mathcal D}_{\text{F}\alpha\beta}(k)$} 
and the component \mbox{$\tilde {\mathcal D}_{\alpha\beta}^{(++)}(k)$} reads
\beq
     \tilde {\mathcal D}_{\alpha\beta}^{(++)}(k)
     =
     \big[ 1 + \eta n(\omega_+) \big]
     \tilde {\mathcal D}_{\text{F}\alpha\beta}(k)
     +
     \eta n(\omega_+)
     \tilde {\mathcal D}_{\text{F}\alpha\beta}^*(k) \,.
\eeq

The insertion of \mbox{$\vb M_\eta$} and its inverse in the Schwinger-Dyson equation 
allows for the identification
\beql{SelfEnergyRelations}
     \vb{\tilde \Pi}
     =
     \mqty(
         \tilde\Pi^{(++)}	& \tilde\Pi^{(+-)}
         \\
         \tilde\Pi^{(-+)}	& \tilde\Pi^{(--)}
     )
     \equiv
     \vb M_\eta^{-1}
     \mqty(\bar\Pi_\text{F}	& 0 \\ 0	& -\bar\Pi_\text{F}^*)
     \vb M_\eta^{-1} \,.
\eeq
The derivation of this is rather lengthy but can be found in Ref.~\cite{LandsmanvanWeert1987} where, most importantly, 
it is argued that $\bar\Pi_\text{F}$ is indeed the analytically continued self-energy of the re-summed propagator
\mbox{$\tilde\Delta_F'$} in \eq{AnalyticContinuationFunction}. For notational convenience, possible spinor and Lorentz 
indices have been suppressed but are generally present in the structure of the Klein-Gordon divisor $d_{\alpha\beta}^{ij}$. 

By inserting the explicit expression for the thermal matrix $\vb M_\eta$, the following relations may be deduced:
\begin{align}
     \tilde\Pi^{(--)} & = -\big[ \tilde\Pi^{(++)} \big]^* \,,
     \\
     \tilde\Pi^{(-+)}
     & =
     \eta e^{-\beta\mu}e^{(\beta - 2\sigma)k_0}
     \tilde\Pi^{(+-)} \,.
\end{align}
It is similarly straightforward to relate \mbox{$\tilde\Pi^{(-+)}$} to \mbox{$\tilde\Pi^{(++)}$}. Hence, a number 
of components has reduced from the initial four to a single independent one. The two relations above hold for 
any $\sigma$, also for \mbox{$\sigma = 0$}. Importantly for the forthcoming calculations, the real and imaginary 
parts of the self-energies are related as
\begin{align}
\label{eq:ImaginaryRealTimeSelfEnergyComponent}
     \Re\tilde\Pi^{(++)} & = \Re\bar\Pi_\text{F} \,,
     \\
     \Im\tilde\Pi^{(++)}
     & =
     \text{sign}(k_0)
     \big[ 1 + 2\eta n(\omega_+) \big] \Im\bar\Pi_\text{F} \,.
\end{align}
The self-energy on the left-hand side will be calculated to the first order for several types of interaction 
vertices in \mbox{Secs.~\ref{sec:pStpSpS}-\ref{sec:FtpSF}}.

%%%%%%%%%%%%%%%%%%%%%%%%%%%%%%%%%%%%%%%%%%%%%%%%%%%%%%%%%%%%%%%%
\section{Thermal decay rates}
\label{sec:ThermalDecayRates}
%%%%%%%%%%%%%%%%%%%%%%%%%%%%%%%%%%%%%%%%%%%%%%%%%%%%%%%%%%%%%%%%

In this section, the self-energies of the previous section will be related to the thermal decay rates. 
Initially, a relation between the thermal quantity that corresponds to the zero-temperature decay rate 
and the self-energy of the imaginary-time formalism due to Weldon~\cite{Weldon1983} will be presented. 
This relation will then be restated in terms of the self-energy components of the real-time formalism 
by making use of the self-energy relations in \sec{SERelations}.

%%%%%%%%%%%%%%%%%%%%%%%%%%%%%%%%%%%%%%%%%%%%%%%%%%%%%%%%%%%%%%%%
\subsection{Thermal rates in Matsubara formalism}
\label{sec:Imagtimeselfenergy}
%%%%%%%%%%%%%%%%%%%%%%%%%%%%%%%%%%%%%%%%%%%%%%%%%%%%%%%%%%%%%%%%

The decay rate $\gamma_D$ for a given process in the zero-temperature theory may be related 
to the discontinuity of the self-energy (i.e. its imaginary part) by means of the optical theorem as
\beq
     \gamma_D
     =
     -\left(\frac{\Im\Pi_0(E_0)}{E_0}\right) \,,
\eeq
where $E_0$ is the energy of the decaying particle. Weldon~\cite{Weldon1983} defined a similar 
quantity $\Gamma(p_0)$ through
\beq
     \Gamma(p_0)
     =
     -\left(\frac{\Im\bar\Pi_\text{F}(p_0)}{p_0}\right) \,,
\eeq
where $\bar\Pi_\text{F}$ is the self-energy of the imaginary-time formalism. 

It is important to reflect upon the physics that this quantity describes. One may assume that 
the distribution of a particle $\Phi$ at some time $t_\text{in}$ is described by a nonequilibrium 
function which Weldon labelled as \mbox{$f(p_0,t_\text{in})$} \cite{Weldon1983}. This function will 
approach the thermal distribution of the equilibrium (Bose-Einstein or Fermi-Dirac) in a simple manner 
if a deviation from equilibrium is small (\mbox{$\partial f/\partial t \ll 1$}). The rate of this approach 
is parametrised by $\Gamma(p_0)$ to the first order. The evolution of this distribution may be 
formulated according to
\beql{ApproachToEquilibrium}
     \pdv{f}{t}
     =
     -f\Gamma_D + (1 + \eta f)\Gamma_I \,.
\eeq
in terms of the quantities $\Gamma_D$, $\Gamma_I$. Here, the first term expresses the loss of 
$\Phi$-particles through decay modes while $\Gamma_I$ takes into account particle production 
by the medium. The thermal medium is filled with particles that couple to $\Phi$ and, hence, 
the second parameter $\Gamma_I$, the \emph{inverse decay rate} (production rate), adds 
the contribution from processes in the medium that produce $\Phi$ particles. As an example, 
production may occur through reactions \mbox{$\phi\phi \to \Phi$}. Weldon comprehensively 
presents an analysis of possible production and decay channels for $\Phi$ through interactions 
with particles in the thermal medium.

The solution to \eq{ApproachToEquilibrium} is
\beq
     f(p_0, t) = n(\omega_+) + c(p_0) e^{-\Gamma(p_0) t} \,,
\eeq
for some function $c(p_0)$ which is constant in time. This solution requires that the temperature $T$ 
is constant over time which is true for equilibrated media. Since $T$ characterises the background 
medium, the deviations of the $\Phi$ distribution from its equilibrium limit are required to be small 
and one may assume that the distribution of the medium particles corresponds to the thermal 
equilibrium distribution. The net decay rate $\Gamma$ of the distribution of $\Phi$ in the medium 
is the rate at which the distribution of $\Phi$ approaches the equilibrium regime and amounts to
\beq
     \Gamma(p_0) = \Gamma_D - \eta\Gamma_I \,.
\eeq
The evaluation of the amplitude for the thermal forward decay $\Gamma_D$ is of interest in this work. 
Weldon provides a thermal relation between the forward and inverse decay rates due to unitarity 
so that one therefore may write the decay rate of $\Phi$ as
\beql{ImaginaryTimeDecayRate}
     \Gamma_D
     =
     -\frac{1}{1 - \eta e^{-\beta (p_0 - \mu)}}
     \frac{\Im\bar\Pi_\text{F}(p_0)}{p_0} \,.
\eeq
The expression provided by Weldon has hereby been extended in order to explicitly take into account 
the role of the chemical potential $\mu$.

%%%%%%%%%%%%%%%%%%%%%%%%%%%%%%%%%%%%%%%%%%%%%%%%%%%%%%%%%%%%%%%%
\subsection{Thermal rates in the real-time formalism}
\label{sec:DecayRateRealTimeSelfEnergy}
%%%%%%%%%%%%%%%%%%%%%%%%%%%%%%%%%%%%%%%%%%%%%%%%%%%%%%%%%%%%%%%%

The relations of \sec{SERelations} connect the self-energy of the Matsubara formalism 
to the components of the real-time self-energy. Hence, \eq{ImaginaryTimeDecayRate} may be 
rewritten as
\beql{RealTimeDecayRate}
     \Gamma_D
     =
     -\text{sign}(p_0)
     \frac{1 + \eta n(\omega_+)}{1 + 2\eta n(\omega_+)}
     \frac{\Im\tilde\Pi^{(++)}(p_0)}{p_0} \,.
\eeq
In the following sections \ref{sec:pStpSpS}-\ref{sec:FtpSF}, the $(++)$-component of the real-time 
self-energy for several types of interactions has been evaluated. The results are used to extract the decay rates 
for fields in an equilibrated thermal medium.

Note that, if \mbox{$\tilde\Pi^{(++)}$} comes with internal or Lorentz indices, one may follow the procedure 
advised by Weldon~\cite{Weldon1983} and define the scalar function
\beq
  \Sigma(p) = \bar\PHI_{\alpha}^{i}(p)\tilde\Pi_{\phantom{(++)\,}\alpha\beta}^{(++)\,ij}(p)\PHI_{\beta}^j(p) \,,
\eeq
which represents a contraction with asymptotic states. This allows for a probabilistic interpretation of $\Sigma$ 
as a decay rate, similar to a Breit-Wigner resonance, for any particle species.

For future convenience in the following sections, the ratio of the thermal decay rate above to 
the zero-temperature limit, as used in the analysis by Ref.~\cite{HoScherrer2015}, may be defined as
\beql{DecayRateRatio}
  R = \frac{\Gamma_D}{\gamma_D} \,.
\eeq
This quantity is a parametrisation for the deviation of thermal theory predictions from those in 
the zero-temperature limit. If this ratio is equal to unity for all temperatures, the medium has 
no effect on decay rates.

%%%%%%%%%%%%%%%%%%%%%%%%%%%%%%%%%%%%%%%%%%%%%%%%%%%%%%%%%%%%%%%%
\section{(Pseudo)scalar decay into two (pseudo)scalars}
\label{sec:pStpSpS}
\setcounter{equation}{0}
%%%%%%%%%%%%%%%%%%%%%%%%%%%%%%%%%%%%%%%%%%%%%%%%%%%%%%%%%%%%%%%%

We are now equipped with the formalism and the relations needed for computation of thermal decay rates. 
This section provides a detailed analysis of the thermal decay rate at which a neutral \mbox{(pseudo)scalar} 
particle $\Phi$ of mass $M$ decays into a pair of neutral \mbox{(pseudo)scalar} particles \mbox{$\phi^1\phi^2$} 
of masses $m_1$, $m_2$. The scalars in the final state may not be identical in the following expressions. 
Note, however, that all scalars considered here are neutral particles. The interaction term responsible 
for such a decay process reads
\beql{StSSInteractionTerm}
     \mathcal L_\text{int} = \kappa\Phi\phi^1\phi^2 \,.
\eeq
The resulting scalar boson self-energy discussed in this section has been previously published in
Ref.~\cite{NishikawaMorimatsuHidaka2003} and represents an important example of the usage of 
the real-time TFT framework.

%%%%%%%%%%%%%%%%%%%%%%%%%%%%%%%%%%%%%%%%%%%%%%%%%%%%%%%%%%%%%%%%
\subsection{Real-time self-energy of the scalar-scalar eye-diagram}
\label{sec:realtimeselfenergyeye}
%%%%%%%%%%%%%%%%%%%%%%%%%%%%%%%%%%%%%%%%%%%%%%%%%%%%%%%%%%%%%%%%

Using the interaction term of \eq{StSSInteractionTerm}, one may draw the following eye-diagram:
\beql{ScalarScalarEyeDiagram}
     \raisebox{-0.47\height}{%
         \includegraphics[scale=0.35]{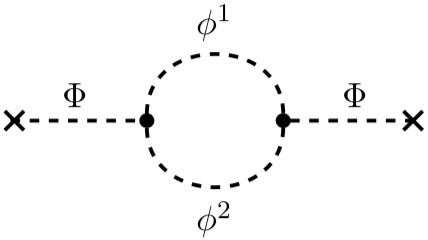}
     }\,.
\eeq
This is the matrix eye-diagram that appears at the first order in the Dyson series representing the full interacting
propagator of $\Phi$. The crosses denote either external legs or connections to vertices. The \mbox{$\Phi$-lines} 
in the loop diagram are amputated in the following self-energy calculation. Note that this diagram represents 
all allowed propagator combinations\footnote{
    The thick lines are drawn in order to indicate the free matrix propagator. Hence, the diagram represents 
    the sum of diagrams over all valid combinations of $\phi_+$, $\phi_-$.
  \begin{align*}
     \raisebox{-0.43\height}{%
         \includegraphics[scale=0.25]{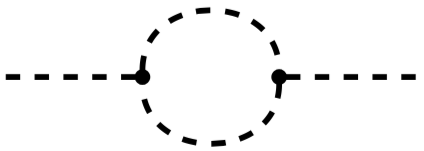}
     }
     & \,=\,
     \raisebox{-0.45\height}{%
         \includegraphics[scale=0.25]{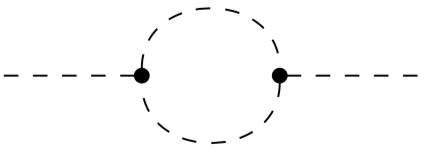}
     }
     \,+\,
     \raisebox{-0.45\height}{%
         \includegraphics[scale=0.25]{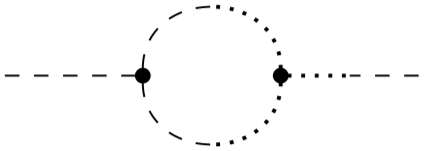}
     }
     \\ & \quad \,+\,
     \raisebox{-0.45\height}{%
         \includegraphics[scale=0.25]{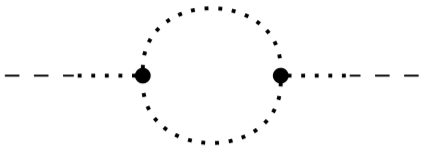}
     }
     \,+\,
     \raisebox{-0.45\height}{%
         \includegraphics[scale=0.25]{figures/Diagrams/minusMinusScalarLoop.png}
     }
  \end{align*}
  }.
The intermediate particles are not restricted to the physical field component $\phi_{+}$ while the external legs are. 
As a consequence, all valid combinations of propagators in the loop must be considered when calculating the total
amplitude for this diagram. The $(+)$- and $(-)$-field components do not mix in the vertices,
cf.~\fig{PhiFourRealTimeVertices}, but propagate into each other due to the off-diagonal propagator elements. 
Besides, each vertex contributes with a factor of \mbox{$\mp i\kappa$}, respectively.

In order to obtain the thermal decay rate, the self-energy of \eq{ScalarScalarEyeDiagram} will be evaluated 
below and it is defined by the loop diagram
\beq
     i\mathcal I_\text{bubble}(p; m_1, m_2)
     =
     \raisebox{-0.48\height}{%
         \includegraphics[scale=0.2]{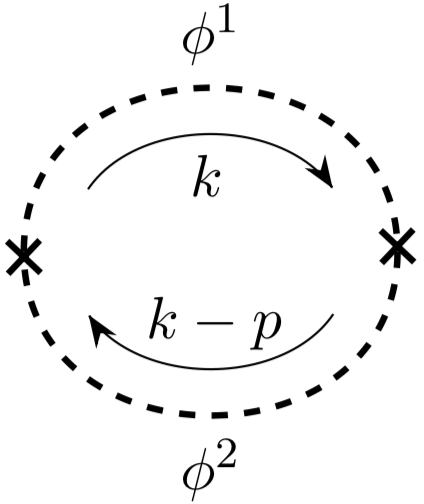}
     }.
\eeq
All matrix components of this self-energy are related through \eq{SelfEnergyRelations}, and evaluation of the \mbox{$(++)$}-component is sufficient in order to obtain the decay rate. This component is
\begin{align}
    i\mathcal I_\text{bubble}^{(++)} & (p; m_1, m_2)
    \nonumber
    \\ & =
    i(1 + \delta_{12})(-i\kappa)^2
    \int \frac{\dd^4 k}{(2\pi)^4}
      i\tilde D^{(++)}(k-p; m_1) i\tilde D^{(++)}(k;m_2)
    \nonumber
    \\ & =
    iI_\text{SS}^{(2)}(p^2; m_1, m_2) + \Big[ iF_\text{SS}^{(2)}(p; m_1, m_2) 
    + (1 \leftrightarrow 2) \Big] + iF_\text{SS}^{(3)}(p; m_1, m_2) \,.
\end{align}
Note here the pre-factor; $\delta_{12}$ is defined as unity if the two particles in the loop are identical and 
zero otherwise. Hence, the correct symmetry factor will be taken into account when considering either identical 
or distinct loop particle species. The two-component structure of the scalar propagator splits the bubble 
into four terms indicated above: a non-thermal (temperature independent) integral $I_\text{SS}^{(2)}$, 
two mixed integrals $F_\text{SS}^{(2)}$ and \mbox{$( 1 \leftrightarrow 2)$}, and one purely thermal 
integral $F_\text{SS}^{(3)}$. The bracketed arrow denotes a na\"ive interchange of indices 1 and 2 
in the second mixed term. The explicit integrals are given as follows
\begin{align}
     &iI_\text{SS}^{(2)}(p^2; m_1, m_2)
     \nonumber
     \\ & =
     i(1 + \delta_{12})(-i\kappa)^2 \!
     \int \! \frac{\dd^4 k}{(2\pi)^4} \,
         \frac{i}{(p-k)^2 - m_1^2 + i\epsilon}
         \,
         \frac{i}{k^2 - m_2^2 + i\epsilon},
     \\ \nonumber \\
     & iF_\text{SS}^{(2)}(p; m_1, m_2)
     \nonumber
     \\ & =
     i(1 + \delta_{12})(-i\kappa)^2 \!
     \int \! \frac{\dd^4 k}{(2\pi)^4} \,
         \frac{i}{(p-k)^2 - m_1^2 + i\epsilon}
         \,
         2\pi \, n_\text{B}\big( \abs{k_0} \big)
         \delta\big( k^2 - m_2^2 \big) \,,
     \\ \nonumber \\
     & iF_\text{SS}^{(3)}(p; m_1, m_2)
     \nonumber
     \\ & =
     i(1 + \delta_{12})(-i\kappa)^2 \!
     \int \! \frac{\dd^4 k}{(2\pi)^4} \,
         2\pi \, n_\text{B}\big( \abs{p_0 \!-\! k_0} \big)
         \delta\big( (p\!-\!k)^2 \!- m_1^2 \big) \,
         2\pi \, n_\text{B}\big( \abs{k_0} \big)
         \delta\big( k^2 \!- m_2^2 \big) \,.
\end{align}
The evaluation of the types of integrals above is discussed in some detail in \sec{loopCalculation}.

%%%%%%%%%%%%%%%%%%%%%%%%%%%%%%%%%%%%%%%%%%%%%%%%%%%%%%%%%%%%%%%%
\subsubsection{Non-thermal self-energy term}
\label{sec:SSnonThermalSE}
%%%%%%%%%%%%%%%%%%%%%%%%%%%%%%%%%%%%%%%%%%%%%%%%%%%%%%%%%%%%%%%%

The non-thermal contribution to the self-energy is well-known from zero-temperature QFT. It may be evaluated 
by dimensional regularisation techniques in \mbox{$D = 4 - 2\epsilon$} dimensions at the renormalisation 
scale $\bar\mu$. The resulting expression for the non-thermal integral is
\beql{SSnonThermalSE}
  iI_\text{SS}^{(2)}(p^2; m_1, m_2)
  =
  (1 + \delta_{12}) \frac{\kappa^2}{16\pi^2}
  \bigg\{
    \!\!-\frac{1}{\tilde\epsilon} - 2
    +
    \frac{x_+}{2}\ln\left[\frac{m_1^2}{\bar\mu^2}\right] -  \frac{x_-}{2}\ln\left[\frac{m_2^2}{\bar\mu^2}\right]
    -I_\text{loop}
  \bigg\} \,.
\eeq
Here, $1/\tilde\epsilon$ parametrises the divergence emerging from the dimensional regularisation scheme and, 
denoting the Euler-Mascheroni constant by $\gamma_\text{E}$, it is defined as \mbox{$\tfrac{1}{\tilde\epsilon} = \tfrac{1}{\epsilon} - \gamma_\text{E} + \ln{4\pi}$}. The last term $I_\text{loop}$ is the term of interest 
to the decay rate since it is complex.
\beq
  I_\text{loop}
  =
  \begin{dcases}
    \phantom{-}
    \frac{\sqrt{C}}{2}
    \left[
      \ln\abs{
        \frac{\big( \sqrt{C} + x_- \big)\big( \sqrt{C} - x_+ \big)}{\big( \sqrt{C} - x_- \big)\big( \sqrt{C} + x_+ \big)}
      }
      +
      2\pi i
    \right]
    \qquad\quad
    \text{if (a),}
    \\ \\
    -\sqrt{D}
    \left[
      \arctan{\frac{x_+}{\sqrt{D}}} - \arctan{\frac{x_-}{\sqrt{D}}}
    \right]
    \hspace{2.23cm} \text{if (b),}
    \\ \\
    \phantom{-}
    \frac{\sqrt{C}}{2}
    \ln\abs{
      \frac{\big( \sqrt{C} + x_- \big)\big( \sqrt{C} - x_+ \big)}{\big( \sqrt{C} - x_- \big)\big( \sqrt{C} + x_+ \big)}
    }
    \hspace{2.35cm}
    \text{if (c).}
  \end{dcases}
\eeq
The cases are defined as the momentum regions
\begin{itemise}
  \item[a)] \mbox{$p^2 \geq (m_1 + m_2)^2$},
  \item[b)] \mbox{$(m_1 - m_2)^2 \leq p^2 < (m_1 + m_2)^2$},
  \item[c)] \mbox{$p^2 < (m_1 - m_2)^2$},
\end{itemise}
and
\beq
     x_\pm = \pm 1 + \frac{m_2^2 - m_1^2}{p^2} \,,
     \qquad
     C = -D =
     \Bigg( 1 - \frac{(m_1 + m_2)^2}{p^2} \Bigg)
     \Bigg( 1 - \frac{(m_1 - m_2)^2}{p^2} \Bigg) \,.
\eeq
The imaginary part of \eq{SSnonThermalSE} reads as follows
\beq
  \imaginary\ \big[ iI_\text{SS}^{(2)}(p^2; m_1, m_2) \big]
  =
  \begin{dcases}
    -(1 + \delta_{12}) \frac{\kappa^2}{16\pi} \sqrt{C}
    \hspace{1.5cm} \text{if (a),}
    \\
    0
    \hspace{3.4cm} \text{otherwise} \,.
  \end{dcases}
\eeq

%%%%%%%%%%%%%%%%%%%%%%%%%%%%%%%%%%%%%%%%%%%%%%%%%%%%%%%%%%%%%%%%
\subsubsection{Mixed self-energy term}
\label{sec:Mixedselfenergy}
%%%%%%%%%%%%%%%%%%%%%%%%%%%%%%%%%%%%%%%%%%%%%%%%%%%%%%%%%%%%%%%%

The mixed self-energy contribution consists of two complex propagator cross terms. 
The evaluation of the imaginary component of those integrals results in
\begin{align}
  & \imaginary\ \big[
     iF_\text{SS}^{(2)} (p; m_1, m_2) + (1 \leftrightarrow 2)
 \big]
  \nonumber
  \\ & =
  -(1 + \delta_{12}) \frac{\kappa^2}{16\pi\abs{\vb p}\beta}
  \begin{dcases}
    \phantom{-}
    \ln\abs{\frac{1 - e^{-\beta\omega_{2, p_+}}}{1 - e^{-\beta\omega_{2, p_-}}}} + (1 \leftrightarrow 2)
    \hspace{2.76cm} \text{if (a)} \,\,\, \vee \,\,\, \text{(c1),}
    \\ \\
    \phantom{-}
    0
    \hspace{7.9cm} \text{if (b),}
    \\ \\
    -\ln\abs{\big( 1 - e^{-\beta\omega_{2, p_+}} \big)\big( 1 - e^{-\beta\omega_{2, p_-}} \big)} + (1 \leftrightarrow 2)
    \hspace{1.55cm} \text{if (c2).}
  \end{dcases}
\end{align}
Here,
\beq
  \omega_{2, p_\pm}
  =
  \frac{1}{2}
  \abs{
    p_0\abs{1 + \frac{m_2^2 \!-\! m_1^2}{p^2}} \pm \abs{\vb p}\sqrt{C}
  } \,.
\eeq
Two further cases have been defined as
\begin{itemise}
  \item[c1)] \mbox{$0 \leq p^2 < (m_1 - m_2)^2$},
  \item[c2)] \mbox{$p^2 < 0$}.
\end{itemise}

The real part of the mixed term is
\begin{align}
  & \real\ \big[
      iF_\text{SS}^{(2)} (p; m_1, m_2) + (1 \leftrightarrow 2)
    \big]
  =
  (1 + \delta_{12}) \frac{\kappa^2}{16\pi^2\abs{\vb p}}
  \int_{m_2}^\infty \dd{\omega_{2, \vb k}}
    n(\omega_{2, \vb k})
    \nonumber
    \\ & \cross \!
    \mathcal P \ln\abs{\frac{
      \big( p^2 \!-\! 2p_0\omega_{2, \vb k} \!+\! 2\abs{\vb p}\sqrt{\omega_{2, \vb k}^2 \!-\! m_2^2} \!+\! m_2^2 \!-\! m_1^2 \big)
      \big( p^2 \!+\! 2p_0\omega_{2, \vb k} \!+\! 2\abs{\vb p}\sqrt{\omega_{2, \vb k}^2 \!-\! m_2^2} \!+\! m_2^2 \!-\! m_1^2 \big)
    }{
      \big( p^2 \!-\! 2p_0\omega_{2, \vb k} \!-\! 2\abs{\vb p}\sqrt{\omega_{2, \vb k}^2 \!-\! m_2^2} \!+\! m_2^2 \!-\! m_1^2 \big)
      \big( p^2 \!+\! 2p_0\omega_{2, \vb k} \!-\! 2\abs{\vb p}\sqrt{\omega_{2, \vb k}^2 \!-\! m_2^2} \!+\! m_2^2 \!-\! m_1^2 \big)
    }}
    \nonumber
    \\ & \,\, +
    (1 \leftrightarrow 2).
\end{align}
Here \mbox{$\omega_{2,\vb k}^2 = \abs{\vb k}^2 + m_2^2$.}

%%%%%%%%%%%%%%%%%%%%%%%%%%%%%%%%%%%%%%%%%%%%%%%%%%%%%%%%%%%%%%%%
\subsubsection{Thermal self-energy term}
\label{sec:SSpureThermalSE}
%%%%%%%%%%%%%%%%%%%%%%%%%%%%%%%%%%%%%%%%%%%%%%%%%%%%%%%%%%%%%%%%

The thermal contribution to the self-energy has no real component 
and is given by the purely imaginary expression
\begin{align}
  &iF_\text{SS}^{(3)} (p; m_1, m_2)
  =
  -i(1 + \delta_{12}) \frac{\kappa^2}{8\pi\abs{\vb p}\beta}
  \cross
  \nonumber
  \\ & \cross
  \begin{dcases}
    \frac{1}{e^{\beta p_0} - 1}
    \ln\abs{\frac{
        \big( 1 - e^{-\beta\omega_{2, p_+}} \big) \big( 1 - e^{\beta(p_0 - \omega_{2, p_-})} \big)
      }{
        \big( 1 - e^{-\beta\omega_{2, p_-}} \big) \big(1 - e^{\beta(p_0 - \omega_{2, p_+})} \big)
      }
    }
    \hspace{4.3cm} \text{if (a),}
    \\ \\
    0
    \hspace{10.9cm} \text{if (b),}
    \\ \\
    \frac{1}{e^{\mp\beta p_0} - 1}
    \Bigg[
      \ln\abs{
        \frac{1 - e^{-\beta\omega_{2, p_+}}}{1 - e^{-\beta\omega_{2, p_-}}}
      }
      -
      e^{\mp\beta p_0}
      \ln\abs{
        \frac{1 - e^{\beta(\pm p_0 - \omega_{2, p_+})}}{1 - e^{\beta(\pm p_0 - \omega_{2, p_-})}}
      }
    \Bigg]
    \hspace{0.73cm} \text{if (c1)} \,\,\, \wedge \,\,\, \text{(d1),}
    \\
    \frac{1}{e^{-\beta p_0} - 1}
    \bigg[
      e^{-\beta p_0} \ln\abs{1 - e^{\beta(p_0 - \omega_{2, p_\pm})}}
      -
      \ln\abs{1 - e^{-\beta\omega_{2, p_\pm}}}
    \bigg]
    \\ \,\, +
    \frac{1}{e^{\beta p_0} - 1}
    \bigg[
      e^{\beta p_0} \ln\abs{1 - e^{-\beta(p_0 + \omega_{2, p_\mp})}}
      -
      \ln\abs{1 - e^{-\beta\omega_{2, p_\mp}}}
    \bigg]
    \hspace{1.04cm} \text{if (c2)} \,\,\, \wedge \,\,\, \text{(d2).}
  \end{dcases}
\end{align}
The further two cases were defined as
\begin{itemise}
  \item[(d1)] \mbox{$m_2 \geq m_1 \,\,\, \vee \,\,\, m_2 < m_1$}. \vspace{5pt}\\
  The leftmost inequality corresponds to the upper sign. The rightmost inequality corresponds 
  to the lower sign.
  \item[(d2)] \mbox{$\bigg\{ \Big[ m_2 \geq m_1 \,\, \wedge \,\, 1 \!+\! 
  \frac{m_2^2 - m_1^2}{p^2}  \geq 0 \Big] \,\,\, \vee \,\,\, m_2 < m_1 \bigg\}
    \, \vee \,
    \bigg\{ m_2 \geq m_1 \,\, \wedge \,\, 1 \!+\! \frac{m_2^2 - m_1^2}{p^2}  < 0 \bigg\}$}. \vspace{5pt}\\
    The leftmost curly bracket corresponds to the upper sign. The rightmost curly bracket 
    corresponds to the lower sign.
\end{itemise}
It is important to notice that in the above expression for a purely thermal contribution, the mass 
parameters $m_1$, $m_2$ in $\omega_{2, p_\pm}$ should be replaced by expressions 
\mbox{$\text{min}\{ m_1, m_2 \}$} and \mbox{$\text{max}\{ m_1, m_2 \}$}, respectively.

The combined \mbox{$(++)$}-self-energy component of the eye-diagram given 
in \mbox{Sec.~\ref{sec:SSnonThermalSE}-\ref{sec:SSpureThermalSE}} was provided 
by Nishikawa et al.~\cite{NishikawaMorimatsuHidaka2003} as part of the evaluation 
of the sunset diagram and has been included here for completeness.

%%%%%%%%%%%%%%%%%%%%%%%%%%%%%%%%%%%%%%%%%%%%%%%%%%%%%%%%%%%%%%%%
\subsection{Decay rate of \texorpdfstring{$\Phi \to \phi\phi$}{TEXT}}
\label{sec:Phitophiphi}
%%%%%%%%%%%%%%%%%%%%%%%%%%%%%%%%%%%%%%%%%%%%%%%%%%%%%%%%%%%%%%%%

This section presents the thermal decay rate of a scalar particle into two scalar particles with 
the rate given by \eq{RealTimeDecayRate}, in \sec{DecayRateRealTimeSelfEnergy}.
In the case of identical loop masses, the expression for the self-energy simplifies significantly 
due to vanishing terms in the mass function $\sqrt{C}$ and $\omega_{2, p_\pm}$. 
The ratio of \eq{DecayRateRatio} may be obtained by considering the limiting case 
of \mbox{$p^2 = M^2 \gg (m_1 + m_2)^2$}, rendering the particles in the loop effectively massless. 
The plotted ratio can be seen in \mbox{Fig.~\ref{fig:StSSDecayRatioHoScherrer}a-c} for a non-relativistic, 
relativistic and ultra-relativistic incoming particle $\Phi$. The figures reproduce the findings 
of Ho and Scherrer \cite{HoScherrer2015} who considered the case of identical loop masses 
\mbox{$m_1 = m_2 = 0$} evaluated in the imaginary-time formalism and their result has hereby 
been verified in the real-time formalism. The limit as \mbox{$T\to0$} for the ratio is 
$R_{\Phi \to \phi\phi} \to 1$ as required.
%%%%%%%%%%%%%%%%%%%%%%%%%%%%%%%%%%%%%%%%%%%%%%%%%%%%%%%%%%%%%%%%
\begin{figure}[htbp]
  \begin{minipage}{0.48\textwidth}
    \centering
    \resizebox{1.0\columnwidth}{!}{
      \input{figures/StSS-nonRel/StSS-nonRel}
    }
  \end{minipage}
  \hfill
  \begin{minipage}[c]{0.48\textwidth}
    \centering
    \resizebox{1.0\columnwidth}{!}{
      \input{figures/StSS-sliRel/StSS-sliRel}
    }
  \end{minipage}
  \\
  \begin{minipage}[t]{0.48\textwidth}
    \vspace{0pt}
    \centering
    \resizebox{1.0\columnwidth}{!}{
      \input{figures/StSS-higRel/StSS-higRel}
    }
  \end{minipage}
  \hfill
  \begin{minipage}[t]{0.45\textwidth}
    \vspace{8pt}
    \caption{Ratio of the thermal decay rate to the zero-temperature limit for \mbox{$\Phi \to \phi\phi$}. 
    The ratio \mbox{$R = \Gamma_D/\gamma_D$} is plotted for \mbox{$p^2 = M^2 \gg 4m^2$}, with $m=m_1=m_2$. Subfigures display ratios for varying \mbox{$\epsilon = \abs{\vb p}/M$} that parametrises the three-momentum of $\Phi$ 
    and are plotted here for non-relativistic (a), relativistic (b) and ultra-relativistic (c) particles. 
    These real-time results agree with the results of Ref.~\cite{HoScherrer2015} that were evaluated in the Matsubara formalism. Note that \mbox{$R \to 1$ when $T$ vanishes.}
    \label{fig:StSSDecayRatioHoScherrer}}
  \end{minipage}
\end{figure}
%%%%%%%%%%%%%%%%%%%%%%%%%%%%%%%%%%%%%%%%%%%%%%%%%%%%%%%%%%%%%%%%

The deviation from Ref.~\cite{HoScherrer2015} when loop masses $m_1$, $m_2$ are not identical was investigated. 
Such deviations are clearly relevant only in the case of \mbox{$M \sim m_1, m_2$}, close to the equality 
of \mbox{$M^2 = (m_1 + m_2)^2$}. An analysis of the self-energy presented earlier in this section shows 
that the behaviour is qualitatively similar to \mbox{Fig.~\ref{fig:StSSDecayRatioHoScherrer}a-c}. Furthermore, 
in the mass region \mbox{$0 \leq M^2 < (m_1 - m_2)^2$} and for virtual $\Phi$ with \mbox{$p^2 < 0$}, 
the behaviour of \mbox{$\Gamma_{\Phi \to \phi\phi}$} is also quadratically growing with temperature. 
Hence, these regions exhibit a qualitatively similar behaviour compared to the plotted cases.

%%%%%%%%%%%%%%%%%%%%%%%%%%%%%%%%%%%%%%%%%%%%%%%%%%%%%%%%%%%%%%%%
\section{Scalar decay into a fermion-antifermion pair}
\label{sec:StFaF}
\setcounter{equation}{0}
%%%%%%%%%%%%%%%%%%%%%%%%%%%%%%%%%%%%%%%%%%%%%%%%%%%%%%%%%%%%%%%%

This section presents the thermal rate at which a neutral scalar particle $\Phi$ decays 
into a fermion-antifermion pair \mbox{$\psi^2\bar\psi^1$}. One model that gives 
rise to such decay process is
\beql{StFaFInteractionTerm}
     \mathcal L_\text{int} = a\Phi\bar\psi^1\psi^2 \,.
\eeq
The particles are associated with masses $M$ for $\Phi$ and $m_i$ for $\psi^i$.

%%%%%%%%%%%%%%%%%%%%%%%%%%%%%%%%%%%%%%%%%%%%%%%%%%%%%%%%%%%%%%%%
\subsection{Real-time self-energy of the fermion-antifermion eye-diagram}
\label{sec:realtimeSEFFbareye}
%%%%%%%%%%%%%%%%%%%%%%%%%%%%%%%%%%%%%%%%%%%%%%%%%%%%%%%%%%%%%%%%

In analogy to the case of the scalar-scalar loop considered in the previous section (\sec{pStpSpS}), the self-energy 
of the eye-diagram is related to the thermal decay rate of the process \mbox{$\Phi \to \psi^2\bar\psi^1$}. Also here, 
all valid combinations of field components must be taken into account but only one self-energy component is independent, 
see \eq{SelfEnergyRelations}, and the evaluation of the self-energy \mbox{$(++)$}-component is sufficient to extract 
the thermal decay rate. This component reads
\begin{align}
    i\mathcal I_\text{bubble}^{(++)} & (p; m_1, m_2)
    \nonumber
    \\ & =
    i(-ia)^2 (-1)
    \int \frac{\dd^4 k}{(2\pi)^4}
      \tr[
        i\tilde S_\text{F}^{(++)}(k-p; m_1) i\tilde S_\text{F}^{(++)}(k;m_2)
      ]
    \nonumber
    \\ & =
    i(-ia)^2 (-1)
    \int \frac{\dd^4 k}{(2\pi)^4}
      \tr[
        -i\tilde S_{\bar{\text{F}}}^{(++)}(p-k; m_1) i\tilde S_\text{F}^{(++)}(k;m_2)
      ]
    \nonumber
    \\ & =
    iI_{\text{F} \bar {\text{F}}}^{(2)}(p^2; m_1, m_2) + 
    \Big[ iF_{\text{F} \bar{\text{F}}}^{(2)}(p; m_1, m_2) + 
    (\text{F} \leftrightarrow \bar{\text{F}}) \Big] + 
    iF_{\text{F} \bar {\text{F}}}^{(3)}(p; m_1, m_2) \,.
\end{align}
Note, the overall negative sign appears together the trace over spinor indices in the case of a fermion loop. 
The bracketed arrow represents the second mixed integral that arises from propagator cross terms. 
As in the case of the scalar-scalar loop, the bubble has split into four terms and the explicit integrals are
\begin{align}
  iI_{\text{F} \bar {\text{F}}}^{(2)} & (p^2; m_1, m_2)
  \nonumber
  \\ & \! =
  i(-ia)^2 \!\! \int \! \frac{\dd^4 k}{(2\pi)^4} \,
    \tr[
      \frac{i\big( (\fsl{p} - \fsl{k}) - m_1 \big)}{(p-k)^2 - m_1^2 + i\epsilon} \, \frac{i\big( \fsl{k} + m_2 \big)}{k^2 - m_2^2 + i\epsilon}
    ]\!,
  \\ \nonumber \\
  iF_{\text{F} \bar {\text{F}}}^{(2)} & (p; m_1, m_2)
  \nonumber
  \\ & \! =
  i(-ia)^2 \!\! \int \! \frac{\dd^4 k}{(2\pi)^4} \,
    \tr[
      \frac{i\big( (\fsl{p} - \fsl{k}) - m_1 \big)}{(p-k)^2 - m_1^2 + i\epsilon}
      (\!-2\pi) \big( \fsl{k} + m_2 \big) n_\text{F}(k_0) \delta\big( k^2 \!-\! m_2^2 \big)
    ]\!,
\end{align}
\begin{align}
  iF_{\text{F} \bar {\text{F}}}^{(3)} & (p; m_1, m_2)
  \nonumber
  \\ & \! =
  i(-ia)^2 \!\! \int \! \frac{\dd^4 k}{(2\pi)^4} \,
    \tr\Big[
      (\!-2\pi) \big( (\fsl{p} - \fsl{k}) - m_1 \big) n_{\bar{\text{F}}}(p_0 \!-\! k_0) \delta\big( (p\!-\!k)^2 - m_1^2 \big)
      \cross\hspace{1.25cm}
      \nonumber
      \\ & \hspace{4.1cm} \cross
      (\!-2\pi) \big( \fsl{k} + m_2 \big) n_\text{F}(k_0) \delta\big( k^2 \!-\! m_2^2 \big)
    \Big]\!.
\end{align}
The trace is common to all integrals and evaluates to
\beq
  \tr\Big[ \big( (\fsl{p} - \fsl{k}) - m_1 \big)\big( \fsl{k} + m_2 \big) \Big]
  =
  -4\big( m_1m_2 - p\cdot k + k^2 \big) \,.
\eeq

%%%%%%%%%%%%%%%%%%%%%%%%%%%%%%%%%%%%%%%%%%%%%%%%%%%%%%%%%%%%%%%%
\subsubsection{Non-thermal self-energy term}
\label{sec:FaFnonThermalSE}
%%%%%%%%%%%%%%%%%%%%%%%%%%%%%%%%%%%%%%%%%%%%%%%%%%%%%%%%%%%%%%%%

The non-thermal contribution is again known from the zero-temperature theory and the imaginary part 
of interest to the decay rate calculation may be extracted. The imaginary part of 
\mbox{$iI_{\text{F} \bar {\text{F}}}^{(2)}(p^2; m_1, m_2)$} reads
\begin{align}
  & \imaginary\ \big[
     iI_{\text{F} \bar {\text{F}}}^{(2)}(p^2; m_1, m_2)
\big]
  =
  \frac{a^2}{8\pi} \sqrt{C}
  \cross
  \nonumber
  \\ & \quad \cross \!
  \begin{dcases}
      \frac{p^2}{2}
      \bigg[
        C - \left( 1 \!+\! \frac{m_2^2 \!-\! m_1^2}{p^2} \right)^{\!2}
      \bigg]
      -
      p^2\!\left( 1 \!+\! \frac{m_2^2 \!-\! m_1^2}{p^2} \right)
      +
      2m_2\big[ m_1 \!+\! 2m_2 \big]
    \hspace{1.38cm}\text{if (a),}
    \\
    0 \hspace{10cm} \text{otherwise}\,,
  \end{dcases}
\end{align}
where $C$ is identical to the definition in \sec{pStpSpS}. In the case of identical loop masses, 
the expression reduces to
\beq
  \imaginary\ \big[
     iI_{\text{F} \bar {\text{F}}}^{(2)}(p^2; m_1, m_2)
 \big]
  =
  \begin{dcases}
    -\frac{a^2 \big( p^2 - 4m^2 \big)}{8\pi} \sqrt{1 - \frac{4m^2}{p^2}}
    \hspace{1.92cm} \text{if (a),}
    \\
    \phantom{-}0
    \hspace{4.8cm} \text{otherwise}.
  \end{dcases}
\eeq
This matches the corresponding result in the literature, see e.g.~Ref.~\cite{Minahan2011}.

%%%%%%%%%%%%%%%%%%%%%%%%%%%%%%%%%%%%%%%%%%%%%%%%%%%%%%%%%%%%%%%%
\subsubsection{Mixed self-energy term}
\label{sec:FaFmixedThermalSE}
%%%%%%%%%%%%%%%%%%%%%%%%%%%%%%%%%%%%%%%%%%%%%%%%%%%%%%%%%%%%%%%%

The mixed self-energy contribution consists of two complex cross terms. The evaluation 
of the imaginary component of those terms results in
\begin{align}
  & \imaginary\ \big[
      iF_{\text{F} \bar {\text{F}}}^{(2)} (p; m_1, m_2)
      +
      (\text{F} \leftrightarrow \bar{\text{F}})
    \big]
  =
  -\frac{a^2}{8\pi \abs{\vb p} \beta}
  \Big( p^2 - (m_2 + m_1)^2 \Big)
  \cross
  \nonumber
  \\ & \cross\!
  \begin{dcases}
    \ln\!\left[
      \frac{
        \big( e^{-\beta(\omega_{2, p_+} - \mu)} + 1 \big)\big( e^{-\beta(\omega_{1, p_+} + \mu)} + 1 \big)
      }{
        \big( e^{-\beta(\omega_{2, p_-} - \mu)} + 1 \big)\big( e^{-\beta(\omega_{1, p_-} + \mu)}  + 1 \big)
      }
    \right] \hspace{2.1cm} \text{if (a) } \vee \,\, \Big\{ \text{(c1)} \, \wedge \, m_2 \geq m_1 \Big\} \,,
    \\ \\
    0
    \hspace{11.72cm} \text{if (b),}
    \\ \\
    \ln\!\left[
      \frac{
        \big( e^{-\beta(\omega_{2, p_+} + \mu)} + 1 \big)\big( e^{-\beta(\omega_{1, p_+} - \mu)} + 1 \big)
      }{
        \big( e^{-\beta(\omega_{2, p_-} + \mu)} + 1 \big)\big( e^{-\beta(\omega_{1, p_-} - \mu)}  + 1 \big)
      }
    \right]
    \hspace{3.47cm} \text{if (c1) } \wedge \,\, m_2 < m_1 \,,
    \\ \\
    \!-\! \ln\!\Big[
      \big( e^{-\beta(\omega_{2, p_+} \!- \mu)} \!+\! 1 \big)
      \big( e^{-\beta(\omega_{1, p_+} \!+ \mu)} \!+\! 1 \big)
      \big( e^{-\beta(\omega_{2, p_-} \!- \mu)} \!+\! 1 \big)
      \big( e^{-\beta(\omega_{1, p_-} \!+ \mu)} \!+\! 1 \big)
    \Big] \hspace{0.5cm} \text{if (c2).}
  \end{dcases}
\end{align}
Here, $\omega_{1, p_\pm}$ has been defined analogously to $\omega_{2, p_\pm}$ in \sec{pStpSpS} 
with interchanged masses.

In compact notation, the real part may be written as
\begin{align}
     & \real\ \big[
         iF_{\text{F} \bar {\text{F}}}^{(2)} (p; m_1, m_2)
         + (\text{F} \leftrightarrow \bar{\text{F}})
     \big]
     \nonumber
  \\ & =
  \frac{a^2}{4\pi^2\abs{\vb p}} \sum_{s=\pm}
    \cross \nonumber
    \\ & \quad \cross \Bigg\{
    \int_{m_2}^\infty \dd{\omega_{2, \vb k}}
        n_{\text{F}, 2}^{s}
        \Bigg[
          C_{2}^{s}
          \mathcal P \ln\abs{\frac{A_{21, +}^{(s)}}{A_{21, -}^{(s)}}}
          +
          \mathcal P\Bigg(
            2\abs{\vb p}\sqrt{\omega_{2, \vb k}^2 \!-\! m_2^2}
            \!+\!
            D_{21}^{s}\ln\abs{\frac{A_{21, -}^{(s)}}{A_{21, +}^{(s)}}}
          \Bigg)
        \Bigg]
        \nonumber%
    \\ & \hspace{.87cm} +
    \int_{m_1}^\infty \dd{\omega_{1, \vb k}}
        n_{\text{F}, 1}^{-s}
        \Bigg[
          C_{1}^{s}
          \mathcal P \ln\abs{\frac{A_{12, +}^{(s)}}{A_{12, -}^{(s)}}}
          +
          \mathcal P\Bigg(
            2\abs{\vb p}\sqrt{\omega_{1, \vb k}^2 \!-\! m_1^2}
            \!+\!
            D_{12}^{s}\ln\abs{\frac{A_{12, -}^{(s)}}{A_{12, +}^{(s)}}}
          \Bigg)
        \Bigg]
  \Bigg\} \,,
\end{align}
where
\begin{align}
  \omega_{1, \vb k}^2 & = \abs{\vb k}^2 + m_1^2 \,,
  \\
  n_{\eta, i}^\pm & = \frac{1}{e^{\beta(\omega_{i, \vb k} \pm \mu)} - \eta},
  \qquad \eta = \pm 1 \,,
  \\
  C_i^\pm & = m_im_j \pm p_0\omega_{i, \vb k} + m_i^2 \,,
  \\
  D_{ij}^\pm & = \frac{1}{2}\big( p^2 \pm 2p_0\omega_{i, \vb k} + m_i^2 - m_j^2 \big) \,,
  \\
  A_{ij, \pm}^{(\pm)} & = p^2 (\pm) 2p_0\omega_{i, \vb k} \pm 2\abs{\vb p}
  \sqrt{\omega_{i, \vb k}^2 - m_i^2} + m_i^2 - m_j^2 \,.
\end{align}

%%%%%%%%%%%%%%%%%%%%%%%%%%%%%%%%%%%%%%%%%%%%%%%%%%%%%%%%%%%%%%%%
\subsubsection{Thermal self-energy term}
\label{sec:FaFpureThermalSE}
%%%%%%%%%%%%%%%%%%%%%%%%%%%%%%%%%%%%%%%%%%%%%%%%%%%%%%%%%%%%%%%%

The purely thermal contribution to the self-energy has no real component 
and may be found as
\begin{align}
  & iF_{\text{F} \bar {\text{F}}}^{(3)} (p; m_1, m_2)
  =
  -i \frac{a^2}{4\pi\abs{\vb p}\beta} \big( p^2 - (m_1 + m_2)^2 \big)
  \cross\nonumber
  \\ & \cross\!
  \begin{dcases}
    \frac{1}{e^{\beta p_0} \!-\! 1}
    \ln\!\left[
      \frac{
        \big( e^{-\beta(\omega_{2, p_+} \!- \mu)} + 1 \big)
        \big( e^{\beta(p_0 - \omega_{2, p_-} \!+ \mu)} + 1 \big)
      }{
        \big( e^{-\beta(\omega_{2, p_-} \!- \mu)} + 1 \big)
        \big( e^{\beta(p_0 - \omega_{2, p_+} \!+ \mu)} + 1 \big)
      }
    \right]
    \hspace{3.9cm} \text{if (a),}
    \\ \\%
    0
    \hspace{11.5cm} \text{if (b),}
    \\ \\%
    \frac{1}{e^{\mp\beta p_0} \!-\! 1}\!
    \Bigg\{
      e^{\mp\beta p_0} \ln\!\left[
        \frac{e^{-\beta(\omega_{2, p_+} \!\mp (p_0 + \mu))} \!+\! 1}{e^{-\beta(\omega_{2, p_-} \!\mp (p_0 + \mu)} \!+\! 1}
      \right]
      +
      \ln\!\left[
        \frac{e^{-\beta(\omega_{2, p_+} \!\mp \mu)} \!+\! 1}{e^{-\beta(\omega_{2, p_-} \!\mp \mu)} \!+\! 1}
      \right]\!\!
    \Bigg\}
    \hspace{0.5cm}\text{if (c1)} \, \wedge \, \text{(d1),}
    \\ \\%
    \frac{1}{e^{-\beta p_0} \!-\! 1}
    \bigg\{
      e^{-\beta p_0} \ln\!\left[
        e^{-\beta(\omega_{2, p_{\pm}} \!- p_0 - \mu)} \!+\! 1
      \right]
      -
      \ln\!\left[
        e^{-\beta(\omega_{2, p_{\pm}} \!- \mu)} \!+\! 1
      \right]\!\!
    \bigg\}
    \\%
    \quad +
    \frac{1}{e^{\beta p_0} \!-\! 1}
    \bigg\{
      e^{\beta p_0} \ln\!\left[
        e^{-\beta(\omega_{2, p_{\mp}} \!+ p_0 + \mu)} \!+\! 1
      \right]
      -
      \ln\!\left[
        e^{-\beta(\omega_{2, p_{\mp}} \!+ \mu)} \!+\! 1
      \right]\!\!
    \bigg\}
    \hspace{0.8cm} \text{if (c2)} \, \wedge \, \text{(d2).}
  \end{dcases}
\end{align}
Note that in this final expression for the thermal contribution, the mass parameters $m_1$, $m_2$ 
should again be replaced by \mbox{$\text{min}\{ m_1, m_2 \}$} and \mbox{$\text{max}\{ m_1, m_2 \}$}, 
respectively; cf.~\sec{SSpureThermalSE}.

The explicit \mbox{$(++)$}-component of the real-time self-energy of the fermionic eye-diagram presented above in \mbox{Secs.~\ref{sec:FaFnonThermalSE}-\ref{sec:FaFpureThermalSE}} has not been found 
in the literature to the best of our knowledge.

%%%%%%%%%%%%%%%%%%%%%%%%%%%%%%%%%%%%%%%%%%%%%%%%%%%%%%%%%%%%%%%%
\subsection{Decay rate of \texorpdfstring{$\Phi \to \psi\bar\psi$}{TEXT}}
\label{sec:Phipsipsibar}
%%%%%%%%%%%%%%%%%%%%%%%%%%%%%%%%%%%%%%%%%%%%%%%%%%%%%%%%%%%%%%%%

The thermal decay rate of a neutral scalar is given by \eq{RealTimeDecayRate} 
in \sec{DecayRateRealTimeSelfEnergy}. Similar to the decay rate treated in \sec{pStpSpS}, 
the particles in the loop were considered to have the same mass $m$ and the limit 
of \mbox{$p^2 = M^2 \gg 4m^2$} has been plotted in \fig{StFaFDecayRatioHoScherrer}. 
In this mass region, the ratio of \eq{DecayRateRatio} may be obtained since 
the zero-temperature decay rate does not vanish here. The plotted ratio can be 
seen for non-relativistic, relativistic and ultra-relativistic $\Phi$. The figure 
agrees with the findings of Ho and Scherrer \cite{HoScherrer2015} who evaluated 
the loop in the Matsubara formalism. The limit as \mbox{$T\to0$} is 
$R_{\Phi \to \psi\bar\psi} \to 1$ as expected.
%%%%%%%%%%%%%%%%%%%%%%%%%%%%%%%%%%%%%%%%%%%%%%%%%%%%%%%%%%%%%%%%%
\begin{figure}[htbp]
    \centering
    \resizebox{1.0\columnwidth}{!}{
      \input{figures/StFaF/StFaF}
    }
    \caption{Ratio of the thermal decay rate to the zero-temperature limit for \mbox{$\Phi \to \psi\bar\psi$.} 
    The ratio \mbox{$R = \Gamma_D/\gamma_D$} is plotted for \mbox{$p^2 = M^2 \gg 4m^2$}. 
    Varying \mbox{$\epsilon = \abs{\vb p}/M$}, ratios are plotted for a non-relativistic 
    (solid and dashed), relativistic (dotted) and ultra-relativistic (dash-dotted) $\Phi$. 
    The figure agrees with the result of Ref.~\cite{HoScherrer2015} evaluated in 
    the Matsubara formalism. For any value of $\epsilon$, the transition to 
    the low-temperature limit is smooth as can be seen in (b).
    \label{fig:StFaFDecayRatioHoScherrer}}
\end{figure}
%%%%%%%%%%%%%%%%%%%%%%%%%%%%%%%%%%%%%%%%%%%%%%%%%%%%%%%%%%%%%%%%%

%%%%%%%%%%%%%%%%%%%%%%%%%%%%%%%%%%%%%%%%%%%%%%%%%%%%%%%%%%%%%%%%
\section{Pseudoscalar decay into a fermion-antifermion pair}
\label{sec:pStFaF}
\setcounter{equation}{0}
%%%%%%%%%%%%%%%%%%%%%%%%%%%%%%%%%%%%%%%%%%%%%%%%%%%%%%%%%%%%%%%%

This section presents the thermal rate at which a neutral pseudoscalar particle $\Phi_5$ 
decays into a \mbox{fermion-antifermion} pair \mbox{$\psi^2\bar\psi^1$} as calculated 
in this work in the real-time formalism. This case was explicitly considered in order 
to contrast the pseudoscalar result with the result for a scalar particle provided in \sec{StFaF}. 
The corresponding interaction term is
\beql{PStFaFInteractionTerm}
     \mathcal L_\text{int} = b\Phi_5\bar\psi^1\gamma_5\psi^2 \,.
\eeq
Here, the vertex factor of $\gamma_5$ requires some care when evaluating 
the loop diagrams in comparison to the previous section. In what follows, 
masses have been assigned according to $M_5$ for $\Phi_5$ and $m_i$ for $\psi^i$.

%%%%%%%%%%%%%%%%%%%%%%%%%%%%%%%%%%%%%%%%%%%%%%%%%%%%%%%%%%%%%%%%
\subsection{Real-time self-energy of the fermion-antifermion eye-diagram}
\label{sec:realtimeselfenergypsipsibar}
%%%%%%%%%%%%%%%%%%%%%%%%%%%%%%%%%%%%%%%%%%%%%%%%%%%%%%%%%%%%%%%%

In analogy to the case of the scalar-scalar loop in \sec{pStpSpS}, the self-energy of the eye-diagram 
is related to the thermal decay rate of the process \mbox{$\Phi_5 \to \psi^1\bar\psi^2$}. Also here, 
all valid combinations of field components must be taken into account but only one self-energy 
component is independent, see \eq{SelfEnergyRelations}, and the evaluation of the self-energy 
$(++)$-component is sufficient to extract the thermal decay rate. This component is found as follows
\begin{align}
    i\mathcal I_\text{bubble}^{(++)} & (p; m_1, m_2)
    \nonumber
    \\ & =
    i(-ib)^2 (-1)
    \int \frac{\dd^4 k}{(2\pi)^4}
      \tr[-\gamma_5 i\tilde S_{\bar{\text{F}}}^{(++)}(p-k; m_1) \gamma_5 i\tilde S_\text{F}^{(++)}(k;m_2)]
    \nonumber
    \\ & =
    iI_{\text{F} \bar {\text{F}}}^{(4)}(p^2; m_1, m_2) + \Big[ iF_{\text{F} \bar {\text{F}}}^{(4)}(p; m_1, m_2) 
    + (\text{F} \leftrightarrow \bar{\text{F}}) \Big] + iF_{\text{F} \bar {\text{F}}}^{(5)}(p; m_1, m_2) \,.
\end{align}
As for the \mbox{$\Phi \to \psi^2\bar\psi^1$}-process considered in the previous section, an overall minus 
sign appears as well as the trace over \mbox{$\gamma$-matrices.} The loop integration splits into four terms, 
all similar to the previous section but for the addition of the two vertex factors of $\gamma_5$. The bracketed 
arrow denotes the second mixed cross term. The emerging trace may be taken in the na\"ive dimensional 
regularisation scheme since the diagram appears at one-loop level\footnote{Calculations potentially require 
a more robust scheme at higher loop orders. For example, in the evaluation of three-body decays 
another scheme is required \cite{Buras1998}.}. The trace evaluated in the na\"ive dimensional 
regularisation scheme reads
\beq
     \tr\Big[
         \gamma_5 \big( (\fsl{p} - \fsl{k}) - m_1 \big)
         \gamma_5 \big( \fsl{k} + m_2 \big)
     \Big]
     =
     4\big( -m_1m_2 - p\cdot k + k^2 \big) \,.
\eeq
The sign differences compared to the case of a decaying \emph{scalar} particle are an overall positive sign 
together with the minus sign in front of $m_1 m_2$. The overall sign change may be reverted by imposing hermiticity 
of the \L\ and thereby force the pseudoscalar coupling to \mbox{$\bar\psi^1\gamma_5\psi^2$} to become purely imaginary, 
i.e. \mbox{$b = i\abs{b}$}. The sign difference of the mass term is more intriguing and signifies a fundamental 
difference between the scalar and pseudoscalar decay rates.

Apart from the mentioned differences in relative signs, the resulting self-energy integrals for the pseudoscalar case 
are identical to those for the scalar case. To obtain the correct non-thermal, mixed and purely thermal self-energy 
integrals it is sufficient to replace \mbox{$m_1 \to -m_1$} in the integrals from the previous section.

%%%%%%%%%%%%%%%%%%%%%%%%%%%%%%%%%%%%%%%%%%%%%%%%%%%%%%%%%%%%%%%%
\subsubsection{Non-thermal self-energy term}
\label{sec:PFaFnonThermalSE}
%%%%%%%%%%%%%%%%%%%%%%%%%%%%%%%%%%%%%%%%%%%%%%%%%%%%%%%%%%%%%%%%

The non-thermal contribution in the case of an external pseudoscalar particle is related 
to the case of an external \emph{scalar} particle as
\beq
     iI_{\text{F} \bar {\text{F}}}^{(4)} \big( p^2; m_1, m_2 \big)
     =
     -iI_{\text{F} \bar {\text{F}}}^{(2)} \big( p^2; - m_1, m_2 \big) \,.
\eeq
The right-hand side is the result found in \sec{FaFnonThermalSE}.

%%%%%%%%%%%%%%%%%%%%%%%%%%%%%%%%%%%%%%%%%%%%%%%%%%%%%%%%%%%%%%%%
\subsubsection{Mixed self-energy term}
\label{sec:MixedSEFFbar}
%%%%%%%%%%%%%%%%%%%%%%%%%%%%%%%%%%%%%%%%%%%%%%%%%%%%%%%%%%%%%%%%

The mixed thermal contribution in the case of an external pseudoscalar particle 
is related to the case of an external \emph{scalar} particle as
\beq
     iF_{\text{F} \bar {\text{F}}}^{(4)} (p; m_1, m_2)
     +
     (\text{F} \leftrightarrow \bar{\text{F}})
     =
     -iF_{\text{F} \bar {\text{F}}}^{(2)} (p; -m_1, m_2)
     +
     (\text{F} \leftrightarrow \bar{\text{F}}) \,.
\eeq
The right-hand side is the result found in \sec{FaFmixedThermalSE}.

%%%%%%%%%%%%%%%%%%%%%%%%%%%%%%%%%%%%%%%%%%%%%%%%%%%%%%%%%%%%%%%%
\subsubsection{Thermal self-energy term}
\label{sec:FaFpureThermalSEPS}
%%%%%%%%%%%%%%%%%%%%%%%%%%%%%%%%%%%%%%%%%%%%%%%%%%%%%%%%%%%%%%%%

The purely thermal contribution in the case of an external pseudoscalar particle 
is related to the case of an external \emph{scalar} particle as
\beq
     iF_{\text{F} \bar {\text{F}}}^{(5)} (p;m_1, m_2)
     =
     -iF_{\text{F} \bar {\text{F}}}^{(3)} (p;-m_1, m_2) \,.
\eeq
The right-hand side is the result found in \sec{FaFpureThermalSE}.

%%%%%%%%%%%%%%%%%%%%%%%%%%%%%%%%%%%%%%%%%%%%%%%%%%%%%%%%%%%%%%%%
\subsection{Decay rate of \texorpdfstring{$\Phi_5 \to \psi\bar\psi$}{TEXT}}
\label{sec:Phi5FFbar}
%%%%%%%%%%%%%%%%%%%%%%%%%%%%%%%%%%%%%%%%%%%%%%%%%%%%%%%%%%%%%%%%

The thermal decay rate of a neutral pseudoscalar is given by \eq{RealTimeDecayRate} 
in \sec{DecayRateRealTimeSelfEnergy}. Similar to the decay rate treated in Secs.~\ref{sec:pStpSpS} 
and \ref{sec:StFaF}, the particles in the loop were considered to have the same mass $m$ and 
the pseudoscalar was assumed to fulfill the on-shell condition \mbox{$p^2 = M_5^2 \gg 4m^2$}. 
The decay rate behaves qualitatively similar to that of a scalar decay (see \fig{StFaFDecayRatioHoScherrer}) 
as stated by Ho and Scherrer~\cite{HoScherrer2015} while one should note a discrepancy between 
the two cases close to the equality \mbox{$M_5^2, M^2 = 4m^2$}. This is a non-thermal effect 
due to the mentioned sign differences in the above subsection and it is presented in \fig{PStSDecayRateRatio} 
for clarity, at \mbox{$T=0$}, with normalised masses $M$, $M_5$ given in units of $m$. 
The ratio \mbox{$\Gamma_{\Phi_5 \to \psi\bar\psi}/\Gamma_{\Phi \to \psi\bar\psi}$} is plotted as 
a function of the mass of the external (pseudo)scalar. Close to the equality \mbox{$M^2 = 4m^2$}, 
the expression diverges which implies that the scalar decay rate vanishes quicker than the pseudoscalar 
decay rate. Normalising the fermion mass to \mbox{$m=1$}, the ratio diverges at threshold $M=2m$.
%%%%%%%%%%%%%%%%%%%%%%%%%%%%%%%%%%%%%%%%%%%%%%%%%%%%%%%%%%%%%%
\begin{figure}[htbp]
  \begin{minipage}[t]{0.56\textwidth}
    \vspace{0pt}
    \centering
    \resizebox{\textwidth}{!}{
      \input{figures/RRpStS/RRpStS}
    }
  \end{minipage}
  \hfill
  \begin{minipage}[t]{0.42\textwidth}
    \vspace{8pt}
    \caption{Ratio of the pseudoscalar decay rate to the scalar ditto. The ratio \mbox{$\Gamma_{\Phi_5 \to \psi\bar\psi}/
    \Gamma_{\Phi \to \psi\bar\psi}$} is plotted for \mbox{$M^2 \geq 4m^2$}, the fermion masses in the loop being equal, 
    as a function of the mass of the decaying scalar. Here, \mbox{$M = M_5$} for comparison and normalised in units of $m$. 
    The pseudoscalar decay rate is enhanced relative to the scalar quantity close to the equality \mbox{$M^2 = 4m^2$}. 
    This behaviour was verified for a broad range of both \mbox{$\epsilon = \abs{\vb p}/M\in [0.001, 100]$} 
    and temperatures \mbox{$T/M\in[0, 1000]$} and the effect is non-thermal. 
    \label{fig:PStSDecayRateRatio}}
  \end{minipage}
\end{figure}

%%%%%%%%%%%%%%%%%%%%%%%%%%%%%%%%%%%%%%%%%%%%%%%%%%%%%%%%%%%%%%%%
\section{Emission of a (pseudo)scalar off a fermion pair}
\label{sec:FtpSF}
\setcounter{equation}{0}
%%%%%%%%%%%%%%%%%%%%%%%%%%%%%%%%%%%%%%%%%%%%%%%%%%%%%%%%%%%%%%%%

The interaction term of \eq{StFaFInteractionTerm} in \sec{StFaF} allows for the emission of a scalar off 
a fermion line. The amplitude for such emission may be evaluated by considering (pseudo)scalar corrections 
to the fermion line. In this section, the thermal self-energy of the scalar-fermion one-loop diagram is presented.

%%%%%%%%%%%%%%%%%%%%%%%%%%%%%%%%%%%%%%%%%%%%%%%%%%%%%%%%%%%%%%%%
\subsection{Real-time self-energy of the scalar-fermion eye-diagram}
\label{sec:RealtimeSESFeye}
%%%%%%%%%%%%%%%%%%%%%%%%%%%%%%%%%%%%%%%%%%%%%%%%%%%%%%%%%%%%%%%%

The scalar-fermion interaction term of \eq{StFaFInteractionTerm} 
gives rise to the eye-diagram
\beq
     \raisebox{-0.26\height}{%
         \includegraphics[scale=0.5]{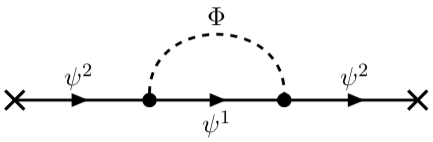}
     }.
\eeq
As in the previous sections, the thick lines of the loop denote the full real-time thermal propagator 
in its matrix structure with components of \mbox{Eqs.~\ref{eq:ExplicitRealTimeScalarPropagatorComponents}, \ref{eq:ExplicitRealTimeFermionPropagatorComponents}}. The \mbox{$\psi^2$-lines} may connect either 
to vertices or external legs. This diagram in itself is in the matrix form and represents the full 
sum of thermal diagrams, similar to the case of the scalar bubble in \sec{pStpSpS}, including both 
physical chronological and unphysical antichronological field components. Using the same convention 
as for the previous calculations, the self-energy matrix is defined as follows
\beq
     i \mathcal I_\text{bubble}(p; m_1, M)
     =
     \raisebox{-0.18\height}{%
         \includegraphics[scale=0.2]{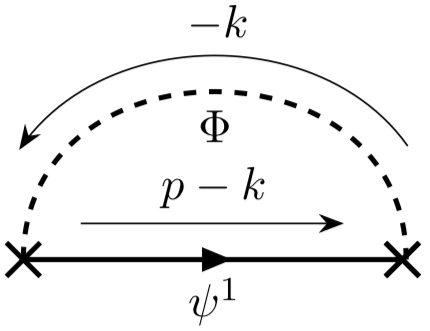}
     }
\eeq
where, for the case of interest, the vertices couple to the fermion lines on external legs. 
The $(++)$-component of this diagram is
\begin{align}
i\mathcal I_\text{bubble}^{(++)} & (p; m_1, M)
\nonumber
\\ & =
i(-ia)^2 \!
\int \! \frac{\dd^4 k}{(2\pi)^4} \,
	i\tilde S_{\text{F}}^{(++)}(p-k; m_1) i\tilde D^{(++)}(-k; M) ]
\nonumber
\\ & =
iI_{\text{SF}}^{(2)}\big( p; m_1, M \big)
+
\Big[ iF_{\text{SF}}^{(2)}\big( p; m_1, M \big) + (\text{S} \leftrightarrow \text{F}) \Big]
+
iF_{\text{SF}}^{(3)}\big( p; m_1, M \big) \,.
\end{align}
The four integrals extracted above correspond to the non-thermal contribution, the mixed thermal contribution 
and the purely thermal contribution to the self-energy. The second mixed term, expressed through the bracketed arrow, 
is similar to its preceding term but with the thermal term of the fermion propagator and the non-thermal term 
of the boson propagator interchanged. The explicit forms are
\begin{align}
  iI_{\text{SF}}^{(2)} & (p; m_1, M)
  \nonumber
  \\ & \! =
  i(-ia)^2 \!\! \int \! \frac{\dd^4 k}{(2\pi)^4} \,
    \frac{i\big( (\fsl{p} - \fsl{k}) + m_1 \big)}{(p-k)^2 - m_1^2 + i\epsilon} \,
    \frac{i}{k^2 - M^2 + i\epsilon}, \label{eq:SFnonThermalSE}
  \\ \nonumber \\
  iF_{\text{SF}}^{(2)} & (p; m_1, M)
  \nonumber
  \\ & \! =
  i(-ia)^2 \!\! \int \! \frac{\dd^4 k}{(2\pi)^4} \,
    \frac{i\big( (\fsl{p} - \fsl{k}) + m_1 \big)}{(p-k)^2 - m_1^2 + i\epsilon}
    2\pi \, n_\text{B}(\abs{k_0}) \delta\big( k^2 - M^2 \big) \,, \label{eq:SFmixedThermalSE}
  \\ \nonumber \\
  (\text{S} & \leftrightarrow \text{F})
  \nonumber
  \\ & \! =
  i(-ia)^2 \!\! \int \! \frac{\dd^4 k}{(2\pi)^4} \,
    (\!-2\pi) \big( (\fsl{p} \!-\! \fsl{k}) + m_1 \big) \, n_\text{F}(p_0 \!-\! k_0) \delta\big( (p\!-\!k)^2 \!- m_1^2 \big)
    \frac{i}{k^2 \!-\! M^2 \!+\! i\epsilon} \,, \label{eq:SFmixedThermalSE2nd}
  \\ \nonumber \\
  iF_{\text{SF}}^{(3)} & (p; m_1, M)
  \nonumber
  \\ & \! =
  i(-ia)^2 \!\! \int \! \frac{\dd^4 k}{(2\pi)^4} \,
    (\!-2\pi) \big( (\fsl{p} - \fsl{k}) + m_1 \big) n_\text{F}(p_0 \!-\! k_0) \delta\big( (p-k)^2 \!-\! m_1^2 \big)
    \cross
    \nonumber
    \\ & \hspace{4.1cm} \cross
    2\pi n_\text{B}(\abs{k_0}) \delta\big( k^2 - M^2 \big) \,. \label{eq:SFpureThermalSE}
  \end{align}

%%%%%%%%%%%%%%%%%%%%%%%%%%%%%%%%%%%%%%%%%%%%%%%%%%%%%%%%%%%%%%%%
\subsubsection{Non-thermal self-energy term}
\label{sec:SFnonThermalSE}
%%%%%%%%%%%%%%%%%%%%%%%%%%%%%%%%%%%%%%%%%%%%%%%%%%%%%%%%%%%%%%%%

Relevant to the emission is the imaginary part of the non-thermal contribution to the $(++)$-component 
of the self-energy. It may be extracted by a similar procedure performed in \sec{FaFnonThermalSE}. 
Evaluated by the dimensional regularisation techniques, the imaginary part of \eq{SFnonThermalSE} 
is found to be
\beq
  \imaginary\ \big[ iI_\text{SF}^{(2)}(p; m_1, M) \big]
  =
  -\frac{a^2}{32\pi}\sqrt{C'} \cross
  \begin{dcases}
    \fsl{p}\bigg( 1 + \frac{m_1^2 \!-\! M^2}{p^2} \bigg) + 2m_1
    \hspace{0.82cm} \text{if } p^2 \geq (m_1 + M)^2 \,,
    \\
    0
    \hspace{5.5cm} \text{otherwise.}
  \end{dcases}
\eeq
Here,
\beq
  C'
  =
  \Bigg(\!
    1 - \frac{(m_1 \!+\! M)^2}{p^2}
  \Bigg)
  \!
  \Bigg(\!
    1 - \frac{(m_1 \!-\! M)^2}{p^2}
  \Bigg) \,.
\eeq

%%%%%%%%%%%%%%%%%%%%%%%%%%%%%%%%%%%%%%%%%%%%%%%%%%%%%%%%%%%%%%%%
\subsubsection{Mixed self-energy term}
\label{sec:SFmixedThermalSE}
%%%%%%%%%%%%%%%%%%%%%%%%%%%%%%%%%%%%%%%%%%%%%%%%%%%%%%%%%%%%%%%%

The mixed self-energy contribution consists of two complex cross terms from the product of the propagators: 
the integrals in \mbox{Eqs.~(\ref{eq:SFmixedThermalSE}) and (\ref{eq:SFmixedThermalSE2nd})}. The resulting 
imaginary component of the sum of the integrals is a rather long expression when written down explicitly. 
For clarity, the result is therefore presented for each momentum region separately.

%%%%%%%%%%%%%%%%%%%%%%%%%%%%%%%%%%%%%%%%%%%%%%%%%%%%%%%%%%%%%%%%
\paragraph{If \texorpdfstring{$p^2 \geq (m_1 + M)^2$}{TEXT}:}
%%%%%%%%%%%%%%%%%%%%%%%%%%%%%%%%%%%%%%%%%%%%%%%%%%%%%%%%%%%%%%%%

\begin{align}
  & \imaginary\ \big[
     iF_\text{SF}^{(2)} (p; m_1, M) + (\text{S} \leftrightarrow \text{F})
\big]
  \nonumber
  \\ & \,\, =
  -\frac{a^2}{16\pi \abs{\vb p} \beta}
  \left( \fsl{p} + \gamma^3\frac{-p^2 \!+\! m_1^2 \!-\! M^2}{2\abs{\vb p}} + m_1 \right)
  \ln\abs{\frac{
      \big( e^{-\beta\omega_{M, p_+}} \!-\! 1 \big)\big( e^{\beta(\omega_{M, p_-} \!- p_0 + \mu)} \!+\! 1 \big)
    }{
      \big( e^{-\beta\omega_{M, p_-}} \!-\! 1 \big)\big( e^{\beta(\omega_{M, p_+} \!- p_0 + \mu)}  \!+\! 1 \big)
    }
  }
  \nonumber
  \\ & \quad\,\,
  -\frac{a^2}{16\pi\abs{\vb p}\beta^2} \left( \!-\gamma^0 + \gamma^3\frac{p_0}{\abs{\vb p}} \right)
  \cross
  \Bigg\{
    \!-\!\frac{\beta^2}{2} \big( \omega_{M, p_+}^2 \! - \omega_{M, p_-}^2 \big)
    \nonumber
    \\ & \quad\quad +
    \text{Li}_2\big( e^{\beta\omega_{M, p_+}} \! \big)
    -
    \text{Li}_2\big( e^{\beta\omega_{M, p_-}} \! \big)
    -
    \text{Li}_2\big( \!\!-\!e^{\beta(\omega_{M, p_+} \!- p_0 + \mu)} \big)
    +
    \text{Li}_2\big( \!\!-\!e^{\beta(\omega_{M, p_-} \!- p_0 + \mu)} \big)
    \nonumber
    \\ & \quad\qquad\,\, +
    \beta\omega_{M, p_+}\!\ln\bigg[\frac{
      1 - e^{\beta\omega_{M, p_+}}
    }{
      e^{\beta(\omega_{M, p_+} \!- p_0 + \mu)} \!+\! 1
    }\bigg]
    -
    \beta\omega_{M, p_-}\!\ln\bigg[\frac{
      1 - e^{\beta\omega_{M, p_-}}
    }{
      e^{\beta(\omega_{M, p_-} \!- p_0 + \mu)
    } \!+\! 1}\bigg]
  \Bigg\} \,,
  \label{eq:SFmixedThermalSEFirstRegion}
\end{align}
where
\beq
     \omega_{M, p_\pm}
     =
     \frac{1}{2}
     \abs{p_0\abs{1 + \frac{M^2 \!-\! m_1^2}{p^2}} \pm \abs{\vb p}\sqrt{C'}} \,.
\eeq
  $\text{Li}_2(z)$ is the analytic continuation of Spence's function, the dilogarithm
\beq
     \hat{\text{Li}}_2(z)
     =
     -\!\int_0^z {\dd\tau}
         \frac{\ln[1 - \tau]}{\tau} \,,
     \quad z\in\mathbb C \,,
\eeq
which cancels exactly the imaginary part that arises due to the negative sign of 
\mbox{$1 \!-\! e^{\beta\omega_{M, p_\pm}}$} in the two analytically continued 
logarithmic terms in \eq{SFmixedThermalSEFirstRegion}.

%%%%%%%%%%%%%%%%%%%%%%%%%%%%%%%%%%%%%%%%%%%%%%%%%%%%%%%%%%%%%%%%
\paragraph{If \texorpdfstring{$(m_1 - M)^2 \leq p^2 < (m_1 + M)^2$}{TEXT}:}
%%%%%%%%%%%%%%%%%%%%%%%%%%%%%%%%%%%%%%%%%%%%%%%%%%%%%%%%%%%%%%%%

\beq
  \imaginary\ \big[
     iF_\text{SF}^{(2)} (p; m_1, M) + (\text{S} \leftrightarrow \text{F})
\big]
  =
  0 \,.
\eeq

%%%%%%%%%%%%%%%%%%%%%%%%%%%%%%%%%%%%%%%%%%%%%%%%%%%%%%%%%%%%%%%%
\paragraph{If \texorpdfstring{$0 \leq p^2 < (m_1 - M)^2$}{TEXT}:}
\label{sec:SFmixedThermalSEc1}
%%%%%%%%%%%%%%%%%%%%%%%%%%%%%%%%%%%%%%%%%%%%%%%%%%%%%%%%%%%%%%%%

\begin{align}
     & \imaginary\ \big[
         iF_\text{SF}^{(2)} (p; m_1, M)
         +
         (\text{S} \leftrightarrow \text{F})
     \big]
     \nonumber
     \\ & \,\, =
     -\frac{a^2}{16\pi \abs{\vb p} \beta}
     \left(
         \fsl{p} + \gamma^3\frac{-p^2 \!+\! m_1^2 \!-\! M^2}{2\abs{\vb p}} + m_1
     \right)
  \ln\abs{
    \frac{
      \big( e^{-\beta\omega_{M, p_+}} \!-\! 1 \big)\big( e^{\beta(\omega_{M, p_-} \!\mp (p_0 - \mu))} \!+\! 1 \big)
    }{
      \big( e^{-\beta\omega_{M, p_-}} \!-\! 1 \big)\big( e^{\beta(\omega_{M, p_+} \!\mp (p_0 - \mu))}  \!+\! 1 \big)
    }
  }
  \nonumber
  \\ & \quad\,\,
  \mp\frac{a^2}{16\pi\abs{\vb p}\beta^2} \left( \!-\gamma^0 + \gamma^3\frac{p_0}{\abs{\vb p}} \right)
  \cross
  \nonumber
  \\ & \quad\quad \cross \!\! \Bigg\{
    \text{Li}_2\big( e^{\beta\omega_{M, p_+}} \! \big)
    -
    \text{Li}_2\big( e^{\beta\omega_{M, p_-}} \! \big)
    +
    \text{Li}_2\big( \!\!-\!e^{\beta(\omega_{M, p_+} \!\mp (p_0 - \mu))} \big)
    -
    \text{Li}_2\big( \!\!-\!e^{\beta(\omega_{M, p_-} \!\mp (p_0 + \mu))} \big)
    \nonumber
    \\ & \quad\qquad\,\, +
    \beta\omega_{M, p_+}\!\ln\bigg[
      \big( e^{-\beta\omega_{M, p_+}} \!-\! 1 \big)
      \big( e^{\beta(\omega_{M, p_+} \!\mp (p_0 - \mu))} \!+\! 1 \big)
    \bigg]
    \nonumber
    \\ & \quad\qquad\,\, -
    \beta\omega_{M, p_-}\!\ln\bigg[
      \big( e^{-\beta\omega_{M, p_-}} \!-\! 1 \big)
      \big( e^{\beta(\omega_{M, p_-} \!\mp (p_0 - \mu))} \!+\! 1 \big)
    \bigg]
  \Bigg\} \,.
\end{align}
Here, the upper sign applies if \mbox{$M \geq m_1$} and the lower sign if \mbox{$M < m_1$}. 
Compare to the similar case (d1) in \sec{SSpureThermalSE}.

%%%%%%%%%%%%%%%%%%%%%%%%%%%%%%%%%%%%%%%%%%%%%%%%%%%%%%%%%%%%%%%%
\paragraph{If \texorpdfstring{$p^2 < 0$}{TEXT}:}
%%%%%%%%%%%%%%%%%%%%%%%%%%%%%%%%%%%%%%%%%%%%%%%%%%%%%%%%%%%%%%%%

\begin{align}
  & \imaginary\ \big[
      iF_\text{SF}^{(2)} (p; m_1, M)
      +
      (\text{S} \leftrightarrow \text{F})
  \big]
  \nonumber
  \\ & \,\, =
  +\frac{a^2}{16\pi \abs{\vb p} \beta}
  \left( \fsl{p} + \gamma^3\frac{-p^2 \!+\! m_1^2 \!-\! M^2}{2\abs{\vb p}} + m_1 \right)
  \cross\nonumber
  \\ &\qquad\,\, \cross \!
  \ln\abs{
    \big( e^{-\beta\omega_{M, p_\pm}} \!-\! 1 \big)
    \big( e^{-\beta\omega_{M, p_\mp}} \!-\! 1 \big)
    \big( e^{-\beta(\omega_{M, p_\pm} \!- p_0 + \mu)} \!+\! 1 \big)
    \big( e^{-\beta(\omega_{M, p_\mp} \!+ p_0 - \mu)}  \!+\! 1 \big)
  }
  \nonumber
  \\ & \quad\,\,\, +
  \frac{a^2}{16\pi\abs{\vb p}\beta^2} \left( \!-\gamma^0 + \gamma^3\frac{p_0}{\abs{\vb p}} \right)
  \cross
  \nonumber
  \\ & \quad\quad\,\, \cross \!\!
  \Bigg\{
    \text{Li}_2\big( e^{\beta\omega_{M, p_\pm}} \! \big)
    -
    \text{Li}_2\big( e^{\beta\omega_{M, p_\mp}} \! \big)
    +
    \text{Li}_2\big( \!\!-\!e^{\beta(\omega_{M, p_\pm} \!- p_0 + \mu)} \big)
    -
    \text{Li}_2\big( \!\!-\!e^{\beta(\omega_{M, p_\mp} \!+ p_0 - \mu)} \big)
    \nonumber
    \\ & \quad\qquad\,\, +
    \beta\omega_{M, p_\pm}\!\ln\bigg[
      \big( e^{-\beta\omega_{M, p_\pm}} \!-\! 1 \big)
      \big( e^{\beta(\omega_{M, p_\pm} \!- p_0 + \mu)} \!+\! 1 \big)
    \bigg]
    \nonumber
    \\ & \quad\qquad\,\, -
    \beta\omega_{M, p_\mp}\!\ln\bigg[
      \big( e^{-\beta\omega_{M, p_\mp}} \!-\! 1 \big)
      \big( e^{\beta(\omega_{M, p_\mp} \!+ p_0 - \mu)} \!+\! 1 \big)
    \bigg]
  \Bigg\} \,.
\end{align}
Here, the upper and lower signs correspond to the modified case (d2) defined in \sec{SSpureThermalSE}. 
The modification is analogous to the modification of the case \mbox{(d1)} in the section above 
for \mbox{$0 \leq p^2 < (m_1 - M)^2$} and, as a consequence, the result depends on the mass 
hierarchy of $M$, $m_1$.

%%%%%%%%%%%%%%%%%%%%%%%%%%%%%%%%%%%%%%%%%%%%%%%%%%%%%%%%%%%%%%%%
\subsubsection{The purely thermal self-energy term}
\label{sec:SFpureThermalSE}
%%%%%%%%%%%%%%%%%%%%%%%%%%%%%%%%%%%%%%%%%%%%%%%%%%%%%%%%%%%%%%%%

The thermal self-energy contribution in \eq{SFpureThermalSE} is purely imaginary, cf. previous self-energy 
evaluations in \mbox{Secs.~\ref{sec:SSpureThermalSE},} \ref{sec:FaFpureThermalSE} and \ref{sec:FaFpureThermalSEPS}. 
The resulting expression for this term, after momentum integration, is again rather long. Therefore, different 
regions of $p^2$ are presented separately below for clarity.

%%%%%%%%%%%%%%%%%%%%%%%%%%%%%%%%%%%%%%%%%%%%%%%%%%%%%%%%%%%%%%%%
\paragraph{If \texorpdfstring{$p^2 \geq (m_1 + M)^2$}{TEXT}:}
%%%%%%%%%%%%%%%%%%%%%%%%%%%%%%%%%%%%%%%%%%%%%%%%%%%%%%%%%%%%%%%%

\begin{align}
\label{eq:p2Greatest}
  & iF_\text{SF}^{(3)} (p; m_1, M)
  \nonumber
  \\ & =
  i\frac{a^2}{8\pi \abs{\vb p} \beta}
  \left( \fsl{p} + \gamma^3\frac{-p^2 \!+\! m_1^2 \!-\! M^2}{2\abs{\vb p}} + m_1 \right)
  \frac{1}{e^{\beta(p_0 - \mu)} \!+\! 1}
  \ln\abs{\frac{
      \big( 1 \!-\! e^{-\beta\omega_{M, p_+}} \big)
      \big( e^{\beta(p_0 - \omega_{M, p_-} \!- \mu)} \!+\! 1 \big)
    }{
      \big( 1 \!-\! e^{-\beta\omega_{M, p_-}} \big)
      \big( e^{\beta(p_0 - \omega_{M, p_+} \!- \mu)}  \!+\! 1 \big)
    }
  }
  \nonumber
  \\ & \quad +
  i\frac{a^2}{8\pi\abs{\vb p}\beta^2} \left( \gamma^0 - \gamma^3\frac{p_0}{\abs{\vb p}} \right)
  \frac{1}{e^{\beta(p_0 - \mu)} \!+\! 1}
  \cross
  \nonumber
  \\ & \quad\quad \cross \!\!
  \Bigg\{
    \text{Li}_2\big( e^{\beta\omega_{M, p_+}} \! \big)
    -
    \text{Li}_2\big( e^{\beta\omega_{M, p_-}} \! \big)
    -
    \text{Li}_2\big( \!\!-\!e^{\beta(\omega_{M, p_+} \!- p_0 + \mu)} \big)
    +
    \text{Li}_2\big( \!\!-\!e^{\beta(\omega_{M, p_-} \!- p_0 + \mu)} \big)
    \nonumber
    \\ & \quad\qquad +
    \beta\omega_{M, p_+}\!\ln\!\bigg[\frac{
        1 \!-\! e^{\beta\omega_{M, p_+}}
      }{
        1 \!+\! e^{\beta(\omega_{M, p_+} \!- p_0 + \mu)}
      }
    \bigg]
    -
    \beta\omega_{M, p_-}\!\ln\!\bigg[\frac{
        1 \!-\! e^{\beta\omega_{M, p_-}}
      }{
        1 \!+\! e^{\beta(\omega_{M, p_-} \!- p_0 + \mu)}
      }
    \bigg]
  \Bigg\} \,.
\end{align}

%%%%%%%%%%%%%%%%%%%%%%%%%%%%%%%%%%%%%%%%%%%%%%%%%%%%%%%%%%%%%%%%
\paragraph{If \texorpdfstring{$(m_1 - M)^2 \leq p^2 < (m_1 + M)^2$}{TEXT}:}
%%%%%%%%%%%%%%%%%%%%%%%%%%%%%%%%%%%%%%%%%%%%%%%%%%%%%%%%%%%%%%%%

\beq
  iF_\text{SF}^{(3)} (p; m_1, M)
  =
  0 \,.
\eeq

%%%%%%%%%%%%%%%%%%%%%%%%%%%%%%%%%%%%%%%%%%%%%%%%%%%%%%%%%%%%%%%%
\paragraph{If \texorpdfstring{$0 \leq p^2 < (m_1 - M)^2$}{TEXT}:}
%%%%%%%%%%%%%%%%%%%%%%%%%%%%%%%%%%%%%%%%%%%%%%%%%%%%%%%%%%%%%%%%

\begin{align}
  & iF_\text{SF}^{(3)} (p; m_1, M)
  \nonumber
  \\ & =
  i\frac{a^2}{8\pi \abs{\vb p} \beta}
  \left( \fsl{p} + \gamma^3\frac{-p^2 \!+\! m_1^2 \!-\! M^2}{2\abs{\vb p}} + m_1 \right)
  \frac{1}{e^{\mp\beta(p_0 - \mu)} \!+\! 1}
  \cross\nonumber
  \\ & \qquad \cross \!
  \Bigg[
    \ln\abs{
      \frac{
        \big( 1 \!-\! e^{-\beta\omega_{M, p_+}} \big)
      }{
        \big( 1 \!-\! e^{-\beta\omega_{M, p_-}} \big)
      }
    }
    +
    e^{\mp\beta(p_0 - \mu)}
    \ln\abs{
      \frac{
        \big( e^{-\beta(\omega_{M, p_+} \!\mp (p_0 - \mu))}  \!+\! 1 \big)
      }{
        \big( e^{-\beta(\omega_{M, p_-} \!\mp (p_0 - \mu))} \!+\! 1 \big)
      }
    }
  \Bigg]
  \nonumber
  \\ & \quad \mp
  i\frac{a^2}{8\pi\abs{\vb p}\beta^2}\left( \gamma^0 - \gamma^3\frac{p_0}{\abs{\vb p}} \right)
  \frac{1}{e^{\mp\beta(p_0 - \mu)} \!+\! 1}
  \cross\nonumber
  \\ & \qquad \cross \!\! \Bigg\{
    \Big( e^{\mp\beta(p_0 - \mu)} \!+\! 1 \Big)
    \frac{\beta^2}{2} \big( \omega_{M, p_-}^2 \! - \omega_{M, p_+}^2 \big)
    +
    \beta\omega_{M, p_+}\!\ln\!\big[
      1 \!-\! e^{\beta\omega_{M, p_+}}
    \big]
    -
    \beta\omega_{M, p_-}\!\ln\!\big[
      1 \!-\! e^{\beta\omega_{M, p_-}}
    \big]
    \nonumber
    \\ & \quad\qquad +
    e^{\mp\beta(p_0 - \mu)} \bigg[
      \beta\omega_{M, p_+}\!\ln\!\Big[
        e^{\beta(\omega_{M, p_+} \!\mp (p_0 - \mu))} \!+\! 1
      \Big]
      -
      \beta\omega_{M, p_-}\!\ln\!\Big[
        e^{\beta(\omega_{M, p_-} \!\mp (p_0 - \mu))} \!+\! 1
      \Big]
     \bigg]
    \nonumber
    \\ & \quad\qquad +
    \text{Li}_2\big( e^{\beta\omega_{M, p_+}} \! \big)
    -
    \text{Li}_2\big( e^{\beta\omega_{M, p_-}} \! \big)
    \nonumber
    \\ & \quad\qquad +
    e^{\mp\beta(p_0 - \mu)} \Big[
      \text{Li}_2\big( \!\!-\!e^{\beta(\omega_{M, p_+} \!\mp (p_0 - \mu))} \big)
      -
       \text{Li}_2\big( \!\!-\!e^{\beta(\omega_{M, p_-} \!\mp (p_0 - \mu))} \big)
     \Big]
  \Bigg\} \,.
\end{align}
Again, the upper and lower signs correspond to the modified case \mbox{(d1)} discussed in \sec{SFmixedThermalSE} 
in order to take into account the hierarchy between $M$, $m_1$.

%%%%%%%%%%%%%%%%%%%%%%%%%%%%%%%%%%%%%%%%%%%%%%%%%%%%%%%%%%%%%%%%
\paragraph{If \texorpdfstring{$p^2 < 0$}{TEXT}:}
%%%%%%%%%%%%%%%%%%%%%%%%%%%%%%%%%%%%%%%%%%%%%%%%%%%%%%%%%%%%%%%%

\begin{align*}
  & iF_\text{SF}^{(3)} (p; m_1, M)
  \\ & =
  -i\frac{a^2}{8\pi \abs{\vb p} \beta}
  \left( \fsl{p} + \gamma^3\frac{-p^2 \!+\! m_1^2 \!-\! M^2}{2\abs{\vb p}} + m_1 \right)
  \cross
  \\ &\qquad \cross \!\! \Bigg\{
    \frac{1}{e^{-\beta(p_0 - \mu)} \!+\! 1} \bigg[
      \ln\abs{
        \big( e^{-\beta\omega_{M, p_\pm}} \!-\! 1 \big)
      }
      +
      e^{-\beta(p_0 - \mu)} \ln\abs{
        \big( e^{-\beta(\omega_{M, p_\pm} \!- p_0 + \mu)} \!+\! 1 \big)
      }
    \bigg]
    \\ & \hspace{1.2cm} +
    \frac{1}{e^{\beta(p_0 - \mu)} \!+\! 1} \bigg[
      \ln\abs{
        \big( e^{-\beta\omega_{M, p_\mp}} \!-\! 1 \big)
      }
      +
      e^{\beta(p_0 - \mu)} \ln\abs{
        \big( e^{-\beta(\omega_{M, p_\mp} \!+ p_0 - \mu)} \!+\! 1 \big)
      }
    \bigg]
  \Bigg\}
  \\ & \quad -
  i\frac{a^2}{8\pi\abs{\vb p}\beta^2} \left( \!-\gamma^0 + \gamma^3\frac{p_0}{\abs{\vb p}} \right)
  \frac{1}{e^{-\beta(p_0 - \mu)} \!+\! 1}
  \cross
  \nonumber
  \\ & \quad\quad \cross \!\!
  \Bigg\{
    \!\!-\!\frac{\pi^2}{3} + \frac{\pi^2}{6}e^{-\beta(p_0 - \mu)} + \frac{\beta^2}{2}(p_0 - \mu)^2e^{-\beta(p_0 - \mu)}
    -\frac{1}{2}\Big( 1 \!+\! e^{-\beta(p_0 - \mu)} \Big) \beta^2\omega_{M, p_\pm}^2
    \nonumber
    \\ & \quad\qquad
    +
    \beta\omega_{M, p_\pm} \bigg[
      \ln\!\big[
        1 \!-\! e^{\beta\omega_{M, p_\pm}}
      \big]
      +
      e^{-\beta(p_0 - \mu)}\ln\!\abs{
        e^{\beta(\omega_{M, p_\pm} \!- p_0 + \mu)} \!+\! 1
      }
    \bigg]
    \nonumber
    \\ & \quad\qquad +
    \text{Li}_2\big( e^{\beta\omega_{M, p_\pm}} \! \big)
    +
    e^{-\beta(p_0 - \mu)} \text{Li}_2\big( \!\!-\!e^{\beta(\omega_{M, p_\pm} \!- p_0 + \mu)} \big)
  \Bigg\}
%%%%%%%%
\end{align*}
\begin{align}
  & \quad -
  i\frac{a^2}{8\pi\abs{\vb p}\beta^2} \left( \!-\gamma^0 + \gamma^3\frac{p_0}{\abs{\vb p}} \right)
  \frac{1}{e^{\beta(p_0 - \mu)} \!+\! 1}
  \cross
  \nonumber
  \\ & \quad\quad \cross \!\!
  \Bigg\{
    \!\!-\!\frac{\pi^2}{3} + \frac{\pi^2}{6}e^{\beta(p_0 - \mu)} + \frac{\beta^2}{2}(p_0 - \mu)^2e^{\beta(p_0 - \mu)}
    -\frac{1}{2}\Big( 1 \!+\! e^{\beta(p_0 - \mu)} \Big) \beta^2\omega_{M, p_\mp}^2
    \hspace{0.7cm}\nonumber
    \\ & \quad\qquad +
    \beta\omega_{M, p_\mp} \bigg[
      \ln\!\big[
      1 \!-\! e^{\beta\omega_{M, p_\mp}}
      \big]
      +
      e^{\beta(p_0 - \mu)} \ln\!\abs{
        \big( e^{\beta(\omega_{M, p_\mp} \!+ p_0 - \mu)} \!+\! 1 \big)
      }
    \bigg]
    \nonumber
    \\ & \quad\qquad +
    \text{Li}_2\big( e^{\beta\omega_{M, p_\mp}} \! \big)
    +
    e^{\beta(p_0 - \mu)} \text{Li}_2\big( \!\!-\!e^{\beta(\omega_{M, p_\mp} \!+ p_0 - \mu)} \big)
  \Bigg\} \,.
  \label{eq:p2Negative}
\end{align}
  Also here, the sign cases correspond to a modified \mbox{(d2)}.

Note that in the final expression for the purely thermal contribution, \mbox{Eqs.~(\ref{eq:p2Greatest})-(\ref{eq:p2Negative})}, 
the mass parameters $m_1$, $M$ should be replaced by \mbox{$\text{min}\{ m_1, M \}$} and \mbox{$\text{max}\{ m_1, M \}$}, 
respectively; cf.~\sec{SSpureThermalSE}.

The explicit \mbox{$(++)$}-component of the real-time self-energy of the fermionic eye-diagram presented in \mbox{Secs.~\ref{sec:SFnonThermalSE}-\ref{sec:SFpureThermalSE}} above is not present 
in the literature to the best of our knowledge.

%%%%%%%%%%%%%%%%%%%%%%%%%%%%%%%%%%%%%%%%%%%%%%%%%%%%%%%%%%%%%%%%
\subsection{Pseudoscalar emission off the fermion line}
%%%%%%%%%%%%%%%%%%%%%%%%%%%%%%%%%%%%%%%%%%%%%%%%%%%%%%%%%%%%%%%%

Analogously to \sec{pStFaF}, the pseudoscalar interaction term \mbox{$\bar\psi\gamma_5\psi$} provides 
an additional vertex factor of $\gamma_5$ compared to the case of \emph{scalar} emission considered 
in the previous three sections. In the na\"ive dimensional regularisation scheme, $\gamma_5$ anticommutes 
with the \mbox{$\gamma$-matrix} in the numerator of the fermion propagator and thereby provides an overall 
sign change together with a sign change of the numerator mass $m_1$. These sign changes are stated relative 
to the obtained results of \mbox{Secs.~\ref{sec:SFnonThermalSE}-\ref{sec:SFpureThermalSE}} and a conclusion 
completely analogous to \sec{pStFaF} is reached.

%%%%%%%%%%%%%%%%%%%%%%%%%%%%%%%%%%%%%%%%%%%%%%%%%%%%%%%%%%%%%%%%
\subsection{Scalar emission rate of the process \texorpdfstring{$\psi^2 \to \psi^1\Phi$}{TEXT}}
%%%%%%%%%%%%%%%%%%%%%%%%%%%%%%%%%%%%%%%%%%%%%%%%%%%%%%%%%%%%%%%%

The thermal rate for the scalar emission provided in this section is given by \eq{RealTimeDecayRate} 
in \sec{DecayRateRealTimeSelfEnergy}. In order to extract an observable from the self-energy presented in \mbox{Secs.~\ref{sec:SFnonThermalSE}-\ref{sec:SFpureThermalSE}}, that carries spin structure, 
it is advisable to follow the strategy provided by Weldon~\cite{Weldon1983} to define
\beq
     \Sigma(p)
     =
     \bar u(p) i\mathcal I_\text{bubble}^{(++)}(p;m_1, M) u(p) \,.
\eeq
This is the self-energy contracted with incoming and outgoing states. These states satisfy the Dirac equation 
\mbox{($\fsl{p} \!-\! m)u(p)=0$} as well as \mbox{$\bar u(p)u(p) = 2\sqrt{p^2}$}. In terms of $\Sigma$, 
the discontinuity over the real axis of this new scalar function may be related to the thermal decay rate. 
More specifically, the contraction of the three different pre-factors that appear in the self-energy 
and which contain explicit \mbox{$\gamma$-matrices,} is of interest. Assuming the external $\psi^2$ 
to be on-shell \mbox{($p^2 = m_2^2$)}, the contraction results in
\begin{align}
     \bar u(p)\bigg( \fsl{p}\Big( 1 + \tfrac{m_1^2 \!-\! M^2}{p^2} \Big) + 2m_1 \bigg)u(p)
     & =
     2\Big[ p^2 + m_1^2 + 2m_1\sqrt{p^2} - M^2 \Big]
     \nonumber
     \\ & =
     -2M^2 \bigg( 1 - \frac{(m_1 \!+\! m_2)^2}{M^2} \bigg) \,,
     \\
     \bar u(p)\bigg( \fsl{p} + \gamma^3\frac{-p^2 \!+\! m_1^2 \!-\! M^2}{2\abs{\vb p}} + m_1 \bigg)u(p)
     & =
     p^2 + m_1^2 + 2m_1\sqrt{p^2} - M^2
     \nonumber
     \\ & =
     -M^2 \bigg( 1 - \frac{(m_1 + m_2)^2}{M^2} \bigg),
     \\
     \bar u(p) \Big( \!-\!\gamma^0 + \gamma^3\frac{p_0}{\abs{\vb p}} \Big) u(p)
     & \equiv
     0 \,.
\end{align}

The ratio of \eq{DecayRateRatio} may be obtained for example in the mass region of \mbox{$p^2 = m_2^2 \gg (m_1 + M)^2$} 
effectively rendering the particles in the loop massless. The plotted ratio can be seen in \fig{FtSFDecayRatio} for 
a non-relativistic, relativistic and ultra-relativistic incoming particle $\psi^2$. In the limit \mbox{$T \to 0$} 
the decay ratio \mbox{$R_{\psi^2 \to \psi^1\Phi} \to 1$}.
%%%%%%%%%%%%%%%%%%%%%%%%%%%%%%%%%%%%%%%%%%%%%%%%%%%%%%%%%%%
\begin{figure}[htbp]
  \begin{minipage}[t]{0.56\textwidth}
    \vspace{0pt}
    \centering
    \resizebox{\textwidth}{!}{
      \input{figures/FtSF/FtSF}
    }
  \end{minipage}
  \hfill
  \begin{minipage}[t]{0.42\textwidth}
    \vspace{8pt}
    \caption{Ratio to the zero-temperature limit of the emission rate of a scalar off a fermion line. 
    The ratio \mbox{$R = \Gamma_D/\gamma_D$} is plotted for \mbox{$m_2^2 \gg (m_1 + M)^2$}. 
    Curves display ratios for varying \mbox{$\epsilon = \abs{\vb p}/m_2$} parameter 
    for a non-relativistic \mbox{$\epsilon \in \{0.001, 1\}$},
    relativistic \mbox{$\epsilon=10$} and ultra-relativistic \mbox{$\epsilon=100$} incoming $\psi^2$. Note that the ratio \mbox{$R \to 1$} when \mbox{$T \to 0$.}
    \label{fig:FtSFDecayRatio}}
  \end{minipage}
\end{figure}
%%%%%%%%%%%%%%%%%%%%%%%%%%%%%%%%%%%%%%%%%%%%%%%%%%%%%%%%%%%

The linear dependence on $T$ of the self-energy seen in \fig{FtSFDecayRatio} is initially surprising in comparison 
to the quadratic temperature dependence of the self-energy of the scalar-scalar loop seen \fig{StSSDecayRatioHoScherrer}. 
This difference in behaviour may be explained in the high-temperature limit. The dependence on temperature 
of the free propagator comes from the term proportional to the thermal distribution function $n(\abs{k_0})$, see
\eq{ExplicitRealTimeScalarPropagatorComponents}. Expanding this function in the high-temperature limit (or small $\beta$) 
results in different powers of $T$ of the leading term for bosons and fermions, respectively. For bosons, 
the temperature-dependence of the propagator is proportional to $T$ while the leading term for the fermion 
propagator is constant in $T$. Hence, it is seen that for \mbox{$T \to \infty$}, the scalar-scalar loop 
of \sec{pStpSpS} contains a term that is quadratic in $T$ (the purely thermal part) while the scalar-fermion 
loop of \sec{FtpSF} contains terms proportional at most only to $T$.

%%%%%%%%%%%%%%%%%%%%%%%%%%%%%%%%%%%%%%%%%%%%%%%%%%%%%%%%%%%%%%%%
\section{Summary and outlook on thermal decays}
\label{sec:outlook}
\setcounter{equation}{0}
%%%%%%%%%%%%%%%%%%%%%%%%%%%%%%%%%%%%%%%%%%%%%%%%%%%%%%%%%%%%%%%%

Thermal effects on key observables in particle physics have caught a strong interest and attention 
of both theorists and experimentalists over past few decades. The key results of the thermal field 
theory (TFT) appear to be scattered over many different sources in the literature, and a complete 
and consistent compendium of these results is yet missing. This review provides a pedagogical
overview of the most important elements of the equilibrium TFT in both imaginary- and real-time 
formulations, together with the most immediate and relevant applications for all possible particle decays 
in thermal medium in a generalised Yukawa theory. A special attention is given to the real-time 
formulation and its predictions for particle two-body decay rates that are compared and verified against 
the existing results in the imaginary-time (Matsubara) framework, widely used in the literature. 
This represents an important step towards a proper generalisation of TFT towards in-equilibrium 
particle dynamics that can be consistently achieved in the real-time framework.

Starting from a theoretical overview of the very general and basic principles of TFT formulated 
by Wagner~\cite{Wagner1991}, we present a collection of possible decay rates for a theory containing 
generic (pseudo)scalar bosons and fermions. Several thermal one-loop calculations were performed within 
the scope of this work in the real-time framework. Components of the self-energy were related to 
a quantity interpreted as a thermal decay rate of a field in an equilibrated medium which is considered 
in various possible kinematical domains. 

A (pseudo)scalar state decaying into a pair of (pseudo)scalar particles was considered as well as a (pseudo)scalar
decaying into a fermion-antifermion pair. The thermal rate of a (pseudo)scalar emission off a fermion 
line is also provided. For all these processes, except for the \mbox{(pseudo)scalar} decay into a
fermion-antifermion pair, an enhancement of the thermal decay rate was quantified relative 
to the corresponding zero-temperature decay rate. In the case of a \mbox{(pseudo)scalar} 
decaying into two distinct \mbox{(pseudo)scalars}, this enhancement is quadratically growing 
with temperature, see \mbox{Fig.~\ref{fig:StSSDecayRatioHoScherrer}a-c}. Considering instead 
a scalar boson emission off a fermion line, the \mbox{$T$-dependence} is linear at high temperatures, 
see \fig{FtSFDecayRatio}.

As a practically important example, we notice a relative thermal suppression present in the case 
of a fermion-antifermion final state emerging from an incoming \mbox{(pseudo)scalar}. The rate 
of this process was shown to be 0.25 times the zero-temperature quantity at high temperatures. 
The real-time self-energy is proportional to a factor of \mbox{($1-n_{\text{F}/\bar{\text{F}}}$)} 
for each outgoing fermion/antifermion in the same way as in the imaginary-time formalism~\cite{Weldon1983}.
Hence, the outgoing particles are accompanied by Pauli suppression through the distribution function
$n_\text{F}$; at high temperatures, this distribution approaches \mbox{$1/2$} and the outgoing particles 
are Pauli blocked by medium particles with this probability as is manifested in
\fig{StFaFDecayRatioHoScherrer}.

The one-loop self-energy for \mbox{(anti)commuting} fields had previously been written down in the
imaginary-time formalism by Weldon~\cite{Weldon1983}. Ho and Scherrer\cite{HoScherrer2015} extracted 
decay rates for the processes of \mbox{Secs.~\ref{sec:pStpSpS}-\ref{sec:StFaF}} using the results 
of Weldon. This work confirms the thermal decay rate of Ref.~\cite{HoScherrer2015} in the referenced 
sections within the real-time formalism. The decay rates have been further extended beyond the results 
of Ref.~\cite{HoScherrer2015} to include the chemical potentials, and, more importantly, the masses of
particles in the loop have been allowed to take on different values. Hence, a new momentum region 
\mbox{$0 \leq p^2 < (m_1 - m_2)^2$} has been investigated for a (pseudo)scalar decaying into 
two (pseudo)scalars, a region in which decays are kinematically forbidden in the zero-temperature theory.

The explicit real-time self-energy of the scalar-scalar loop has been published previously in
Ref.~\cite{NishikawaMorimatsuHidaka2003} and was reproduced in this work. The fermion-fermion 
self-energy has been discussed in Ref.~\cite{Weldon1983} in the imaginary-time formalism.
The calculated decay rate of \fig{StFaFDecayRatioHoScherrer} reproduces the imaginary-time 
result of Ref.~\cite{HoScherrer2015} if the masses of virtual states are taken to be identical.
The explicit real-time self-energy of the scalar-fermion self-energy does not exist in the literature 
to the best of our knowledge. The thermal decay rate has been extracted from such a self-energy, 
see \fig{FtSFDecayRatio}.

%%%%%%%%%%%%%%%%%%%%%%%%%%%%%%%%%
\section*{Acknowledgements}
%%%%%%%%%%%%%%%%%%%%%%%%%%%%%%%%%

We acknowledge inspiring discussions with Hugo Ser\^odio, Johan Bijnens and Malin Sjödahl.
The work has been performed in the framework of COST Action CA15213 ``Theory of hot matter 
and relativistic heavy-ion collisions'' (THOR). R.P.~is supported in part by the Swedish 
Research Council grants, contract numbers 621-2013-4287 and 2016-05996, as well as by 
the European Research Council (ERC) under the European Union's Horizon 2020 research 
and innovation programme (grant agreement No 668679). T.L.~is supported by Knut and 
Alice Wallenberg foundation, contract no 2017.0036.

\appendix

%%%%%%%%%%%%%%%%%%%%%%%%%%%%%%%%%%%%%%%%%%%%%%%%%%%%%%%%%%%%%%%%
\section{An example of bubble cancellations in the imaginary-time formalism}
\label{sec:bubble-example}
\setcounter{equation}{0}
%%%%%%%%%%%%%%%%%%%%%%%%%%%%%%%%%%%%%%%%%%%%%%%%%%%%%%%%%%%%%%%%

  In this section, the argument for why $Z[0]$ contains the sum of all closed diagrams, i.e. diagrams with no external legs is presented. Further, it is shown pictorially why the cumulant expansion of $W[0]$ only contains fully connected bubbles.

%%%%%%%%%%%%%%%%%%%%%%%%%%%%%%%%%%%%%%%%%%%%%%%%%%%%%%%%%%%%%%%%
\subsection{Bubbles}
\label{sec:bubbles}
%%%%%%%%%%%%%%%%%%%%%%%%%%%%%%%%%%%%%%%%%%%%%%%%%%%%%%%%%%%%%%%%

  With no source, no external fields can be created. The sum of all closed diagrams (bubbles) are therefore given by the generating functional with the source $j$ set to zero.
  \beql{bubblesExpansionSeries}
    Z[0]
    =
    Z_0[0] \big \{
	  1 + \langle -S_I + j\phi \rangle_0 + \tfrac{1}{2} \langle (-S_I + j\phi)^2 \rangle_0 + \ldots
    \big \} \big\vert_{j=0}
    =
    Z_0[0] \sum_{n=0}^\infty
	  \frac{(-\kappa)^n}{n!} \langle (S_I)^n \rangle_0.
  \eeq
  Here, $\kappa$ has explicitly been extracted from the interaction-term $S_I$ of the action to visualise the order of expansion in the perturbation series. To each order in $\kappa$ it is then possible to calculate the contribution to $Z[0]$ from the vacuum bubbles by computing
  \beq
    \langle (S_I)^n \rangle_0
    =
    \frac{
	  \int \! \mathcal D\phi \,
	    e^{-S_0} (S_I)^n
      }{
	  \int \! \mathcal D\phi \,
	    e^{-S_0}
    }.
  \eeq

%%%%%%%%%%%%%%%%%%%%%%%%%%%%%%%%%%%%%%%%%%%%%%%%%%%%%%%%%%%%%%%%
\subsection{Cumulant expansion}
\label{sec:cumulantExpansion}
%%%%%%%%%%%%%%%%%%%%%%%%%%%%%%%%%%%%%%%%%%%%%%%%%%%%%%%%%%%%%%%%

  Recall the generating functional for the connected Green's functions, \mbox{$W[j] = \log Z[j]$} and the cumulant expansion from Sec.~\ref{sec:GeneratingFunctional}. Using these expressions, one arrives at an alternative expansion for the generating functional at $j=0$ as
  \beql{connectedExpansion}
    Z[0]
    \equiv
    Z_0[0] \exp{
	  \sum_{n=1}^\infty
		\frac{(-\kappa)^n}{n!} \langle (S_I)^n \rangle_\text{con}
    }.
  \eeq
  This is the defining relation for the \emph{connected averages} \mbox{$\langle (S_I)^n \rangle_\text{con}$}. By combination of \eq{bubblesExpansionSeries} and \eq{connectedExpansion}, the following identification can be made
  \beql{cumulantSeriesEquivalence}
    \log
	  \sum_{n=0}^\infty
	    \frac{(-\kappa)^n}{n!} \langle (S_I)^n \rangle_0
    \equiv
    \sum_{n=1}^\infty
	  \frac{(-\kappa)^n}{n!}
	  \langle (S_I)^n \rangle_\text{con}.
  \eeq
  This can be considered as a perturbative series expansion in $\kappa$ of the logarithm on the left hand side. The expression on the right-hand side contains, apart from powers of $\kappa$, the newly defined connected averages, the \emph{cumulants} \mbox{$\langle (S_I)^n \rangle_\text{con}$}. These are defined explicitly order by order in $\kappa$ through the full expansion of the sums in the above expression. In order to obtain the explicit expansion of \eq{cumulantSeriesEquivalence} below, one may exponentiate the left- and right-hand sides and expand the resulting exponential and the two sums to identify the cumulants in each order in $\kappa$:
  \begin{align}
    1
    -
    \kappa & \langle S_I \rangle_0
    +
    \frac{\kappa^2}{2!} \langle (S_I)^2 \rangle_0
    -
    \frac{\kappa^3}{3!} \langle (S_I)^3 \rangle_0
    +
    \ldots
    =
    \nonumber
    \\ & =
    1
    +
    \frac{1}{2!} \sum_{n}^\infty
      \frac{(-\kappa)^n}{n!}
      \langle (S_I)^n \rangle_\text{con}
    +
    \sum_{n=1,m=1}^\infty
      \frac{(-\kappa)^{n+m}}{n! m!}
      \langle (S_I)^n \rangle_\text{con}
      \langle (S_I)^m \rangle_\text{con}
    +
    \ldots
    \nonumber
    \\ & =
    1
    +
    \left (
      -\kappa \langle S_I \rangle_\text{con}
      +
      \frac{\kappa^2}{2!} \langle (S_I)^2 \rangle_\text{con}
      -
      \frac{\kappa^3}{3!} \langle (S_I)^3 \rangle_\text{con}
      +
      \ldots
    \right )
    \nonumber
    \\ & \qquad +
    \frac{1}{2!} \left (
      \kappa^2 \langle S_I \rangle_\text{con}^2
      +
      2 \frac{\kappa^3}{2!}
      \langle S_I \rangle_\text{con}
      \langle (S_I)^2 \rangle_\text{con}
      +
      \ldots
    \right )
    \nonumber
    \\ & \qquad\qquad +
    \frac{1}{3!} \left (
      \kappa^3 \langle S_I \rangle_\text{con}^3
      +
      \ldots
    \right )
    +
    \ldots
  \end{align}
  The cumulants can now be identified order by order in $\kappa$ as:
  
  \vspace{0.5cm}
  {\centering
  \begin{tabular}{@{}cc@{}} \toprule
      Order	& Cumulant definition \\ \colrule %\midrule
      $\kappa^0$	& $1 \equiv 1$ \\
      $\kappa^1$	& $\langle S_I \rangle_0 \equiv \langle S_I \rangle_\text{con}$ \\
      $\kappa^2$	& $\langle (S_I)^2 \rangle_0 \equiv \langle (S_I)^2 \rangle_\text{con} + \langle S_I \rangle_\text{con}^2$ \\
      $\kappa^3$	& $\langle (S_I)^3 \rangle_0 \equiv \langle (S_I)^3 \rangle_\text{con} + 3 \langle S_I \rangle_\text{con}  \langle (S_I)^2 \rangle_\text{con} + \langle S_I \rangle_\text{con}^3$ \\
      \vdots	& \vdots \\
      \botrule %\bottomrule
  \end{tabular}
  \par}% End centering
  \vspace{0.5cm}
  
  The zeroth-order relation provides nothing but a trivial identity. By making use of the relation at order $\kappa$, the second-order relation can be solved for the square-cumulant \mbox{$\langle (S_I)^2 \rangle_\text{con}$} in terms of ordinary averages:
  \beq
    \langle (S_I)^2 \rangle_\text{con}
    =
    \langle (S_I)^2 \rangle_0 - \langle S_I \rangle_0^2.
  \eeq
  This relation together with the definition at order $\kappa$ further gives the third order cubed connected term as
  \beq
    \langle (S_I)^3 \rangle_\text{con}
    =
    \langle (S_I)^3 \rangle_0
    -
    3 \langle S_I \rangle_0 \langle (S_I)^2 \rangle_0
    +
    2 \langle S_I \rangle_0^3.
  \eeq
  Collecting
  
  \vspace{0.5cm}
  {\centering
  \begin{tabular}{@{}cc@{}} \toprule
      Order	& Cumulant definition \\ \colrule %\midrule
      $\kappa^0$	& $1 \equiv 1$ \\
      $\kappa^1$	& $\langle S_I \rangle_\text{con}
      =
      \langle S_I \rangle_0$ \\
      $\kappa^2$	& $\langle (S_I)^2 \rangle_\text{con}
      =
      \langle (S_I)^2 \rangle_0 - \langle S_I \rangle_0^2$ \\
      $\kappa^3$	& $\langle (S_I)^3 \rangle_\text{con}
      =
      \langle (S_I)^3 \rangle_0
      -
      3 \langle S_I \rangle_0 \langle (S_I)^2 \rangle_0
      +
      2 \langle S_I \rangle_0^3$ \\
      \vdots	& \vdots \\ \botrule %\bottomrule
  \end{tabular}
  \par}% End centering
  \vspace{0.5cm}
  
  With this scheme, \mbox{$\langle (S_I)^n \rangle_\text{con}$} to any order $n$ may be defined iteratively. By explicitly calculation of the cumulant at the first four orders (up to $\kappa^3$), their diagrammatical interpretation is clear.

%%%%%%%%%%%%%%%%%%%%%%%%%%%%%%%%%%%%%%%%%%%%%%%%%%%%%%%%%%%%%%%%
\paragraph{Order $\kappa^0$}
%%%%%%%%%%%%%%%%%%%%%%%%%%%%%%%%%%%%%%%%%%%%%%%%%%%%%%%%%%%%%%%%

  Trivial case, $1 \equiv 1$.

%%%%%%%%%%%%%%%%%%%%%%%%%%%%%%%%%%%%%%%%%%%%%%%%%%%%%%%%%%%%%%%%
\paragraph{Order $\kappa^1$}
%%%%%%%%%%%%%%%%%%%%%%%%%%%%%%%%%%%%%%%%%%%%%%%%%%%%%%%%%%%%%%%%

  This calculation may straightforwardly be performed.
\begin{align*}
     \langle S_I \rangle_\text{con}
     & =
     \langle S_I \rangle_0
     =
     \frac{\kappa^2}{(4!)^2}
     \int_0^\beta \! \dd \tau \,
         \int \! \dd^3 x \,
             \langle
                 \phi(x)\phi(x)\phi(x)\phi(x)
             \rangle_0
     =
     \frac{\kappa^2}{(4!)^2}
     \int_0^\beta \! \dd \tau \,
         \int \! \dd^3 x \,
             3\Delta^2(x-x)
     \\ & =
     3 \cdot
     \raisebox{-0.42\height}{%
         \includegraphics[scale=0.15]{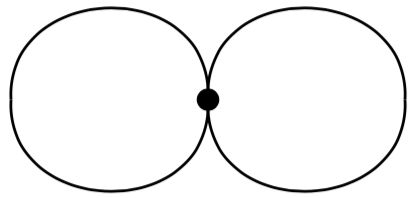}
     }
     =
     \sum \text{(all connected diagrams)}.
    \end{align*}

%%%%%%%%%%%%%%%%%%%%%%%%%%%%%%%%%%%%%%%%%%%%%%%%%%%%%%%%%%%%%%%%
\paragraph{Order $\kappa^2$}
%%%%%%%%%%%%%%%%%%%%%%%%%%%%%%%%%%%%%%%%%%%%%%%%%%%%%%%%%%%%%%%%

  At second order, calculations are here shown explicitly.
\begin{align*}
     \langle (S_I)^2 \rangle_\text{con}
     & =
     \langle (S_I)^2 \rangle_0 - \langle S_I \rangle_0^2
     \\ & =
     \frac{\kappa^2}{(4!)^2}
     \int_0^\beta \! \dd \tau \dd \tau' \,
         \int \! \dd^3 x \dd^3 y \,
             \langle
                 \phi(x)\phi(x)\phi(x)\phi(x)
                 \phi(y)\phi(y)\phi(y)\phi(y)
            \rangle_0
     \\ & \qquad -
     \frac{\kappa^2}{(4!)^2} \Bigg(
         \int_0^\beta \! \dd \tau \,
             \int \! \dd^3 x \,
                 \langle \phi^4(x) \rangle_0
     \Bigg )
     \Bigg (
         \int_0^\beta \! \dd \tau' \,
             \int \! \dd^3 y \,
                 \langle \phi^4(y) \rangle_0
     \Bigg )
     \\ & =
     \frac{\kappa^2}{(4!)^2}
     \int_0^\beta \! \dd \tau \dd \tau' \,
         \int \! \dd^3 x \dd^3 y \,
             \Big \{
                 3 \Delta(x-x) \cdot 3 \Delta(y-y) +
                 \\ & \hspace{3cm} +
                 4^2 \Delta(x-y) \cdot 3^2 \Delta(x-y)
                 \cdot \Delta(x-x) \cdot \Delta(y-y) +
                 \\ & \hspace{3cm} +
                 4^2 \Delta(x-y) \cdot 3^2 \Delta(x-y)
                 \cdot 2^2 \Delta(x-y) \cdot \Delta(x-y)
             \Big \}
     \\ & \qquad -
     \frac{\kappa^2}{(4!)^2} \Bigg(
         \int_0^\beta \! \dd \tau \,
             \int \! \dd^3 x \,
                 3 \Delta^2(x-x)
     \Bigg )
     \Bigg (
         \int_0^\beta \! \dd \tau' \,
             \int \! \dd^3 y \,
                 3 \Delta^2(y-y)
     \Bigg )
     \\ & =
     3^2 \cdot
     \raisebox{-0.48\height}{%
         \includegraphics[scale=0.15]{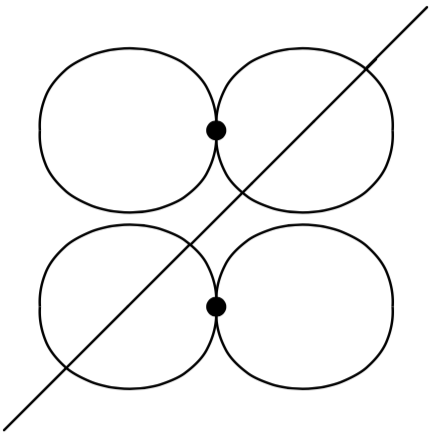}
     }
     +
     4^2 \cdot 3^2 \cdot
     \raisebox{-0.48\height}{%
         \includegraphics[scale=0.14]{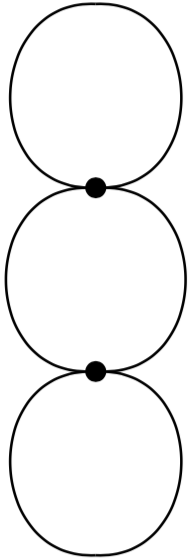}
     }
     +
     4^2 \cdot 3^2 \cdot 2^2 \cdot
     \raisebox{-0.48\height}{%
         \includegraphics[scale=0.11]{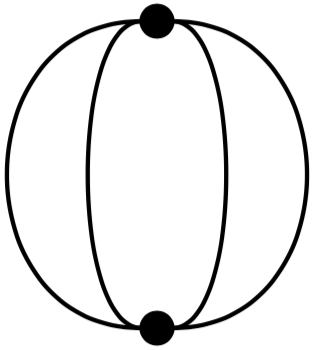}
     }
     -%\\ & \qquad -
     3^2 \cdot
     \raisebox{-0.48\height}{%
         \includegraphics[scale=0.15]{figures/Diagrams/order2-1.png}
     }
     \\ & =
     \sum \text{(all connected diagrams)}.
  \end{align*}
\vfill

%%%%%%%%%%%%%%%%%%%%%%%%%%%%%%%%%%%%%%%%%%%%%%%%%%%%%%%%%%%%%%%%
\paragraph{Order $\kappa^3$}
%%%%%%%%%%%%%%%%%%%%%%%%%%%%%%%%%%%%%%%%%%%%%%%%%%%%%%%%%%%%%%%%

  At third order in $\kappa$, the calculation is more cumbersome but not any less straight forward.
  \begin{align*}
     \langle (S_I)^3 \rangle_\text{con}
     & =
     \langle (S_I)^3 \rangle_0
     -
     3\langle S_I \rangle_0 \langle (S_I)^2 \rangle_0
     +
     2 \langle S_I \rangle_0^3
     \\ & =
     \frac{\kappa^3}{(4!)^3}
     \int_0^\beta \! \dd \tau \, \dd \tau' \, \dd \tau'' \,
         \int \! \dd^3 x \, \dd^3 y \, \dd^3 z \,
             \\ & \qquad\qquad \cross
             \langle
                 \phi(x)\phi(x)\phi(x)\phi(x)
                 \phi(y)\phi(y)\phi(y)\phi(y)
                 \phi(z)\phi(z)\phi(z)\phi(z)
             \rangle_0
     \\ & \qquad -
     3 \frac{\kappa^3}{(4!)^3} \Bigg (
         \int_0^\beta \! \dd \tau \,
             \int \! \dd^3 x \,
                 \langle
                     \phi(x)\phi(x)\phi(x)\phi(x)
                 \rangle_0
     \Bigg ) \cross
     \\ & \qquad\qquad \cross
     \Bigg (
         \int_0^\beta \! \dd \tau \, \dd \tau' \,
             \int \! \dd^3 x \, \dd^3 y \,
                 \langle
                     \phi(x)\phi(x)\phi(x)\phi(x)
                     \phi(y)\phi(y)\phi(y)\phi(y)
                 \rangle_0
     \Bigg )
     \\ & \qquad +
     2 \frac{\kappa^3}{(4!)^3} \Bigg (
         \int_0^\beta \! \dd \tau \,
             \int \! \dd^3 x \,
                 \langle
                     \phi(x)\phi(x)\phi(x)\phi(x)
                 \rangle_0
     \Bigg ) \cross
     \\ & \qquad\qquad\cross
     \Bigg (
         \int_0^\beta \! \dd \tau \,
             \int \! \dd^3 x \,
                 \langle
                     \phi(x)\phi(x)\phi(x)\phi(x)
                 \rangle_0
     \Bigg )
     \\ & \qquad\qquad\qquad \cross
     \Bigg (
         \int_0^\beta \! \dd \tau \,
             \int \! \dd^3 x \,
                 \langle
                     \phi(x)\phi(x)\phi(x)\phi(x)
                 \rangle_0
     \Bigg )
     \\ & =
     \frac{\kappa^3}{(4!)^3}
     \int_0^\beta \! \dd \tau \dd \tau' \dd \tau'' \,
         \int \! \dd^3 x \, \dd^3 y \, \dd^3 z \,
             \Big \{
                 3 \Delta^2(x-x) \cdot 3 \Delta^2(y-y)
                 \cdot 3 \Delta^2(z-z)
                 \\ & \qquad\qquad +
                 3 \cdot 4^2 \Delta(x-y) \cdot 3^2 \Delta(x-y)
                 \cdot \Delta(x-x) \cdot \Delta(y-y)
                 \cdot 3 \Delta^2(z-z)
                 \\ & \qquad\qquad +
                 3 \cdot 4^2 \Delta(x-y) \cdot 3^2 \Delta(x-y)
                 \cdot 2^2 \Delta(x-y) \cdot \Delta(x-y)
                 \cdot 3 \Delta^2(z-z)
                 \\ & \qquad\qquad +
                 2 \cdot 4^2 \Delta(x-y) \cdot 3^2 \Delta(x-y)
                 \cdot \Delta(x-x) \cdot 2 \cdot 4 \Delta(y-z)
                 \cdot 3 \Delta(y-z) \cdot \Delta(z-z)
                 \\ & \qquad\qquad +
                 4^2 \Delta(x-y) \cdot 4 \cdot 3 \Delta(x-z)
                 \cdot \Delta(x-x) \cdot 3 \Delta(y-y)
                 \cdot 3 \Delta(y-z) \cdot \Delta(z-z)
                 \\ & \qquad\qquad +
                 4^2 \Delta(x-y) \cdot 4 \cdot 3 \Delta(x-z)
                 \cdot \Delta(x-x) \cdot 3^2 \Delta(y-z)
                 \cdot 2^2 \Delta(y-z) \Delta(y-z)
                 \\ & \qquad\qquad +
                 4^2 \Delta(x-y) \cdot 3^2 \Delta(x-y)
                 \cdot 3 \Delta(y-y) \cdot 4 \cdot 2 \Delta(x-z)
                 \cdot 3 \Delta(x-z) \Delta(z-z)
                 \\ & \qquad\qquad +
                 4^2 \Delta(x-y) \cdot 3^2 \Delta(x-y) \cdot 4
                 \cdot 2 \Delta(x-z) \cdot 3 \Delta(x-z)
                 \cdot 2^2 \Delta(y-z) \cdot \Delta(y-z)
             \Big \}
     \\ & \qquad -
     3 \frac{\kappa^3}{(4!)^3} \Bigg (
         \int_0^\beta \! \dd \tau \,
             \int \! \dd^3 x \,
                 3 \Delta^2(x-x)
     \Bigg )
     \cross
     \\ & \hspace{1.6cm} \cross
     \Bigg (
         \int_0^\beta \! \dd \tau \, \dd \tau' \,
         \int \! \dd^3 x \, \dd^3 y \,
             \Big \{
                 3 \Delta^2(x-x) \cdot \Delta^2(y-y)
                 +
                 \\ & \hspace{4cm} +
                 4^2 \Delta(x-y) \cdot 3^2 \Delta(x-y)
                 \cdot \Delta(x-x) \cdot \Delta(y-y)
                 +
                 \\ & \hspace{4cm} +
                 4^2 \Delta(x-y) \cdot 3^2 \Delta(x-y)
                 \cdot 2^2 \Delta(x-y) \cdot \Delta(x-y)
             \Big \}
     \Bigg )
     \\ & \qquad +
     2 \frac{\kappa^3}{(4!)^3} \Bigg (
         \int_0^\beta \! \dd \tau \,
             \int \! \dd^3 x \,
                 3 \Delta^2(x-x)
     \Bigg )
     \cross
     \\ & \hspace{2cm} \cross
     \Bigg (
         \int_0^\beta \! \dd \tau \,
             \int \! \dd^3 x \,
                 3 \Delta^2(x-x)
     \Bigg )
     \cross
     \\ & \hspace{3cm} \cross
     \Bigg (
         \int_0^\beta \! \dd \tau \,
             \int \! \dd^3 x \,
                 3 \Delta^2(x-x)
     \Bigg ).
  \end{align*}
  
  Diagrammatically this expression is
\begin{align*}
     \langle (&S_I)^3 \rangle_\text{con}
     = 
     %\\ & =
     3^3 \cdot
     \raisebox{-0.48\height}{%
         \includegraphics[scale=0.15]{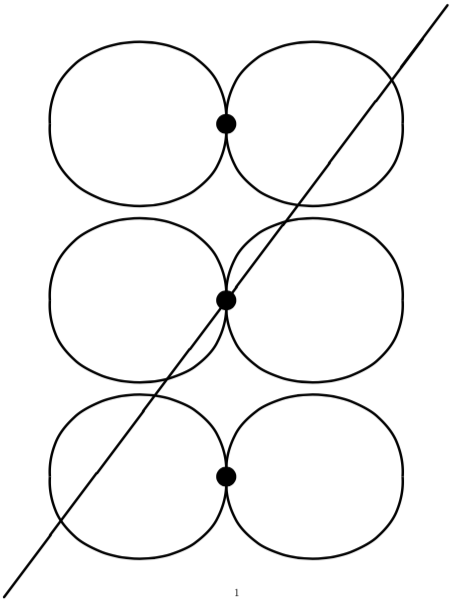}
     }
     +
     4^2 \cdot 3^4 \cdot
     \raisebox{-0.48\height}{%
         \includegraphics[scale=0.2]{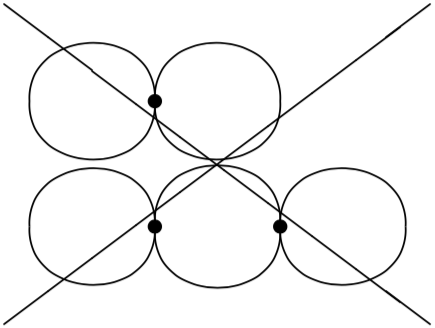}
     }
     +
     4^2 \cdot 3^4 \cdot 2^2 \cdot
     \raisebox{-0.48\height}{%
         \includegraphics[scale=0.2]{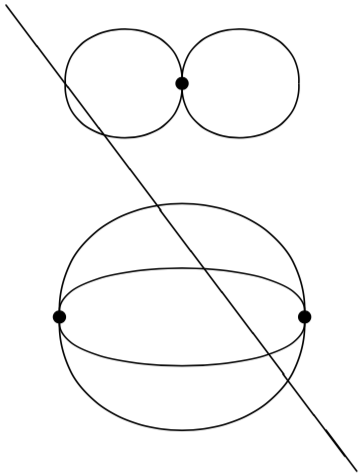}
     }
     \\ & +
     4^3 \cdot 3^3 \cdot 2^2 \cdot
     \raisebox{-0.48\height}{%
         \includegraphics[scale=0.18]{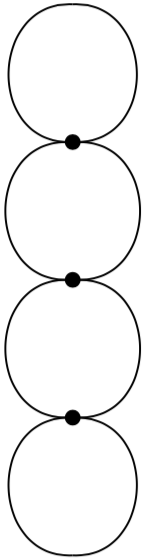}
     }
     +
     4^3 \cdot 3^3 \cdot
     \raisebox{-0.48\height}{%
         \includegraphics[scale=0.18]{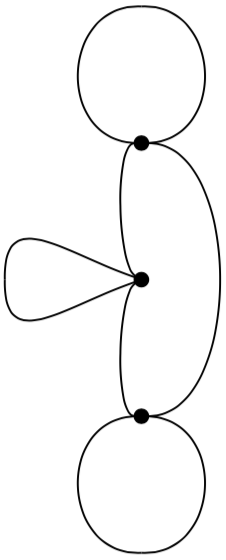}
     }
     +
     4^3 \cdot 3^3 \cdot 2^2 \cdot
     \raisebox{-0.48\height}{%
         \includegraphics[scale=0.14]{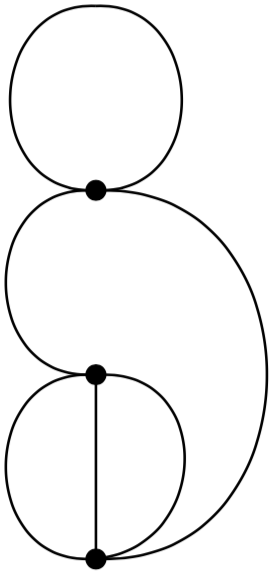}
     }
     \\ & +
     4^3 \cdot 3^4 \cdot 2 \cdot
     \raisebox{-0.48\height}{%
         \includegraphics[scale=0.18]{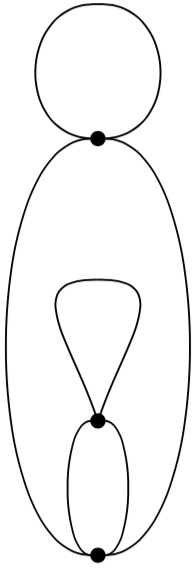}
     }
     +
     4^3 \cdot 3^3 \cdot 2^3 \cdot
     \raisebox{-0.48\height}{%
         \includegraphics[scale=0.11]{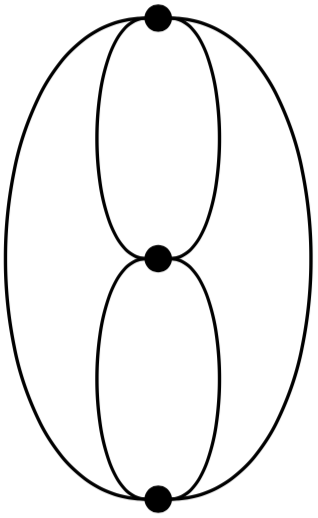}
     }
     -
     3^4 \cdot
     \raisebox{-0.48\height}{%
         \includegraphics[scale=0.15]{figures/Diagrams/order3-1.png}
     }
     \\ & -
     4^2 \cdot 3^4 \cdot
     \raisebox{-0.48\height}{%
         \includegraphics[scale=0.2]{figures/Diagrams/order3-2.png}
     }
     -
     4^2 \cdot 3^4 \cdot 2^2 \cdot
     \raisebox{-0.48\height}{%
         \includegraphics[scale=0.2]{figures/Diagrams/order3-3.png}
     }
     +
     3^3 \cdot 2 \cdot
     \raisebox{-0.48\height}{%
         \includegraphics[scale=0.15]{figures/Diagrams/order3-1.png}
     }
     \\ & =
     \sum \text{(all connected diagrams)}.
  \end{align*}

It is clear from the diagrammatic calculations above that the cumulants precisely correspond to the fully connected diagrams at each order in $\kappa$. This can be shown in general to any order, see for example Refs.~\cite{PeskinSchroeder1995, LandsmanvanWeert1987}.

%%%%%%%%%%%%%%%%%%%%%%%%%%%%%%%%%%%%%%%%%%%%%%%%%%%%%%%%%%%%%%%%
\section{An example of a thermal loop calculation in the real-time formalism}
\label{sec:loopCalculation}
\setcounter{equation}{0}
%%%%%%%%%%%%%%%%%%%%%%%%%%%%%%%%%%%%%%%%%%%%%%%%%%%%%%%%%%%%%%%%

  In this section, the evaluation of a thermal loop integral is provided as a reference problem. The scenario considered is that of \sec{pStpSpS} and the diagram is generated by the interaction term
\beq
     \mathcal L_\text{int} = \kappa\Phi\phi^1\phi^2.
\eeq
  The explicit integrals of interest are
\begin{align}
     \label{eq:nonThermalIntegral}
     &iI_\text{SS}^{(2)}(p^2; m_1, m_2)
     \nonumber
     \\ & =
     i(1 + \delta_{12})(-i\kappa)^2 \!
     \int \! \frac{\dd^4 k}{(2\pi)^4} \,
         \frac{i}{(p-k)^2 - m_1^2 + i\epsilon}
         \,
         \frac{i}{k^2 - m_2^2 + i\epsilon},
     \\ \nonumber \\
     \label{eq:mixedThermalIntegral}
     & iF_\text{SS}^{(2)}(p; m_1, m_2)
     \nonumber
     \\ & =
     i(1 + \delta_{12})(-i\kappa)^2 \!
     \int \! \frac{\dd^4 k}{(2\pi)^4} \,
         \frac{i}{(p-k)^2 - m_1^2 + i\epsilon}
         \,
         2\pi \, n_\text{B}\big( \abs{k_0} \big)
         \delta\big( k^2 - m_2^2 \big),
     \\ \nonumber \\
     \label{eq:purelThermalIntegral}
     & iF_\text{SS}^{(3)}(p; m_1, m_2)
     \nonumber
     \\ & =
     i(1 + \delta_{12})(-i\kappa)^2 \!
     \int \! \frac{\dd^4 k}{(2\pi)^4} \,
         2\pi \, n_\text{B}\big( \abs{p_0 \!-\! k_0} \big)
         \delta\big( (p\!-\!k)^2 \!- m_1^2 \big) \,
         2\pi \, n_\text{B}\big( \abs{k_0} \big)
         \delta\big( k^2 \!- m_2^2 \big).
\end{align}

%%%%%%%%%%%%%%%%%%%%%%%%%%%%%%%%%%%%%%%%%%%%%%%%%%%%%%%%%%%%%%%%
\subsection{\texorpdfstring{$iI_\text{SS}^{(2)}$}{TEXT}-integral}
\label{sec:nonThermalExlicitIntegration}
%%%%%%%%%%%%%%%%%%%%%%%%%%%%%%%%%%%%%%%%%%%%%%%%%%%%%%%%%%%%%%%%

  The first term in \eq{nonThermalIntegral} is the cross-term of the non-thermal part of the two propagators and for the loop of interest it is
\beq
     iI_\text{SS}^{(2)}(p^2; m_1, m_2)
     =
     i(1 \!\!+\! \delta_{12}) (-i\kappa^2) \!\!
     \int \! \frac{\dd^4 k}{(2 \pi)^4} \,
         \frac{i}{(p-k)^2 - m_1^2 + i\epsilon}
         \frac{i}{k^2 - m_2^2 + i\epsilon}.
\eeq
  This integral is recognised from the zero-temperature theory and includes no reference to the thermal parameter $\beta$. One may in this instance appreciate the separation of the thermal part from the zero-temperature contribution since it is immediately realised that the calculation reduces to the zero-temperature result if \mbox{$T \to 0$}. In this limit, the above integral is the only remaining term since the others contain the thermal distribution function which vanishes when \mbox{$T \to 0$}.

  The evaluation of this integral is a standard procedure from zero-temperature QFT and the result may be found in any textbook on loop calculations (e.g. \cite{PeskinSchroeder1995}). Therefore, the calculation of this integral will not be repeated in any detail here. It is enough to say that the procedure involves the Feynman parametrisation of the two fractions followed by a shift of the integration momentum $k$. By a subsequent Wick rotation of the energy component according to
\beq
     k^0 \to ik_\text{E}^0,
     \qquad  \Rightarrow \qquad
     k^2 \to -k_\text{E}^2
     =
     -\big( k_\text{E}^0 \big)^2 - \abs{\vb k}^2,
\eeq
  the integral may be recast into a Euclidean integral on polar form, generally in $D$ dimensions as
\begin{align}
     iI_\text{SS}^{(2)}(p^2; m_1, m_2)
     & =
     -(1 + \delta_{12}) \kappa^2
     \int_0^1 \! \dd x \,
         \int \! \frac{\dd^D k_\text{E}}{(2\pi)^D} \,
             \frac{1}{%
                 \big[
                     k_\text{E}^2 + \Delta(x) - i\epsilon
                 \big]^n
             }.
\end{align}
  Generally, integrals on this form may be UV divergent at \mbox{$D=4$} and a scheme for regulating such divergences must be deployed. A common approach is \emph{dimensional regularisation techniques} in which the dimensionality of the above momentum integral is taken to be \mbox{$D = 4-2\delta$} with \mbox{$\delta \in (0,1)$}. The innermost integral may be solved in the general case by shifting to polar co\"ordinates and solving the angular and radial contributions separately. The well known result is
\beq
     \int \! \frac{\dd^D k_\text{E}}{(2\pi)^D} \,
         \frac{\big(k_\text{E}^2\big)^r}{%
             \big[
                 k_\text{E}^2 + \Delta(x) - i\epsilon
             \big]^n
         }
     =
     \frac{(-1)^{N-r}}{(4\pi)^{D/2}}
     \frac{\Gamma(r+D/2)\Gamma(n-r-D/2)}{\Gamma(D/2)\Gamma(n)}
     \big[ \Delta(x) - i\epsilon \big]^{D/2 - n},
\eeq
  where the $\Gamma$-function parameterises the divergence. Poles arise for \mbox{$D \to 2m, m \in \mathbbm{N}$} such that arguments of the $\Gamma$-functions becomes negative whole numbers. The divergence may be explicitly extracted as
\beq
     \Gamma(\delta)
     =
     \frac{1}{\delta} - \gamma_E + \mathcal O(\delta).
\eeq
  $\gamma_E$ is the Euler-Mascheroni constant.

  The above momentum integral is an important relation since many loop integrals may be expressed on this form. Hence, there is seldom any need for explicit momentum integration\footnote{
    An additional complication is of course potential tensor structures of the propagators but also in this case, solution at zero temperature exists in the literature (see e.g.~Refs.~\cite{BrownFeynman1952, PassarinoVeltman1979, Davydychev1991, Melrose1965}). The case is further complicated in thermal theories by the presence of the medium in the sense that the space of basis tensors is larger than in zero-temperature theory and for this case no well established procedure is in place.
  }.
  
  In the case of interest to the evaluation of the integral \mbox{$iI_\text{SS}^{(2)}$} \mbox{$n = 2$} s.t.
\beq
     \int \! \frac{\dd^D k_\text{E}}{(2 \pi)^D} \,
         \frac{1}{
             \big[
                 k_\text{E}^2 + \Delta(x) - i\epsilon
             \big]^2
         }
     \underset{D \to 4}{=}
     \frac{1}{(4 \pi)^2}
     \left(
         \frac{1}{\delta} - \ln\big[\Delta(x) - i\epsilon\big]
         -
         \gamma_E + \ln{4 \pi} + \mathcal O(\delta)
     \right).
\eeq

  It is important to note a consequence of modifying the dimension \mbox{$D = 4 \to 4 - 2\delta$} when regularising the loop integral. The action have dimension $0$ when the \L\ density is integrated over space-time. When modifying the dimensionality s.t. the space-time integration over the \L\ \mbox{$\int \! \dd^4 x \to \int \! \dd^{D} x$}, this requires the missing dimensionality to be accounted for elsewhere. In the interaction term, it is common to extend the dimension of the coupling constant in order to absorb extra dimensions. This is usually done s.t. \mbox{$\kappa \to \kappa\bar\mu^{p}$} where $\bar\mu$ has dimension 1 and is defined as the \emph{renormalisation point} to some power $p$. The exponent $x$ is determined through bookkeeping of dimensionality. 
  
  Introducing the renormalised coupling constant \mbox{$\kappa \bar\mu^p$} squared as a pre-factor to the above integral, one may move the resulting ${\big(\bar\mu^2\big)}^\epsilon$ into the logarithm. The final step is to evaluate the resulting integral in $x$ over the logarithm
\beq
     (1 \!\!+\! \delta_{12}) \frac{\kappa^2}{16 \pi^2}
     \int_0^1 \! \dd x \,
         \ln{\frac{\Delta(x) - i\epsilon}{\bar\mu^2}}
\eeq
  which is proportional to the following form
\beq
     \int_0^1 \! \dd x \,
         \ln{\big[ax^2 + bx + c\big]}.
\eeq
  In the case of interest, the constants in $x$ are
\beql{coeffDefs}
     a = \frac{p^2}{\bar\mu^2},
     \qquad
     b = \frac{-p^2 + m_1^2 - m_2^2}{\bar\mu^2},
     \qquad
     c = \frac{m_2^2}{\bar\mu^2} - i\epsilon.
\eeq
  The integral may be evaluated through factorisation of the polynomial as
\beq
     ax^2 + bx + c
     =
     a \left[
         x - \left( \frac{\sqrt{b^2 - 4ac} - b}{2a} \right)
     \right]
     \left[
         x + \left( \frac{\sqrt{b^2 - 4ac} + b}{2a} \right)
     \right].
\eeq
  All but the last four terms of the evaluation simplify trivially s.t.
\begin{align}
     \int_0^1 & \! \dd x \,
         \ln{\big[ax^2 + bx + c\big]}
     =
     \\ & =
     -2
     +
     \ln{ \left[a + b + c \right] }
     +
     \frac{b}{2a}
     \ln{ \left[ \frac{a + b + c}{c} \right] }
     \nonumber
     \\ & \qquad +
     \frac{1}{2a}
     \sqrt{b^2 - 4ac}
     \Bigg\{
         \ln{\left[-\left(
             \frac{\sqrt{b^2 - 4ac} - b}{\bcancel{2a}}
         \right)\right]}
         -%	\\ & \qquad \qquad \qquad \qquad \,\,\,\, -
         \ln{\left[\left(
             \frac{\sqrt{b^2 - 4ac} + b}{\bcancel{2a}}
         \right)\right]}
         \nonumber
         \\ & \qquad \qquad \qquad \qquad \,\,\,\, +
         \ln{\left[
             \cancel{\frac{1}{2a}}
             \left( \sqrt{b^2 - 4ac} + (2a + b) \right)
         \right]}
         -
         \ln{\left[
             -\cancel{\frac{1}{2a}}
             \left( \sqrt{b^2 - 4ac} - (2a + b) \right) \right]
         }
     \Bigg\}.
     \nonumber
\end{align}
  The final functional form of this expression depends on the sign of \mbox{$b^2 - 4ac$} which is determined by the sign and magnitude of $p^2$. Disregarding corrections to order $\epsilon$, this expression is \emph{positive} when:
\beq
     p^2 < (m_1 - m_2)^2 \,\, \vee \,\, p^2 > (m_1 + m_2)^2
\eeq
  and \emph{negative} when
\beq
     (m_1 - m_2)^2 < p^2 < (m_1 + m_2)^2.
\eeq
  In the negative case when $p^2$ lies between the difference squared and the sum squared of the two masses, a further simplification can be made.
\begin{align}
     \ln{\left[
         \frac{
             \sqrt{b^2 - 4ac} + b
         }{\sqrt{b^2 - 4ac}}
     \right]}
     -
     \ln{\left[
         \frac{
             \sqrt{b^2 - 4ac} - b
         }{\sqrt{b^2 - 4ac}}
     \right]}
     & =
     -2i
     \arctan{\left( \frac{x_-}{\sqrt{-C}} \right)}.
     \\
     \ln{\left[
         \frac{
             \sqrt{b^2 - 4ac} - (2a + b)
         }{\sqrt{b^2 - 4ac}}
     \right]}
     -
     \ln{\left[
         \frac{
             \sqrt{b^2 - 4ac} + (2a + b)
         }{\sqrt{b^2 - 4ac}}
     \right]}
     & =
     2i \arctan{\left( \frac{x_+}{\sqrt{-C}} \right)}.
\end{align}
  where the three quantities $x_\pm$ and $C$ was defined as
\beq
     x_\pm
     \equiv
     \pm 1 + \frac{m_1^2 - m_2^2}{p^2},
     \qquad
     C
     \equiv
     \left( 1 - \frac{(m_1 + m_2)^2}{p^2} \right)
     \left( 1 - \frac{(m_1 - m_2)^2}{p^2} \right).
\eeq

  In the positive case, $p^2$ is either above or below a certain threshold value: \\
  \mbox{$p^2 < (m_1 - m_2)^2 \,\, \vee \,\, p^2 > (m_1 + m_2)^2$}.
  Each of the four logarithmic terms may acquirer a negative argument thus making it undefined. A convenient approach is the identification, and further expansion, of, the following square roots:
\begin{align}
     \ln{\left[
         \frac{
             \sqrt{b^2 - 4ac} \pm b
         }{\sqrt{b^2 - 4ac}}
     \right]}
     & =
     \ln{\left[
         \sqrt{C + \mathcal O(i\epsilon)} \pm x_-
     \right]},
     \\
     \ln{\left[
         \frac{
             \sqrt{b^2 - 4ac} \mp (2a + b)
         }{\sqrt{b^2 - 4ac}}
     \right]}
     & =
     \ln{\left[
         \sqrt{C + \mathcal O(i\epsilon)} \mp x_+
     \right]}.
\end{align}
  An identity for the principal value of complex logarithms \mbox{$\ln{[x \pm i\epsilon]} = \ln{\abs{x}} \pm i\pi \Theta(-x) + \mathcal{O}(\epsilon)$} may be used in order to expand the logarithms as:
\begin{align}
\label{eq:LogarithmSplitting}
     \ln{\left[
         \sqrt{C + \mathcal O(i\epsilon)} - x_-
     \right]}
     & \approx
     \ln{\abs{ \sqrt{C} - x_- }}
     +
     i\pi \Theta \left( (m_1 - m_2)^2  - p^2 \right),
     \\
     \ln{\left[
         \sqrt{C + \mathcal O(i\epsilon)} + x_-
     \right]}
     & \approx
     \ln{\abs{ \sqrt{C} + x_- }}
     +
     i\pi \Big[
         \Theta \left(
             (m_1 \!-\! m_2)^2  - p^2
         \right)
         +
         \Theta \left(
             p^2 - (m_1 \!+\! m_2)^2 
         \right)
     \Big],
     \\
     \ln{\left[
         \sqrt{C + \mathcal O(i\epsilon)} - x_+
     \right]}
     & \approx
     \ln{\abs{ \sqrt{C} - x_+ }}
     +
     i\pi \Big[
         \Theta \left(
             (m_1 \!-\! m_2)^2  - p^2
         \right)
         +
         \Theta \left(
             p^2 - (m_1 \!+\! m_2)^2 
         \right)
     \Big],
     \\
     \ln{\left[
         \sqrt{C + \mathcal O(i\epsilon)} + x_+
     \right]}
     & \approx
     \ln{\abs{ \sqrt{C} + x_+ }}
     +
     i\pi \Theta \left( (m_1 - m_2)^2  - p^2 \right).
\end{align}

  Collecting the above result, the full non-thermal part of the loop integral of interest to this subsection may be written down:
\beq
     iI_\text{SS}^{(2)}(p^2; m_1, m_2)
     =
     (1 \!\!+\! \delta_{12}) \frac{\kappa^2}{16 \pi^2}
     \left[
         -\frac{1}{\tilde \epsilon}
         -
         2
         +
         \frac{x_+}{2} \ln{\bigg[\frac{m_1^2}{\bar\mu^2}\bigg]}
         -
         \frac{x_-}{2} \ln{\bigg[\frac{m_2^2}{\bar\mu^2}\bigg]}
         -
         I_\text{loop}.
     \right]
\eeq
  The last integral is
\beq
     \! I_\text{loop}
     \! = \!
     \begin{cases}
     \! \frac{\sqrt{C}}{2} \left[
         \ln{\abs{\frac{
             \big( \sqrt{C} + x_- \big)
             \big( \sqrt{C} - x_+ \big)
         }{
             \big( \sqrt{C} - x_- \big)
             \big( \sqrt{C} + x_+ \big)
         }}
         +
         2\pi i}
     \right],
     \hspace{2.35cm} \,\,\, \text{for }
     p^2 \geq (m_1 + m_2)^2,
     \\ \! \\
     \!\! -\sqrt{-C} \left[
         \arctan{\frac{x_+}{\sqrt{-C}}}
         -
         \arctan{\frac{x_-}{\sqrt{-C}}}
     \right]\!,
     \hspace{0.5cm} \text{for }
     (m_1 - m_2)^2 \! \leq \! p^2 \! \leq \! (m_1 + m_2)^2,
     \\ \! \\
     \! \frac{\sqrt{C}}{2} 
         \ln{\abs{\frac{
             \big( \sqrt{C} + x_- \big)
             \big( \sqrt{C} - x_+ \big)
         }{
             \big( \sqrt{C} - x_- \big)
             \big( \sqrt{C} + x_+ \big)
         }}},
         \hspace{3.7cm} \,\, \text{for }
         p^2 \leq (m_1 - m_2)^2.
     \end{cases}
\eeq

%%%%%%%%%%%%%%%%%%%%%%%%%%%%%%%%%%%%%%%%%%%%%%%%%%%%%%%%%%%%%%%%
\subsection{Isolating the imaginary part}
\label{sec:FasterImaginaryEvaluation}
%%%%%%%%%%%%%%%%%%%%%%%%%%%%%%%%%%%%%%%%%%%%%%%%%%%%%%%%%%%%%%%%

  In the case of decay rates, the main point of interest is the imaginary part of the loop diagram. In the above case of the non-thermal part, it proved to be a decent amount of work for essentially a contribution of $2\pi i$ and it should be noted that a quicker calculation is viable in order to extract the imaginary part of the loop.
  
  After dimensional regularisation in the previous subsection, the logarithm to be integrated over the Feynman parametrisation may be written on explicit complex form using the identity preceding \eq{LogarithmSplitting}. Thus, the imaginary part of the loop may be extracted at an early stage:
\beq
     \imaginary\ \big[ iI_\text{SS}^{(2)}(p^2; m_1, m_2) \big]
     =
     -(1 \!\!+\! \delta_{12}) \frac{\kappa^2}{16 \pi^2}
     \int_0^1 \! \dd x \,
         \Theta\big(-(ax^2 + bx + c')\big).
\eeq
  Recall the definition of \mbox{$a, b$} (\eq{coeffDefs}) and note that after expansion, the $i\epsilon$ contribution has been taken into account. Therefore, $c$ now reduces to $c'$ with
\beq
     c' = \frac{m_2^2}{\bar\mu^2}.
\eeq
  The only remaining task is to identify for what regions of $p^2$ the roots of the polynomial in the \mbox{$\Theta$-function} fall inside the integration domain and for which regions the entire interval is to be considered. Factorise
\beq
     ax^2 + bx + c'
     =
     a \big( x - \Lambda_- \big)\big( x - \Lambda_+ \big),
     \qquad
     \Lambda_\pm
     =
     \mp\left( \frac{\sqrt{b^2 - 4ac'} \pm b}{2a} \right).
\eeq
  To exhaust the possibilities of sign combinations, the values of the roots has been tabulated, see \tab{Zeroes}.

\begin{table}[h!]
\centering
\caption{Root values in momentum regions. The sign of the roots of the polynomial is tabulated below. The region with root values denoted with circled numbers has a more intricate boundary structure explained in the main text. The sign and magnitude of $\Lambda_\pm$ determine the result of integration over the Feynman parameter. 
\label{tab:Zeroes}}
\begin{tabular}{@{} ccccc @{}}
     \toprule
     \multicolumn{3}{c}{Parameters}	& \multicolumn{2}{c}{Roots} \\
\cline{1-3}\cline{4-5}     
     %\cmidrule(r){1-3}\cmidrule(l){4-5}
     a	& \multicolumn{2}{c}{$b$}	& $\Lambda_-$	& $\Lambda_+$ \\ \colrule %\midrule
     $a \geq 0$	& $b \geq 0$	& $b^2 \geq 4ac'$	& $\leq 0$	& $\leq 0$ \\
        & 	& $b^2 < 4ac'$	& complex	&  complex \\
		& $-a - c' \leq b < 0$	& $b^2 < 4ac'$	& complex	& complex \\
		& 	& $b^2 \geq 4ac'$	& \circled{1}	& \circled{2} \\
		$a < 0$	& $\forall b \geq -a - c'$	& 	& $< 0$	& $\geq 1$ \\ \colrule % \midrule
	Comment	& \multicolumn{2}{c}{Note $b < -a - c'$}	& \multicolumn{2}{c}{Outside def. of b.} \\
		& \multicolumn{2}{c}{is not allowed.}	& \\
	\botrule %\bottomrule
\end{tabular}
\end{table}
In the case \circled{1} in the table, one may show that
\beq
     \begin{array}{c}
         \Lambda_- \in [0,1],
         \quad p^2 \geq (m_1 + m_2)^2,
         \\
         \Lambda_- \in [1, \infty],
         \quad 0 \leq p^2 \leq (m_1 - m_2)^2
         \,\, \wedge \,\,
         m_2 \geq m_1.
     \end{array}
\eeq

In the case \circled{2} in the table, one may show that \mbox{$\Lambda_+ \leq \Lambda_-$} and
\beq
\begin{array}{c}
     \Lambda_+ \in [0,1], \quad p^2 \geq (m_1 + m_2)^2,
     \\
     \Lambda_+ \in [1, \infty],
     \quad 0 \leq p^2 \leq (m_1 - m_2)^2
     \,\, \wedge \,\,
     m_2 \geq m_1.
\end{array}
\eeq

Hence, only for \mbox{$p^2 \geq (m_1 + m_2)^2$} is the polynomial negative inside the region of integration giving a finite imaginary component. Both roots \mbox{$\Lambda_-, \Lambda_+ \in [0,1]$}, ordered as \mbox{$\Lambda_+ \leq \Lambda_-$}. The result of the previous subsection for the imaginary part is the swiftly recovered:
\beq
     \imaginary\ \big[ iI_\text{SS}^{(2)} (p^2; m_1, m_2) \big]
     =
     \begin{dcases}
         -(1 \!\!+\! \delta_{12}) \frac{\kappa^2}{16 \pi^2}
         \cdot 2\pi\frac{\sqrt{C}}{2}
         \hspace{1cm} \text{if }
         p^2 \geq (m_1 + m_2)^2,
         \\
         0 \hspace{5.28cm} \text{otherwise}.
     \end{dcases}
\eeq

%%%%%%%%%%%%%%%%%%%%%%%%%%%%%%%%%%%%%%%%%%%%%%%%%%%%%%%%%%%%%%%%
\subsection{\texorpdfstring{$iF_\text{SS}$}{TEXT}-integral(s)}
\label{sec:mixedThermalExlicitIntegration}
%%%%%%%%%%%%%%%%%%%%%%%%%%%%%%%%%%%%%%%%%%%%%%%%%%%%%%%%%%%%%%%%

  This section provides an example calculation of the so-called \emph{mixed} first-order contribution to the thermal self-energy. The mixed integrals are due to the cross-terms of the two propagators that mix the thermal and non-thermal propagator terms.
\begin{align}
     iF_\text{SS}^{(2)}(p; m_1, m_2)
     & =
     i (1 + \delta_{12}) (-i\kappa)^2
     \int \! \frac{\dd^4 k}{(2 \pi)^4} \,
         \frac{i}{(p-k)^2 - m_1^2 + i\epsilon} \,
         2 \pi \, n(\lvert k_0 \rvert) \,
         \delta\big(k^2 - m_2^2\big)
     \\
     (1 \leftrightarrow 2)
     & =
     i (1 + \delta_{12}) (-i\kappa)^2
     \int \! \frac{\dd^4 k}{(2 \pi)^4} \,
         \frac{i}{k^2 - m_2^2 + i\epsilon} \,
         2 \pi \, n(\lvert p_0 - k_0 \rvert) \,
         \delta\big((p-k)^2 - m_1^2\big).
\end{align}
  Since the two integrals differ in structure only by a substitution of variables (\mbox{$q = p - k$}) the second integral has simply been denoted as \mbox{$(1 \leftrightarrow 2)$}. This notation implies a na\"ive interchange of indices 1 and 2. The full mixed contribution is
\begin{align}
     i & F_\text{SS}^{(2)}(p; m_1, m_2)
     +
     (1 \leftrightarrow 2)
     =
     i (1 + \delta_{12}) (i\kappa)^2
     \int \! \frac{\dd^4 k}{(2 \pi)^4} \cross
         \nonumber
         \\ & \cross \! \Bigg[
         \frac{i}{(p\!-\!k)^2 - m_1^2 + i\epsilon} \,
         n(\lvert k_0 \rvert) \, 2\pi \,
         \delta\big( k^2 \!\!-\! m_2^2 \big)
         +%\\ & \hspace{3.2cm} +
         \frac{i}{k^2 - m_2^2 + i\epsilon} \,
         n(\lvert p_0 \!-\! k_0 \rvert) \, 2\pi \,
         \delta\big( (p\!-\!k)^2 - m_1^2 \big)
     \Bigg]
\end{align}
  The integrals are evaluated through the Sochocki-Plemelj theorem\footnote{This theorem in its version for the real number line often appears in physics but is rarely mentioned by name. It was proven by Julian Sochocki in 1868 and rediscovered by Josip Plemelj in 1908.} which requires a smooth closed simple curve $C$ in the complex plane. Splitting the Feynman propagator and the \mbox{$\delta$-functions} results is
\begin{align}
     i & F_\text{SS}^{(2)}(p; m_1, m_2)
     +
     (1 \leftrightarrow 2)
     =
     \nonumber
     \\ & =
     (1 + \delta_{12}) \kappa^2
     \int \! \frac{\dd^4 k}{(2 \pi)^3} \,
         n(\lvert k_0 \rvert)
         \cross \nonumber
         \\ & \cross
         \Bigg[
             \mathcal P \left(
                 \frac{1}{(p_0-k_0)^2 - \omega_{1, p-k}^2}
             \right)
             \frac{1}{2\omega_{2, k}}
             \Big\{
                 \delta(k_0 - \omega_{2, k})
                 +
                 \delta(k_0 + \omega_{2,k})
             \Big\}
             -
             \nonumber
             \\ & \qquad -
             \frac{i\pi}{4 \omega_{1, p-k} \omega_{2, k}}
             \Big\{
                 \delta\big( k_0 - \omega_{2, k} \big)
                 +
                 \delta\big( k_0 + \omega_{2, k} \big)
             \Big\}
             \Big\{
                 \delta\big( (p_0 \!-\! k_0) - \omega_{1, p-k} \big)
                 +
                 \delta\big( (p_0 \!-\! k_0) + \omega_{1, p-k} \big)
             \Big\}
             \nonumber
             \\ & \quad +
             (1 \leftrightarrow 2)
         \Bigg].
\end{align}
  Thus, the real and imaginary parts of the loop has been separated and are written out below for clarity.

  The real part:
\begin{align}
     & \real\ \big[
         iF_\text{SS}^{(2)}(p; m_1, m_2)
         +
         (1 \leftrightarrow 2)
     \big]
     =
     \nonumber
     \\ & =
     (1 + \delta_{12}) \kappa^2
     \int \! \frac{\dd^4 k}{(2 \pi)^3} \,
         n(\lvert k_0 \rvert)
         \Bigg[
             \frac{1}{4 \omega_{1, p-k} \omega_{2, k}}
             \bigg\{
                 \mathcal P
                     \frac{1}{(p_0 - k_0) - \omega_{1, p-k}}
                 -
                 \mathcal P
                     \frac{1}{(p_0 - k_0) + \omega_{1, p-k}}
             \bigg\}
             \nonumber
             \\ & \hspace{6.0cm} \cross
             \bigg\{
                 \delta(k_0  - \omega_{2, k})
                 +
                 \delta(k_0  + \omega_{2, k})
             \bigg\}
             +
             (1 \leftrightarrow 2)
         \Bigg],
\end{align}
and the imaginary part of the mixed contribution to the loop:
\begin{align}
\label{eq:ImaginaryScalarBubble}
     & \imaginary\ \big[
         iF_\text{SS}^{(2)}(p; m_1, m_2)
         +
         (1 \leftrightarrow 2)
     \big]
     =
     \nonumber
     \\ & =
     - (1 + \delta_{12}) \pi \kappa^2
     \int \! \frac{\dd^4 k}{(2 \pi)^3} \,
         n(\lvert k_0 \rvert)
         \Bigg[
             \frac{1}{4 \omega_{1, p-k} \omega_{2, k}}
             \bigg\{
                 \delta\big( (p_0 - k_0)  - \omega_{1, p-k} \big)
                 +
                 \delta \big( (p_0 - k_0)  + \omega_{1, p-k} \big)
             \bigg\}
             \nonumber
             \\ & \hspace{6.5cm} \cross
             \bigg\{
                 \delta(k_0  - \omega_{2, k})
                 +
                 \delta(k_0  + \omega_{2, k})
             \bigg\}
             +
             (1 \leftrightarrow 2)
	\Bigg].
\end{align}

%%%%%%%%%%%%%%%%%%%%%%%%%%%%%%%%%%%%%%%%%%%%%%%%%%%%%%%%%%%%%%%%
\subsubsection{The imaginary part}
%%%%%%%%%%%%%%%%%%%%%%%%%%%%%%%%%%%%%%%%%%%%%%%%%%%%%%%%%%%%%%%%

  This subsection considers the imaginary part of the \mbox{$iF_\text{SS}^{(2)}$-integral} on the previous page \\ (\eq{ImaginaryScalarBubble}). $k_0$ may immediately be integrated and each of the four resulting $\delta$-functions specify an energy domain together with the physical condition \mbox{$p_0 > 0$}.
\begin{align}
     \delta(p_0 & - \omega_{2, k} - \omega_{1, p-k})
     \qquad \Rightarrow \qquad
     p_0 = \omega_{2, k} + \omega_{1, p-k}
     \nonumber
     \\ \Rightarrow & \qquad
     p^2
     =
     \abs{\vb k}^2 + m_2^2
     +
     \abs{\vb p - \vb k}^2 + m_1^2
     +
     2\sqrt{\abs{\vb k}^2 + m_2^2}
     \sqrt{\abs{\vb p - \vb k}^2 + m_1^2}
     -
     \abs{\vb p}^2
     \nonumber
     \\ & \qquad \phantom{\Rightarrow} \geq
     (m_1 + m_2)^2
\end{align}
  Similarly
\begin{align}
     \delta(p_0 & - \omega_{2, k} + \omega_{1, p-k})
     \qquad \Rightarrow \qquad
     p_0 = \omega_{2, k} - \omega_{1, p-k}, \quad p_0 \geq 0
     \Rightarrow m_2 \geq m_1
     \nonumber
     \\ \Rightarrow & \qquad
     p^2 \leq (m_2 - m_1)^2
     \text{ if } m_2 \geq m_1\text{, otherwise no solutions.}
\end{align}
\begin{align}
     \delta(p_0 & + \omega_{2, k} - \omega_{1, p-k})
     \qquad \Rightarrow \qquad
     p_0 = -\omega_{2, k} + \omega_{1, p-k},
     \quad p_0 \geq 0 \Rightarrow m_1 \geq m_2
     \nonumber
     \\ \Rightarrow & \qquad
     p^2 \leq (m_2 - m_1)^2
     \text{ if } m_2 \leq m_1\text{, otherwise no solutions.}
\end{align}
  Note that no contributions can come from
\beq
     \delta(p_0 + \omega_{2, k} + \omega_{1, p-k})
     \qquad \Rightarrow \qquad
     p_0 = -\omega_{2, k} - \omega_{1, p-k}.
\eeq
  since the energy domain is forbidden.

  For simplicity, consider $\vb p$ aligned with the $z$-axis. Then
\beq
     \vb p \cdot \vb k = \abs{\vb p} \abs{\vb k} \cos{\theta}
\eeq
  and n polar coordinates
\beq
     \int_0^\infty \! \dd \abs{\vb k} \,
         \abs{\vb k}^2
         \int_0^\pi \! \dd \theta \,
             \sin{\theta}
             \int_0^{2\pi} \! \dd \varphi \,
     \to
     2\pi
     \int_{k_-^{,',''}}^{k_+^{,',''}} \! \dd \abs{\vb k} \,
         \abs{\vb k}^2
         \int_{-1}^1 \! \dd \big(\cos \theta \big) \,.
\eeq
  The limits $k_+^{,',''}$, $k_-^{,',''}$ are given by restrictions from the $\delta$-functions on $\vb k$ listed above.

  Now, the integral of interest may be recast on this polar form:
\begin{align}
     \imaginary\ & \big[
         F_\text{SS}^{(2)}(p; m_1, m_2)
         +
         (1 \leftrightarrow 2)
     \big]
     =
     \nonumber
     \\ & =
     -(1 + \delta_{12}) \frac{\pi \kappa^2}{(2\pi)^2}
     \int_0^\infty \!\!\! \dd \abs{\vb k} \, \abs{\vb k}^2
         \int_{-1}^1 \! \! \dd \big( \cos \theta \big) \,
             \Bigg[
                 \frac{n(\omega_{2, k})}{
                     4 \omega_{1, p-k} \omega_{2, k}
                 }
                 \sum_{r, s = \pm}
                     \delta\big( f^{(rs)}(\cos \theta) \big)
                 +
                 (1 \! \leftrightarrow \! 2)
             \Bigg].
\end{align}	
  where the function \mbox{$f^{(rs)}(x)$} was defined as
\beq
     f^{(rs)}(x)
     =
     p_0 + r\omega_{2,k} + s\omega_{1, p-k}
     =
     p_0
     +
     r\sqrt{\abs{\vb k}^2 + m_2^2}
     +
     s\sqrt{
         \abs{\vb p}^2 + \abs{\vb k}^2
         -
         2\abs{\vb p}\abs{\vb k} \cdot x + m_1^2
     }
\eeq
  with \mbox{$r, s = \pm 1$}. Now, simplify the $\delta$-function using
\beq
     \delta\big(f^{(rs)}(x)\big)
     =
     \sum_\text{roots}
         \frac{\delta(x - x_\text{root})}
           {\abs{\big( f^{(rs)}(x_\text{root}) \big)'}}
\eeq
  where the roots $x_\text{root}$ are points $x_0$ so that \mbox{$f^{(rs)}(x_0) = 0$}. Solving for $x_0$ gives
\beq
     x_0^{(r)}
     =
     \frac{-p^2 - r2p_0\omega_{2, k} + m_1^2 - m_2^2}
       {2 \abs{\vb p} \abs{\vb k}}.
\eeq

  Contributions to the imaginary part may only come from regions of $p^2$ when the $\delta$-functions contribute, i.e. when \mbox{$x - x_0^{(r)} = 0$}, \mbox{$x \in [-1, 1]$}. This restriction limits the integration interval of $\abs{\vb k}$ through
\beq
     -1 \leq x_0^{(r)}\big( \abs{\vb k} \big) \leq 1.
\eeq
  Plug in the roots $x_0^{(r)}$ so that the inequalities become
\beq
     \frac{p^2 - m_1^2 + m_2^2 - 2\abs{\vb p}\abs{\vb k}}{2 p_0}
     \leq
     -r\omega_{2, k}
     \leq
     \frac{p^2 - m_1^2 + m_2^2 + 2\abs{\vb p}\abs{\vb k}}
       {2 p_0}.
\eeq
  The end points of the interval are of primary interest as to determine the limits of integration and one may consider specifically the values of $\abs{\vb k}$ that fulfill the equalities, labelled $\abs{\vb k}_0$. Separately, the two equalities may then be squared and stated as
\beq
     \begin{dcases}
         \frac{\big( p^2 - m_1^2 + m_2^2 \big)^2}{4 p_0^2}
         -
         m_2^2
         -
         \abs{\vb k}_0
         \frac{\abs{\vb p}}{p_0^2}
         \big( p^2 - m_1^2 + m_2^2 \big)
         +
         \abs{\vb k}_0^2
         \left( \frac{\abs{\vb p}^2}{p_0^2} - 1  \right)
         =
         0
         \\
         \frac{\big( p^2 - m_1^2 + m_2^2 \big)^2}{4 p_0^2}
         -
         m_2^2
         +
         \abs{\vb k}_0
         \frac{\abs{\vb p}}{p_0^2}
         \big( p^2 - m_1^2 + m_2^2 \big)
         +
         \abs{\vb k}_0^2
         \left( \frac{\abs{\vb p}^2}{p_0^2} - 1 \right)
         =
         0
     \end{dcases}
\eeq
  The four solutions are
\beq
     \abs{\vb k}_{0, \pm}^{(\pm)}
     =
     (\pm) \frac{\abs{\vb p}}{2}
     \left( 1 + \frac{m_2^2 - m_1^2}{p^2} \right)
     \pm
     \frac{p_0}{2} \sqrt{C}.
\eeq
  These resulting limits on the polar integral for the imaginary part of \mbox{$iF^{(2)}(p; m_1, m_2) + (1 \leftrightarrow 2)$}, indicated above as $k_+^{,',''}$, $k_-^{,',''}$. The limits are $p^2$-dependent and contributions may come from any of the three $\delta$-functions in the integral over \mbox{$\cos{\theta}$}. In \mbox{Tab.~\ref{tab:MomentumRegionsForDeltaFunctions}}, the different regions are presented.
%\afterpage{
%\clearpage
%\thispagestyle{empty}
%\begin{turnpage}
\begin{table}[h!]
\centering
\captionof{table}{Integration limits on the radial variable $\abs{\vb k}$ are displayed below. The total integral of interest contains three terms, each proportional to a $\delta$-function of different arguments. The $\delta$-functions give rise to integration limits on the radial coordinate and labelled accordingly as \mbox{$k_\pm^{,',''}$} to distinguish the three contributions. The integration limits for \mbox{$0 \leq p^2 \leq (m_2 - m_1)^2$} as \mbox{$m_2 < m_1$} and the limits in parenthesis for \mbox{$p^2 < 0$} as \mbox{$m_2 > m_1$} correspond to the case when \mbox{$1 + \frac{m_2^2 - m_1^2}{p^2} < 0$}. Empty entries for the limits imply no contribution from the $\delta$-function in this region.
\label{tab:MomentumRegionsForDeltaFunctions}}
\begin{tabular}{@{}ccccc@{}} \toprule
     \multicolumn{2}{c}{Integration region}	& \multicolumn{3}{c}{Limits} \\ \cline{1-2}\cline{3-5} %\cmidrule(r){1-2}\cmidrule(l){3-5}
     Momentum
     &
     Mass-ordering
     &
     $\delta(p_0 - \omega_{2, k} - \omega_{1, p-k})$
     &
     $\delta(p_0 - \omega_{2, k} + \omega_{1, p-k})$
     &
     $\delta(p_0 + \omega_{2, k} - \omega_{1, p-k})$
     \\ \colrule %\midrule
     $p^2 \geq (m_2 + m_1)^2$	 & any	& $k_+ = \abs{\vb k}_{0,+}^{(+)}$	& 	&
     \\
     & 	& $k_- = \abs{\vb k}_{0,+}^{(-)}$	& 	&
     \\
     $0 \leq p^2 \leq (m_2 - m_1)^2$	& $m_2 > m_1$	& 	& $k_+' = \abs{\vb k}_{0,+}^{(+)}$	&
     \\
     & 	& 	& $k_-' = \abs{\vb k}_{0,+}^{(-)}$ 	&
     \\
     & $m_2 < m_1$	& 	& 	& $k_+'' = \abs{\vb k}_{0,+}^{(-)}$
     \\
     & 	& 	& 	& $k_-'' = \abs{\vb k}_{0,+}^{(+)}$
     \\
     $p^2 < 0$	& $m_2 > m_1$	& 	& $k_+' = +\infty$	& $k_+'' = +\infty$
     \\
     & 	& 	& $k_-' = \abs{\vb k}_{0,+}^{(+)} \, \Big( \abs{\vb k}_{0, -}^{(-)} \Big)$	& $k_-'' = \abs{\vb k}_{0,+}^{(-)} \, \Big( \abs{\vb k}_{0, -}^{(+)} \Big)$
     \\
     & $m_2 < m_1$	& 	& $k_+' = +\infty$	& $k_+'' = +\infty$
     \\
     & 	& 	& $k_-' = \abs{\vb k}_{0,+}^{(+)}$	& $k_-'' = \abs{\vb k}_{0,-}^{(+)}$
     \\ \botrule %\bottomrule
\end{tabular}
%\end{turnpage}
%\clearpage% Flush page
%}
\end{table}

  The integration limits $k_+^{,',''}$, $k_-^{,',''}$ must be positive in order to apply as limits for the polar coordinate $\abs{\vb k}$ and the modulus of the limits may be considered in order to express the primed limits in terms of the unprimed.
  
  First consider \mbox{$p^2 \geq (m_2 + m_1)^2$}:
\beq
     k_\pm
     \equiv
     \abs{\vb k}_{0,+}^\pm
     =
     \frac{1}{2}
     \abs{
         \abs{\vb p}
         \abs{ 1 + \frac{m_2^2 - m_1^2}{p^2} }
         \pm
         p_0 \sqrt{C}
     }.
\eeq

  Secondly consider \mbox{$0 \leq p^2 \leq (m_2 - m_1)^2$}. For \mbox{$m_2 > m_1$} \tab{MomentumRegionsForDeltaFunctions} shows that \mbox{$k_\pm' = k_\pm$} (the limit from the previous case). For \mbox{$m_2 < m_1$}:
\beq
     k_\pm''
     =
     \frac{1}{2}
     \abs{
         \mp
         \abs{\vb p}
         \left( 1 + \frac{m_2^2 - m_1^2}{p^2} \right)
         +
         p_0 \sqrt{C}
     }
     =
     \frac{1}{2}
     \abs{
         \pm
         \abs{\vb p}
         \abs{ 1 + \frac{m_2^2 - m_1^2}{p^2} }
         +
         p_0 \sqrt{C}
     }
     =
     k_\pm.
\eeq

  In the region where \mbox{$p^2 < 0$}, two further cases in the mass hierarchy of \mbox{$m_2 > m_1$} exists. $k_+$ is determined to be \mbox{$+\infty$} and
\begin{align}
     k_-'
     & =
     \begin{cases}
         k_+, \quad 1 + \tfrac{m_2^2 - m_1^2}{p^2} > 0,
         \\ \\
         k_-, \quad 1 + \tfrac{m_2^2 - m_1^2}{p^2} < 0,
     \end{cases}
     , \qquad
     k_-''
     & =
     \begin{cases}
     k_-, \quad 1 + \tfrac{m_2^2 - m_1^2}{p^2} > 0,
     \\ \\
     k_+, \quad 1 + \tfrac{m_2^2 - m_1^2}{p^2} < 0.
     \end{cases}
\end{align}

  Finally, for \mbox{$m_2 < m_1$}, \mbox{$k_-' = k_+$} and \mbox{$k_-'' = k_-$}. With the polar integration limits given by integration over $\delta\big(f^{(rs)}(\cos(\theta))\big)$ identified, mixed contributions \mbox{$\imaginary\ \big[ iF_\text{SS}^{(2)}(p; m_1, m_2)(p; m_1, m_2) + (1 \leftrightarrow 2)\big]$} may be evaluated.

%%%%%%%%%%%%%%%%%%%%%%%%%%%%%%%%%%%%%%%%%%%%%%%%%%%%%%%%%%%%%%%%
\subsubsection{\texorpdfstring{$p^2 \geq (m_2 + m_1)^2$}{TEXT}:}
%%%%%%%%%%%%%%%%%%%%%%%%%%%%%%%%%%%%%%%%%%%%%%%%%%%%%%%%%%%%%%%%

  Contributions to this region of momentum come only from \mbox{$\delta\big( p_0 - \omega_{2, k} - \omega_{1, p-k} \big)$} giving integration limits on \mbox{$\abs{\vb k}$} as shown in \tab{MomentumRegionsForDeltaFunctions}.
\begin{align}
     \imaginary\ & \big[
         i F_{\text{SS}}^{(2)}(p; m_1, m_2)
         +
         (1 \leftrightarrow 2)
     \big]
     \nonumber
     = \\ & =
     -(1 + \delta_{12}) \pi\kappa^2
     \int \! \frac{\dd^3 k}{(2\pi)^3} \,
         \frac{n(\omega_{2, k})}
           {4 \omega_{1, p-k} \omega_{2, k}}
         \delta\big(p_0 - \omega_{2, k} - \omega_{1, p-k} \big)
     +
     (1 \leftrightarrow 2)
     \nonumber
     \\ & \hspace{-7pt} \overset{r = -1}{=} \!\!\!
     -\frac{(1 + \delta_{12}) \pi\kappa^2}{4}
     \int_{0}^{\infty} \!
         \frac{\dd \abs{\vb k} \, \abs{\vb k}^2}{(2\pi)^2} \,
         \frac{n(\omega_{2, k})}
           {\omega_{1, p-k} \omega_{2, k}} \!
         \int_{-1}^1 \! \dd x \,
             \frac{\abs{p_0 - \omega_{2, k}}}
               {\abs{\vb p}\abs{\vb k}}
             \delta\Big(x - x_0^{(-)}\Big)
     +
     (1 \leftrightarrow 2).
\end{align}
  Note that the $\delta$-function imposes the condition
  $p_0 - \omega_{2, k} = \omega_{1, p-k}$.
  Hence
\beq
     \imaginary\ \big[
         iF_\text{SS}^{(2)}(p; m_1, m_2)
         +
         (1 \leftrightarrow 2)
     \big]
     =
     -\frac{(1 + \delta_{12}) \kappa^2}{16 \pi \abs{p}}
     \int_{k_-}^{k_+} \! \dd \abs{\vb k} \,
         \frac{\abs{\vb k}}{\omega_{2, k}}
         \frac{1}{e^{\beta\omega_{2, k}} - 1}
     +
     (1 \leftrightarrow 2),
\eeq
  which may be recast and solved as
\begin{align}
     \imaginary\ \big[
         iF_\text{SS}^{(2)}(p; m_1, m_2) + (1 \leftrightarrow 2)
     \big]
     & =% \\ & =
     -(1 + \delta_{12})\frac{\kappa^2}{16 \pi \abs{p} \beta}
     \int_{y_-}^{y_+} \! \dd y \,
         \frac{1}{e^{y} - 1}
     +
     (1 \leftrightarrow 2)
     \\ & =
     -(1 + \delta_{12}) \frac{\kappa^2}{16 \pi \abs{p} \beta}
     \ln{\left\lvert
         \frac{1 - e^{-\beta\omega_{2, k_+}}}
           {1 - e^{-\beta\omega_{2, k_-}}}
     \right\rvert}
     +
     (1 \leftrightarrow 2).
\end{align}
  The final limits $\omega_{2, k_\pm}$:
\begin{align}
     \omega_{2, k_\pm}
     & =
     \abs{
         \pm \frac{\abs{\vb p}}{2} \sqrt{C}
         +
         \frac{p_0}{2}
         \abs{1 + \frac{m_2^2 - m_1^2}{p2}}
     }.
\end{align}

%%%%%%%%%%%%%%%%%%%%%%%%%%%%%%%%%%%%%%%%%%%%%%%%%%%%%%%%%%%%%%%%
\subsubsection{\texorpdfstring{$(m_2 - m_1)^2 \leq p^2 \leq (m_2 + m_1)^2$}{TEXT}:}
%%%%%%%%%%%%%%%%%%%%%%%%%%%%%%%%%%%%%%%%%%%%%%%%%%%%%%%%%%%%%%%%

  This region of $p^2$ gets no contribution to the imaginary part of the loop.
\beq
     \imaginary\ \big[
         iF_\text{SS}^{(2)}(p; m_1, m_2) + (1 \leftrightarrow 2)
     \big]
     =
     0.
\eeq

%%%%%%%%%%%%%%%%%%%%%%%%%%%%%%%%%%%%%%%%%%%%%%%%%%%%%%%%%%%%%%%%
\subsubsection{\texorpdfstring{$0 \leq p^2 \leq (m_2 - m_1)^2$}{TEXT}:}
%%%%%%%%%%%%%%%%%%%%%%%%%%%%%%%%%%%%%%%%%%%%%%%%%%%%%%%%%%%%%%%%

  The contribution to the loop in the momentum region
  $0 \leq p^2 \leq (m_2 - m_1)^2$ comes from either
  \mbox{$\delta\big(p_0 - \omega_{2, k} + \omega_{1, p-k} \big)$} with \mbox{$r=-1$}
  or
  \mbox{$\delta\big(p_0 + \omega_{2, k} - \omega_{1, p-k} \big)$}
  with with \mbox{$r=+1$}, depending on the mass hierarchy in this region of $p^2$. It turns out, however, that the final result is independent of the mass ordering:
\beq
     \imaginary\ \big[
         iF_\text{SS}^{(2)}(p; m_1, m_2) + (1 \leftrightarrow 2)
     \big]
     =
     -(1 + \delta_{12})
     \frac{\kappa^2}{16\pi \abs{\vb p} \beta}
     \ln{\left\lvert
         \frac{1 - e^{-\beta\omega_{2, k_+}}}
           {1 - e^{-\beta\omega_{2, k_-}}}
     \right\rvert}
     +
     (1 \leftrightarrow 2).
\eeq
Note that any explicit dependence on the polar angle ($\theta$) (e.g. resulting from any Lorentz structure of the numerator) would give a result dependent on the mass orderings.

%%%%%%%%%%%%%%%%%%%%%%%%%%%%%%%%%%%%%%%%%%%%%%%%%%%%%%%%%%%%%%%%
\subsubsection{\texorpdfstring{$p^2 < 0$}{TEXT}:}
%%%%%%%%%%%%%%%%%%%%%%%%%%%%%%%%%%%%%%%%%%%%%%%%%%%%%%%%%%%%%%%%

  Contributions to this region of momentum comes from the two $\delta$-functions
  \mbox{$\delta\big(p_0 - \omega_{2, k} + \omega_{1, p-k} \big)$}
  and
  \mbox{$\delta\big(p_0 + \omega_{2, k} - \omega_{1, p-k} \big)$}.
  Consider initially \mbox{$m_2 > m_1$} with the following two possible cases:
\beq
     1 + \frac{m_2^2 - m_1^2}{p^2} > 0
     \text{ or }
     1 + \frac{m_2^2 - m_1^2}{p^2} < 0.
\eeq
  The limits of integration will be different for the two cases.
\begin{align}
     & \imaginary\ \big[
         iF_\text{SS}^{(2)}(p; m_1, m_2) + (1 \leftrightarrow 2)
     \big]
     \nonumber
     = \\ & =
     -(1 + \delta_{12}) \pi\kappa^2
     \int \! \frac{\dd^3 k}{(2\pi)^3} \,
         \frac{n(\omega_{2, k})}
           {4 \omega_{1, p-k} \omega_{2, k}}
         \Big[
             \delta\big(
                 p_0 - \omega_{2, k} + \omega_{1, p-k}
             \big)
             +
             \delta\big(
                 p_0 + \omega_{2, k} - \omega_{1, p-k}
             \big)
         \Big]
         +
         (1 \leftrightarrow 2)
     \nonumber
     \\ & =
     (1 + \delta_{12}) \frac{\kappa^2}{16\pi \abs{\vb p} \beta}
     \ln{\abs{
         \left( 1 - e^{-\beta\omega_{2, k_+}} \right)
         \!
         \left( 1 - e^{-\beta\omega_{2, k_-}} \right)
     }}
     +
     (1 \leftrightarrow 2).
\end{align}
  The above expression holds also for the case \mbox{$m_2 < m_1$} because of the symmetry in the above result.

%%%%%%%%%%%%%%%%%%%%%%%%%%%%%%%%%%%%%%%%%%%%%%%%%%%%%%%%%%%%%%%%
\subsubsection{The real part}
%%%%%%%%%%%%%%%%%%%%%%%%%%%%%%%%%%%%%%%%%%%%%%%%%%%%%%%%%%%%%%%%

  The real part of the mixed term in the bubble may be evaluated analytically up to a final integration. Utilise $\delta$-functions over $k_0$ and combine fractions of principal values to get
\begin{align}
     & \real\ \big[
         iF_\text{SS}^{(2)}(p; m_1, m_2)
         +
         (1 \leftrightarrow 2)
     \big]
     =
     \nonumber
     \\ & =
     (1 + \delta_{12}) \kappa^2
     \int_0^\infty \!
         \frac{\dd \abs{\vb k} \abs{\vb k}^2}{(2 \pi)^2} \,
         \frac{n(\omega_{2, k})}{2\omega_{2, k}}
         \int_{-1}^1 \! \dd x
             \bigg(
                 \mathcal P \frac{1}{
                     (p_0
                     - \omega_{2, k})^2 - \omega_{1, p-k}^2
                 }
                 +
                 \mathcal P \frac{1}{
                     (p_0
                     +
                     \omega_{2, k})^2 \!-\! \omega_{1, p-k}^2
                 }
             \bigg)
     \nonumber
     \\ & \qquad +
     (1 \leftrightarrow 2).
\end{align}
  Again, taking $\vb p$ in the direction of the $z$-axis, the denominators may be fully expanded and the inner integral(s) to be evaluated:
\beq
     \int_{-1}^1 \! \dd x
         \frac{1}{(p_0 \pm \omega_{2, k})^2 - \omega_{1, p-k}^2}
     =
     \frac{1}{2\abs{\vb p}\abs{\vb k}}
     \ln{\abs{\frac{
         p^2 \pm 2p_0\omega_{2, k} + 2\abs{\vb p}\abs{\vb k}
         +
         m_2^2 - m_1^2
     }{
         p^2 \pm 2p_0\omega_{2, k} - 2\abs{\vb p}\abs{\vb k}
         +
         m_2^2 - m_1^2
     }}}.
\eeq
  Define
\beq
     D_{(\pm)}^{\pm}(p^2; m_1, m_2)
     =
     p^2 \pm 2p_0\omega_{2, k}
     \,(\pm)\, 2\abs{\vb p}\sqrt{\omega_{2,k}^2 - m_2^2}
     +
     m_2^2 - m_1^2
\eeq
  and plug this into the integral for the real part we get
\begin{align}
     & \real\ \big[
         iF_\text{SS}^{(2)}(p; m_1, m_2)
         +
         (1 \leftrightarrow 2)
     \big]
     =
     \nonumber
     \\ & =
     (1 + \delta_{12}) \frac{\kappa^2}{16\pi^2\abs{\vb p}}
     \int_{m_2}^\infty \! \dd \omega_{2, k} \,
         n(\omega_{2, k})
         \mathcal P \ln{\abs{\frac{
             D_{(+)}^{-}(p^2; m_1, m_2)
             D_{(+)}^{+}(p^2; m_1, m_2)
         }{
             D_{(-)}^{-}(p^2; m_1, m_2)
             D_{(-)}^{+}(p^2; m_1, m_2)
         }}}
     +
     (1 \leftrightarrow 2).
\end{align}

%%%%%%%%%%%%%%%%%%%%%%%%%%%%%%%%%%%%%%%%%%%%%%%%%%%%%%%%%%%%%%%%
\subsection{\texorpdfstring{$iF_\text{SS}^{(3)}$}{TEXT}-integral}
\label{sec:pureThermalExlicitIntegration}
%%%%%%%%%%%%%%%%%%%%%%%%%%%%%%%%%%%%%%%%%%%%%%%%%%%%%%%%%%%%%%%%

  The fourth thermal integral of the loop is the product of the thermal terms from each of the propagators.
\begin{align}
     iF_\text{SS}^{(3)}(p; m_1, m_2)
     & =
     i (1 + \delta_{12}) (-i\kappa)^2
     \int \! \frac{\dd^4 k}{(2 \pi)^4} \,
         2 \pi \, n(\lvert p_0 - k_0 \rvert) \,
         \delta\big((p-k)^2 - m_1^2\big)
         2 \pi \, n(\lvert k_0 \rvert) \,
         \delta\big(k^2 - m_2^2\big)
     \nonumber
     \\ & =
     -i (1 + \delta_{12}) \frac{\kappa^2}{4}
     \int \! \frac{\dd^3 k}{(2 \pi)^2} \,
         \frac{n(\omega_{2, k})}{\omega_{1, p-k}\omega_{2, k}}
         \cross
         \nonumber
         \\ & \qquad \qquad \cross
         \Big[
             n(p_0 - \omega_{2, k})
             \big\{
                 \delta(p_0 - \omega_{2, k} - \omega_{1, p-k})
                 +
                 \delta(p_0 - \omega_{2, k} + \omega_{1, p-k})
             \big\}
             \nonumber
             \\ & \qquad \qquad \,\,\,\, +
             n(p_0 + \omega_{2, k})
             \big\{
                 \delta(p_0 + \omega_{2, k} - \omega_{1, p-k})
                 +
                 \delta(p_0 + \omega_{2, k} + \omega_{1, p-k})
             \big\}
         \Big].
\end{align}
  Note that the right-hand side is purely imaginary. The fourth term above proportional to the integrand
  \mbox{$\delta(p_0 + \omega_{2, k} + \omega_{1, p-k})$}
  may be discarded due to the physical condition \mbox{$p_0 \geq 0$}.

%%%%%%%%%%%%%%%%%%%%%%%%%%%%%%%%%%%%%%%%%%%%%%%%%%%%%%%%%%%%%%%%
\subsubsection{\texorpdfstring{$p^2 \geq (m_2 + m_1)^2$}{TEXT}:}
%%%%%%%%%%%%%%%%%%%%%%%%%%%%%%%%%%%%%%%%%%%%%%%%%%%%%%%%%%%%%%%%

  Contributions come from the term containing
  \mbox{$\delta(p_0 - \omega_{2, k} - \omega_{1, p-k})$}.
\begin{align}
     \imaginary\ \big[ iF_\text{SS}^{(3)}(p; m_1, m_2) \big]
     & =
     -(1 + \delta_{12}) \frac{\kappa^2}{4}
     \int \! \frac{\dd^3 k}{(2 \pi)^2} \,
         \frac{
             n(\omega_{2, k}) n(\abs{p_0 - \omega_{2, k}})
         }{\omega_{1, p-k}\omega_{2, k}}
         \delta(p_0 - \omega_{2, k} - \omega_{1, p-k})
     \nonumber
     \\ & =
     -(1 + \delta_{12}) \frac{\kappa^2}{8\pi\abs{\vb p}\beta}
     \int_{y_-}^{y_+} \! \dd y \,
         \frac{1}{e^y - 1}
         \frac{1}{e^{\abs{\beta p_0 - y}} - 1}.
\end{align}

  The $\delta$-function provides an important inequality:
\beq
     \beta p_0 \geq \beta \omega_{2, k} = y
\eeq
  Hence, \mbox{$\abs{\beta p_0  - y} = \beta p_0  - y$} which is is expected due to the essential singularity at
  \mbox{$y = \beta p_0$}. (If the range of integration included this point there would be no hope of convergence.)
  
  Apply
\beq
     \frac{1}{e^y - 1}
     \frac{1}{e^{\beta p_0 - y} - 1}
     =
     \frac{1}{e^{\beta p_0} - 1} \left[
         1
         +
         \frac{1}{e^y - 1}
         +
         \frac{1}{e^{\beta p_0 - y} - 1}
     \right]
\eeq
  in order to evaluate the integral:
\beq
     \imaginary\ \big[ iF_\text{SS}^{(3)}(p; m_1, m_2) \big]
     =
     -(1 + \delta_{12}) \frac{\kappa^2}{8\pi\abs{\vb p}\beta}
     \frac{1}{e^{\beta p_0} - 1}
     \ln{\abs{\frac{
         \big( 1 - e^{-\beta\omega_{2, k_+}} \big)
         \big( e^{\beta(p_0 - \omega_{2, k_-})} - 1 \big)
     }{
         \big( 1 - e^{-\beta\omega_{2, k_-}} \big)
         \big( e^{\beta(p_0 - \omega_{2, k_+})} - 1 \big)
     }}}.
\eeq

%%%%%%%%%%%%%%%%%%%%%%%%%%%%%%%%%%%%%%%%%%%%%%%%%%%%%%%%%%%%%%%%
\subsubsection{\texorpdfstring{$(m_2 - m_1)^2 \leq p^2 \leq (m_2 + m_1)^2$}{TEXT}:}
%%%%%%%%%%%%%%%%%%%%%%%%%%%%%%%%%%%%%%%%%%%%%%%%%%%%%%%%%%%%%%%%

  None of the $\delta$-functions contributes to the region of momentum where $(m_2 - m_1)^2 \leq p^2 \leq (m_2 + m_1)^2$.
\beq
     \imaginary\ \big[ iF_\text{SS}^{(3)} (p; m_1, m_2) \big] = 0
\eeq

%%%%%%%%%%%%%%%%%%%%%%%%%%%%%%%%%%%%%%%%%%%%%%%%%%%%%%%%%%%%%%%%
\subsubsection{\texorpdfstring{$0 \leq p^2 \leq (m_2 - m_1)^2$}{TEXT}:}
%%%%%%%%%%%%%%%%%%%%%%%%%%%%%%%%%%%%%%%%%%%%%%%%%%%%%%%%%%%%%%%%

  In the momentum region where $0 \leq p^2 \leq (m_2 - m_1)^2$, two cases must be considered depending on the hierarchy of the loop masses.
  
  If \mbox{$m_2 > m_1$}, the contribution come from
  \mbox{$\delta\big( p_0 - \omega_{2, k} + \omega_{1, p-k} \big)$}
  while for \mbox{$m_2 < m_1$} the contribution come from
  \mbox{$\delta\big( p_0 + \omega_{2, k} - \omega_{1, p-k} \big)$}. The integral becomes
\beq
     \imaginary\ \big[ iF_\text{SS}^{(3)} (p; m_1, m_2) \big]
     =
     -(1 + \delta_{12}) \frac{\kappa^2}{8\pi\abs{\vb p}\beta}
     \int_{y_-}^{y_+} \! \dd y \,
         \frac{1}{e^y - 1}
         \frac{1}{e^{\abs{\beta p_0 + ry}} - 1}.
\eeq
  For $r=-1$, the $\delta$-function imposes the condition \mbox{$\beta p_0 \leq y$}. Hence the absolute value in the exponent may be replaced by an additional minus sign. The integral may then be solved using
\beq
     \frac{1}{e^y - 1}
     \frac{1}{e^{-\beta p_0 + y} - 1}
     =
     e^{-2y}
     +
     \frac{e^{-2y}}{e^y - 1}
     -
     \frac{e^{-y}}{e^{-\beta p_0 + y} - 1}
\eeq
  as
\begin{align}
     \imaginary\ & \big[ iF_\text{SS}^{(3)} (p; m_1, m_2) \big]
     =
     \nonumber
     \\ & =
     -(1 + \delta_{12}) \frac{\kappa^2}{8\pi\abs{\vb p}\beta}
     \frac{1}{e^{-\beta p_0} - 1}
     \Bigg[
         \ln{\abs{\frac{
             1 - e^{-\beta\omega_{2, k_+}}
         }{
             1 - e^{-\beta\omega_{2, k_-}}
         }}}
         -
         e^{-\beta p_0}
         \ln{\abs{\frac{
             e^{\beta (p_0 - \omega_{2, k_+})} - 1
         }{
             e^{\beta (p_0 - \omega_{2, k_-})} - 1
         }}}
     \Bigg].
\end{align}

  Secondly, consider the case of \mbox{$r=+1$}. In this case, one should take the relevant $\delta$-function as \mbox{$\delta\big( p_0 + \omega_{2, k} - \omega_{1, p-k} \big)$}, the condition imposed by this function is \mbox{$\beta p_0 + y > 0$}. The integration proceed very similar as to the first case in this momentum region using:
\beq
     \frac{1}{e^y - 1}
     \frac{1}{e^{\beta p_0 + y} - 1}
     =
     e^{-2y}
     +
     \frac{e^{-2y}}{e^y - 1}
     -
     \frac{e^{-y}}{e^{\beta p_0 + y} - 1}
\eeq
  Then
\begin{align}
     \imaginary\ & \big[ iF_\text{SS}^{(3)} (p; m_1, m_2) \big]
     =
     \nonumber
     \\ & =
     -(1 + \delta_{12}) \frac{\kappa^2}{8\pi\abs{\vb p}\beta}
     \frac{1}{e^{\beta p_0} - 1}
     \Bigg[
         \ln{\abs{\frac{
             1 - e^{-\beta\omega_{2, k_+}}
         }{
             1 - e^{-\beta\omega_{2, k_-}}
         }}}
         -
         e^{\beta p_0}
         \ln{\abs{\frac{
             e^{-\beta (p_0 + \omega_{2, k_+})} - 1
         }{
             e^{-\beta (p_0 + \omega_{2, k_-})} - 1
         }}}
     \Bigg].
\end{align}

  Note that the above result is identical to the case of \mbox{$m_2 > m_1$} with the heuristic change of sign \mbox{$p_0 \to -p_0$}.

%%%%%%%%%%%%%%%%%%%%%%%%%%%%%%%%%%%%%%%%%%%%%%%%%%%%%%%%%%%%%%%%
\subsubsection{\texorpdfstring{$p^2 \leq 0$}{TEXT}:}
%%%%%%%%%%%%%%%%%%%%%%%%%%%%%%%%%%%%%%%%%%%%%%%%%%%%%%%%%%%%%%%%

  For the momentum region where $p^2 \leq 0$, contributions come from both
  \mbox{$\delta\big( p_0 - \omega_{2, k} + \omega_{1, p-k} \big)$}
  and
  \mbox{$\delta\big( p_0 + \omega_{2, k} - \omega_{1, p-k} \big)$}. The imposed limits of integration on $\abs{\vb k}$ are different in the two terms.

  First consider a mass hierarchy of \mbox{$m_2 > m_1$}. The limits on $\abs{\vb k}$ are different in the cases of
  \mbox{$1 + \frac{m_2^2 - m_1^2}{p^2} > 0$}
  or
  \mbox{$1 + \frac{m_2^2 - m_1^2}{p^2} < 0$}. The integral becomes
\begin{align}
     \imaginary\ & \big[ iF_\text{SS}^{(3)} (p; m_1, m_2) \big]
     =
     \nonumber
     \\ & \qquad =
     -(1 + \delta_{12}) \frac{\kappa^2}{8\pi\abs{\vb p}\beta}
     \Bigg[
         \int_{y_\pm}^{\infty} \! \dd y \,
             \frac{1}{e^y - 1}
             \frac{1}{e^{-\beta p_0 + y} - 1}
         +%\\ & \qquad \qquad \qquad +
         \int_{y_\mp}^{\infty} \! \dd y \,
             \frac{1}{e^y - 1}
             \frac{1}{e^{\beta p_0 + y} - 1}
     \Bigg].
\end{align}
  Integration results in
\begin{align}
     \imaginary\ & \big[ iF^{(3)} (p; m_1, m_2) \big]
     =
     \nonumber
     \\ & =
     -(1 + \delta_{12}) \frac{\kappa^2}{8\pi\abs{\vb p}\beta}
     \Bigg[
         \frac{1}{e^{-\beta p_0} - 1}
         \bigg(
             e^{-\beta p_0}
             \ln{\abs{
                 1 - e^{\beta ( p_0 - \omega_{2, k_\pm})}
             }}
             -
             \ln{\abs{
                 1 - e^{-\beta\omega_{2, k_\pm}}
             }}
         \bigg)
         \nonumber
         \\ & \hspace{3.45cm} +
         \frac{1}{e^{\beta p_0} - 1}
         \bigg(
             e^{\beta p_0}
             \ln{\abs{
                 1 - e^{-\beta (p_0 + \omega_{2, k_\mp})}
             }}
             -
             \ln{\abs{
                 1 - e^{-\beta\omega_{2, k_\mp}}
             }}
         \bigg)
     \Bigg].
\end{align}
  If instead \mbox{$m_2 < m_1$}, note that the limits are the same as for the opposing mass hierarchy with
  \mbox{$1 + \frac{m_2^2 - m_1^1}{p^2} > 0$} and hence, the integration result is identical.

%%%%%%%%%%%%%%%%%%%%%%%%%%%%%%%%%%%%%%%%%%%%%%%%%%%%%%%%%%%%%%%%
\subsection{Summary of the results}
%%%%%%%%%%%%%%%%%%%%%%%%%%%%%%%%%%%%%%%%%%%%%%%%%%%%%%%%%%%%%%%%

  Collecting the results of the thermal loop integration presented in Secs.~\ref{sec:nonThermalExlicitIntegration}-\ref{sec:pureThermalExlicitIntegration} for all different momentum regions are presented in his subsection.

  The non-thermal contribution comes from the term
\beq
     iI_{\text{SS}}^{(2)}(p^2; m_1, m_2)
     =
     (1 + \delta_{12}) \frac{\kappa^2}{16\pi^2}
     \left[
         -\frac{1}{\tilde\epsilon}
         -
         2
         +
         \frac{x_+}{2} \ln{\left[ \frac{m_1^2}{\bar\mu^2} \right]}
         -
         \frac{x_-}{2} \ln{\left[ \frac{m_2^2}{\bar\mu^2} \right]}
         -
         I_\text{loop}
     \right].
\eeq
  Here, the last integral term is
\beq
     I_\text{loop}
     =
     \begin{dcases}
         \phantom{-} \frac{\sqrt{C}}{2}\left[
             \ln{\abs{\frac{
                 \big( \sqrt{C} + x_- \big)
                 \big( \sqrt{C} - x_+ \big)
             }{
                 \big( \sqrt{C} - x_- \big)
                 \big( \sqrt{C} + x_+ \big)
             }}}
             +
             2\pi i
         \right]
         \hspace{2.44cm} \text{for } p^2 \geq (m_2 + m_1)^2,
         \\
         -\sqrt{-C}\left[
             \arctan{\frac{x_+}{\sqrt{-C}}}
             -
             \arctan{\frac{x_-}{\sqrt{-C}}}
         \right]
         \hspace{0.75cm}
         \text{ for } (m_2 - m_1)^2 \leq p^2 \leq (m_2 + m_1)^2,
         \\
         \phantom{-} \frac{\sqrt{C}}{2}\left[
             \ln{\abs{\frac{
                 \big( \sqrt{C} + x_- \big)
                 \big( \sqrt{C} - x_+ \big)
             }{
                 \big( \sqrt{C} - x_- \big)
                 \big( \sqrt{C} + x_+ \big)
             }}}
         \right]
         \hspace{3.17cm} \,
         \text{ for } p^2 \leq (m_2 - m_1)^2,
     \end{dcases}
\eeq
  with
\beq
     x_\pm = \pm 1 + \frac{m_1^2 - m_2^2}{p^2},
     \qquad
     C
     =
     \Bigg(
         1 - \frac{(m_2 + m_1)^2}{p^2}
     \Bigg)
     \Bigg(
         1 - \frac{(m_2 - m_1)^2}{p^2}
     \Bigg).
\eeq

  The mixed thermal-non-thermal integration gives the contributions
\begin{align}
     \imaginary\ & \big[
         iF_{\text{SS}}^{(2)} (p; m_1, m_2)
         +
         (1 \leftrightarrow 2) 
     \big]
     =
     \nonumber
     \\ & =
     -(1 + \delta_{12})
     \frac{\kappa^2}{16\pi \abs{\vb p} \beta}
     \begin{dcases}
         \ln{\abs{\frac{
             1 - e^{-\beta\omega_{2, k_+}}
         }{
             1 - e^{-\beta\omega_{2, k_-}}
         }}}
         +
         (1 \leftrightarrow 2 )
         \hspace{2.45cm} \text{for } p^2 \geq (m_2 + m_1)^2
         \\ \hspace{5.97cm} \vee \,\,\,
         0 \leq p^2 \leq (m_2 - m_1)^2,
         \\ \\
         0
         \hspace{4.1cm} \text{for }
         (m_2 - m_1)^2 \leq p^2 \leq (m_2 + m_1)^2,
         \\ \\
         -\ln{\abs{
             \big(
                 1 - e^{-\beta\omega_{2, k_+}}
             \big)
             \big(
                 1 - e^{-\beta\omega_{2, k_-}}
             \big)
         }}
         +
         (1 \leftrightarrow 2)
         \hspace{1.3cm} \text{for } p^2 < 0.
     \end{dcases}
\end{align}
  and
\begin{align}
     \real\ & \big[
         iF_{\text{SS}}^{(2)}(p; m_1, m_2)
         +
         (1 \leftrightarrow 2)
     \big]
     =
     (1 + \delta_{12}) \frac{\kappa^2}{16\pi^2\abs{\vb p}}
     \int_{m_2}^\infty \! \dd \omega_{2, k} \,
         n(\omega_{2, k})
         \cross \nonumber
         \\ \cross &
         \mathcal P
         \ln{\abs{\frac{
             \big(
                 p^2 - 2p_0\omega_{2, k}
                 +
                 2\abs{\vb p} \sqrt{\omega_{2, k}^2 \!-\! m_2^2}
                 +
                 m_2^2 - m_1^2
             \big)
             \big(
                 p^2 + 2p_0\omega_{2, k}
                 +
                 2\abs{\vb p} \sqrt{\omega_{2, k}^2 \!-\! m_2^2}
                 +
                 m_2^2 - m_1^2
             \big)
         }{
             \big(
                 p^2 - 2p_0\omega_{2, k}
                 -
                 2\abs{\vb p} \sqrt{\omega_{2, k}^2 \!-\! m_2^2}
                 +
                 m_2^2 - m_1^2
             \big)
             \big(
                 p^2 + 2p_0\omega_{2, k}
                 -
                 2\abs{\vb p} \sqrt{\omega_{2, k}^2 \!-\! m_2^2}
                 +
                 m_2^2 - m_1^2
             \big)
         }}}
     \nonumber
     \\ & +
     (1 \leftrightarrow 2).
\end{align}
  Here
\beq
     \omega_{2, k_\pm}
     =
     \frac{1}{2}
     \abs{
         \pm\abs{\vb p}\sqrt{C}
         +
         p_0 \abs{1 + \frac{m_2^2 - m_1^2}{p^2}}
     }.
\eeq

  The pure thermal contribution takes the following form:
\begin{align}
     \imaginary\ & \big[ iF_{\text{SS}}^{(3)}(p; m_1, m_2) \big]
     =
     -(1 + \delta_{12}) \frac{\kappa^2}{8\pi\abs{\vb p}\beta}
     \cross
     \\ & \cross
     \begin{dcases}
         \frac{1}{e^{\beta p_0} - 1}
         \ln{\abs{
             \frac{1 - e^{-\beta\omega_{2, k_+}}}
               {1 - e^{-\beta\omega_{2, k_-}}}
             \frac{1 - e^{\beta(p_0 - \omega_{2, k_-})}}
               {1 - e^{\beta(p_0 - \omega_{2, k_+})}}
         }}
         \hspace{4.05cm} \text{for } p^2 \geq (m_2 + m_1)^2,
         \\
         0
         \hspace{8.0cm} \text{for }
         (m_2 - m_1)^2 \leq p^2 \leq (m_2 + m_1)^2,
         \\
         \frac{1}{e^{\mp\beta p_0} - 1}
         \Bigg[
             \ln{\abs{\frac{
                 1 - e^{-\beta\omega_{2, k_+}}
             }{
                 1 - e^{-\beta\omega_{2, k_-}}}
             }}
             -
             e^{\mp\beta p_0}
             \ln{\abs{\frac{
                 1 - e^{\pm\beta(p_0 \mp \omega_{2, k_+})}
             }{
                 1 - e^{\pm\beta(p_0 \mp \omega_{2, k_-})}}
             }}
         \Bigg]
         \hspace{0.8cm}
         \text{for } 0 \leq p^2 \leq (m_2 - m_1)^2
         \\ \hspace{6.93cm} \wedge \,\,\,
         \big\{
             m_2 > m_1 \,
             (\text{upper})
             \,\,\, \vee \,\,\,
             m_2 < m_1 \,
             (\text{lower})
         \big\},
         \\
         \frac{1}{e^{-\beta p_0} - 1}
         \Big[
             e^{-\beta p_0}
             \ln{\abs{
                 1 - e^{\beta(p_0 - \omega_{2, k_\pm})}
             }}
             -
             \ln{\abs{
                 1 - e^{-\beta\omega_{2, k_\pm}}
             }}
         \Big]
         \\ \quad +
         \frac{1}{e^{\beta p_0} - 1}
         \Big[
             e^{\beta p_0}
             \ln{\abs{
                 1 - e^{-\beta(p_0 + \omega_{2, k_\mp})}
             }}
             -
             \ln{\abs{
                 1 - e^{-\beta\omega_{2, k_\mp}}
             }}
         \Big]
         \hspace{3.1cm} \text{for } p^2 < 0
         \\ \hspace{4.7cm} \wedge \,\,\,
         \Big\{
             \big(
                 m_2 > m_1 \,\,\, \wedge \,\,\,
                 1 + \frac{m_2^2 - m_1^2}{p^2} > 0
             \big)
             \,\,\, \vee \,\,\,
             m_2 < m_1 \, (\text{upper}) 
             \\ \hspace{6.75cm} \vee \,\,\,
             \big(
                 m_2 > m_1 \,\,\, \wedge \,\,\,
                 1 + \frac{m_2^2 - m_1^2}{p^2} < 0
             \big) \,
             (\text{lower})
         \Big\}.
     \end{dcases}
     \nonumber
\end{align}

This section presented, in detail, the type of integrals encountered in the evaluation of thermal loop diagrams by making use of the simplest possible loop of two scalar particles. It is clear from the result that the final expressions simplify further if the two loop masses are identical. Additional complications for the procedure of integration arise when the propagators carry Lorentz structure. It should be noted, however, that both such complications and loop divergences are introduced mainly through the non-thermal integration and, therefore, the thermal terms does provide any major increase in complexity.

%%%%%
\bibliographystyle{unsrt}
\bibliography{thermalBib}

\end{document}